\documentclass[useAMS,usenatbib]{mn2e}
\usepackage{graphicx}
\usepackage{mathptmx}
\usepackage{fixltx2e}
\usepackage{xcolor}

\def \arcsec {$^{\prime\prime}$}
\def \arcmin {$^\prime$}

\def \msol  {\hbox{M$_{\odot}$}}

\def \arcmin {\hbox{$^\prime$}}                         
\def \arcsec {\hbox{$^{\prime\prime}$}}                 

\title[GBS SCUBA-2 Survey: Taurus]{The JCMT Gould Belt Survey: SCUBA-2 observations of circumstellar disks in L\,1495}
\author[The GBS Taurus Team]{J. V. Buckle$^{1, 2}$, E. Drabek-Maunder$^{3}$, J. Greaves$^{4}$, J. S. Richer$^{1, 2}$, B.C. Matthews$^{5, 6}$, \newauthor D. Johnstone$^{7, 5, 6}$,  H. Kirk$^{5}$, S.F. Beaulieu$^{8}$, D.S. Berry$^{7}$, H. Broekhoven-Fiene$^{6}$, \newauthor M.J. Currie$^{7}$, M. Fich$^{8}$, J. Hatchell$^{9}$, T. Jenness$^{7, 10}$, J.C. Mottram$^{11}$,  D. Nutter$^{12}$, \newauthor K. Pattle$^{13}$, J.E. Pineda$^{14,15\dagger}$, C. Salji$^{1, 2}$, S. Tisi$^{8}$,  J. Di Francesco$^{5, 6}$, \newauthor M.R. Hogerheijde$^{11}$, D. Ward-Thompson$^{13}$,  P. Bastien$^{16}$,  H. Butner$^{17}$, \newauthor  M. Chen$^{6}$,  A. Chrysostomou$^{18}$, S. Coude$^{16}$,   C.J. Davis$^{19}$, A. Duarte-Cabral$^{9}$,   \newauthor P. Friberg$^{7}$, R. Friesen$^{20}$, G.A. Fuller$^{15}$, S. Graves$^{7}$, J. Gregson$^{21, 22}$, \newauthor W. Holland$^{23, 24}$, G. Joncas$^{25}$, J.M. Kirk$^{13}$, L.B.G. Knee$^{5}$, S. Mairs$^{6}$, \newauthor K. Marsh$^{12}$,  G. Moriarty-Schieven$^{5}$,   J. Rawlings$^{26}$, E. Rosolowsky$^{27}$, \newauthor D. Rumble$^{9}$, S. Sadavoy$^{28}$,  H. Thomas$^{7}$, N. Tothill$^{29}$, S. Viti$^{30}$, G.J. White$^{21, 22}$,  \newauthor C.D. Wilson$^{31}$, J. Wouterloot$^{7}$, J. Yates$^{26}$, M. Zhu$^{32}$\\
$^{1}$Astrophysics Group, Cavendish Laboratory, J J Thomson Avenue, Cambridge, CB3 0HE\\
$^{2}$Kavli Institute for Cosmology, Institute of Astronomy, University of Cambridge, Madingley Road, Cambridge, CB3 0HA, UK\\
$^{3}$Imperial College London, Blackett Laboratory, Prince Consort Rd, London SW7 2BB, UK\\
$^{4}$Physics \& Astronomy, University of St Andrews, North Haugh, St Andrews, Fife KY16 9SS, UK\\
$^{5}$NRC Herzberg Astronomy and Astrophysics, 5071 West Saanich Rd, Victoria, BC, V9E 2E7, Canada\\
$^{6}$Department of Physics and Astronomy, University of Victoria, Victoria, BC, V8P 1A1, Canada\\
$^{7}$Joint Astronomy Centre, 660 N. A`oh\={o}k\={u} Place, University Park, Hilo, Hawaii 96720, USA\\
$^{8}$Department of Physics and Astronomy, University of Waterloo, Waterloo, Ontario, Canada  N2L 3G1\\
$^{9}$Physics and Astronomy, University of Exeter, Stocker Road, Exeter EX4 4QL, UK\\
$^{10}$Department of Astronomy, Cornell University, Ithaca, NY 14853, USA\\
$^{11}$Leiden Observatory, Leiden University, PO Box 9513, 2300 RA Leiden, The Netherlands\\
$^{12}$School of Physics and Astronomy, Cardiff University, The Parade, Cardiff, CF24 3AA, UK\\
$^{13}$Jeremiah Horrocks Institute, University of Central Lancashire, Preston, Lancashire, PR1 2HE, UK\\
$^{14}$European Southern Observatory (ESO), Garching, Germany\\
$^{15}$Jodrell Bank Centre for Astrophysics, Alan Turing Building, School of Physics and Astronomy, University of Manchester, Oxford Road, Manchester, M13 9PL, UK\\
$^{16}$Universit\'e de Montr\'eal, Centre de Recherche en Astrophysique du Qu\'ebec et d\'epartement de physique, C.P. 6128, succ. centre-ville, \\
Montr\'eal, QC, H3C 3J7, Canada\\
$^{17}$James Madison University, Harrisonburg, Virginia 22807, USA\\
$^{18}$School of Physics, Astronomy \& Mathematics, University of Hertfordshire, College Lane, Hatfield, HERTS AL10 9AB, UK\\
$^{19}$Astrophysics Research Institute, Liverpool John Moores University, Egerton Warf, Birkenhead, CH41 1LD, UK\\
$^{20}$Dunlap Institute for Astronomy \& Astrophysics, University of Toronto, 50 St. George St., Toronto ON M5S 3H4 Canada\\
$^{21}$Dept. of Physical Sciences, The Open University, Milton Keynes MK7 6AA, UK\\
$^{22}$The Rutherford Appleton Laboratory, Chilton, Didcot, OX11 0NL, UK.\\
$^{23}$UK Astronomy Technology Centre, Royal Observatory, Blackford Hill, Edinburgh EH9 3HJ, UK\\
$^{24}$Institute for Astronomy, Royal Observatory, University of Edinburgh, Blackford Hill, Edinburgh EH9 3HJ, UK\\
$^{25}$Centre de recherche en astrophysique du Qu\'ebec et D\'epartement de physique, de g\'enie physique et d'optique, Universit\'e Laval, 1045 avenue de la m\'edecine, \\
Qu\'ebec, G1V 0A6, Canada\\
$^{26}$Department of Physics and Astronomy, UCL, Gower St, London, WC1E 6BT, UK\\
$^{27}$Department of Physics, University of Alberta, Edmonton, AB T6G 2E1, Canada\\
$^{28}$Max Planck Institute for Astronomy, K\"{o}nigstuhl 17, D-69117 Heidelberg, Germany\\
$^{29}$University of Western Sydney, Locked Bag 1797, Penrith NSW 2751, Australia\\
$^{30}$Department of Physics and Astronomy, University College London, Gower Street, London, UK, WC1E 6BT\\
$^{31}$Department of Physics and Astronomy, McMaster University, Hamilton, ON, L8S 4M1, Canada\\
$^{32}$National Astronomical Observatory of China, 20A Datun Road, Chaoyang District, Beijing 100012, China\\
$\dagger$Current address: Institute for Astronomy, ETH Zurich, Wolfgang-Pauli-Strasse 27, CH-8093 Zurich, Switzerland
}
\begin{document}

\date{Accepted  Received; in original form }

\pagerange{\pageref{firstpage}--\pageref{lastpage}} \pubyear{2014}

\maketitle

\label{firstpage}

\begin{abstract}
We present 850\,$\mu$m and 450\,$\mu$m data from the JCMT Gould Belt Survey obtained with SCUBA-2 and characterise the dust attributes of Class I, Class II and Class III disk sources in L\,1495. We detect 23\% of the sample at both wavelengths, with the detection rate decreasing through the Classes from I--III.  The median disk mask is 1.6\,$\times 10^{-3}$\,\msol, and only 7\% of Class II sources have disk masses larger than 20 Jupiter masses. We detect a higher proportion of disks towards sources with stellar hosts of spectral type K than spectral type M. Class II disks with single stellar hosts of spectral type K have higher masses than those of spectral type M, supporting the hypothesis that higher mass stars have more massive disks. Variations in disk masses calculated at the two wavelengths suggests there may be differences in dust opacity and/or dust temperature between disks with hosts of spectral types K to those with spectral type M.
\end{abstract}


\section{Introduction}
One of the fundamental properties of circumstellar disks, particularly in relation to planet formation, is their mass \citep{andrews2013}. The most useful diagnostic of disk mass is the brightness of dust continuum emission at submillimetre wavelengths, where the dust emission is optically thin over much of the disk volume, and is not overly dependent on temperature. Disk masses are thought to decrease during the evolutionary process, either through internal processes, or due to interactions with nearby massive stars.  

The traditional classification of pre-main sequence (PMS) sources is  defined observationally through the Class I--III phases \citep{lada87,adam87}. Class I sources are more evolved embedded young stellar objects (YSOs) with massive accretion disks, while Class II sources are T Tauri stars with gas-rich circumstellar disks where the supply of envelope material has dissipated. Class III sources are stars with tenuous or debris disks, where at least the inner part of the disk has been evacuated. The bulk of disk mass is accreted during the earliest embedded phases, but more evolved sources continue to accrete material onto the disk and stellar host. There is not always a clear distinction between Class I and Class II sources, with detailed modelling demonstrating that embedded Class I sources observed face-on can be classified as Class II, while Class II disks observed edge-on may look like there is still a significant amount of envelope material remaining, and be classified as Class I \citep{whitney2003,crapsi2008,evans2009}. 
\subsection{JCMT GBS}
The Gould Belt Legacy Survey \citep[GBS;][]{wardthompson2007} has been awarded 612\,h of time on the James Clerk Maxwell Telescope (JCMT) to survey nearby star-forming regions (within 500\,pc), using Heterodyne Array Receiver Programme \citep[HARP;][]{buckle2009} and Submillimetre Common-User Bolometer Array 2 \citep[SCUBA-2;][]{holland2013}. The GBS will trace the very earliest stages of star formation,  providing an inventory of all the protostellar objects contained in the nearby molecular clouds of the Gould Belt, covering $\sim$700 deg$^2$. 
The key science goals of the SCUBA-2 survey are:
\begin{itemize}
\item	To calculate the duration of each of the protostellar stages.
\item	To elucidate the nature of the evolution of protostellar collapse.
\item	To discover the origin of the IMF of stars, from intermediate mass stars to substellar objects.
\item	To discern the connection between protostars and the molecular cloud structure from which they formed.
\end{itemize}
\subsection{Taurus}
 The Taurus star-forming complex is young, with typical YSO ages of $\sim$1\,Myr \citep{kenyon1995}. As a result, even its most evolved sources may have remnants of dusty disks surrounding them, since low--mass stars have a median disk lifetime of a few Myr \citep{williams2011}. This is one of the key reference regions for the distribution of disk masses \citep{andrews2005,andrews2013}. It is well-studied, nearby, and contains some of the most complete and well-characterised catalogues of disk sources \citep[e.g.,][]{furlan2011,luhman2010,esplin2014}. The complex has a significant population of YSOs that have no remaining envelope, exposing the disk, and the lack of Class 0 sources provides evidence that even the embedded sources are reasonably evolved. 
The observations of L\,1495 presented here probe dust continuum emission which spans protostellar evolutionary phases. Moreover, the wide-field mapping shown here ensures disk parameters can be examined without selection bias when examining source statistics, something shown to be problematic for previous snapshot surveys \citep{andrews2013}. 

The low stellar density in Taurus \citep[1--10\,stars/pc$^{-2}$,][]{gomez1993,gudel2007}, with the most massive star in L\,1495 an intermediate Ae star, suggests that disk evolution is expected to proceed through dynamical star--disk or, in multiple systems, by disk--disk interactions, rather than evaporation by  radiation from nearby stars. Recent observations \citep{lee2012,dodds2014} indicate that the eventual primitive planetary systems that are the end results of disk evolution may be dependent on the natal environment, a topic which can be investigated using the large-scale maps obtained by the GBS survey. 

 One of the most direct measures of planet-forming capability in disks is the mass that they contain. The submillimetre regime, where thermal dust emission is optically thin throughout the disk, provides one of the best methods of inferring disk mass. At these wavelengths, observations are sensitive to micron- to millimetre-sized grains. With grain growth and dust settling expected to occur in the inner disk, where temperatures are warmer, submillimetre observations are able to trace the mass in disks to later evolutionary phases than infrared observations can \citep{hernandez2007,ricci2010}.  With dual-wavelength observations of all the Class I, II, and III sources in L\,1495, we are able to assess the effects of temperature and dust opacity on measurements of disk mass in a consistent manner.

Sec.\,\ref{sec-dr} describes the observations, data reduction and processing and disk sample. We present analyses of the physical characteristics of the disks in Sec.\,\ref{sec-an}, and discuss  disk evolution in Sec.\,\ref{sec-evolution}. Finally, in Sec.\,\ref{sec-con} we summarise the conclusions of this research.
\section{Observations}
\label{sec-dr}
The SCUBA-2 \citep{holland2013} observations presented here form part 
of the JCMT Gould Belt Survey \cite[GBS,][]{wardthompson2007}. 
Continuum observations at 850\,$\mu$m and 450\,$\mu$m were made using 
fully sampled 30\arcmin\ diameter circular regions \cite[PONG1800 mapping 
mode;][]{chapin2013}.  Also included are observations from the GBS Science verification phase, taken in PONG900 and PONG3600 mapping modes. Larger regions were mosaicked with 
overlapping scans.

The data were reduced using an iterative map-making technique \cite[makemap 
in {\sc smurf;}][]{chapin2013}, and gridded to 6\arcsec\ pixels at 
850\,$\mu$m and 4\arcsec\ pixels at 450\,$\mu$m.  The iterations were halted 
when the map pixels, on average, changed by $<$0.1\% of the estimated 
map rms. The initial reductions of each individual scan were coadded to 
form a mosaic from which a signal-to-noise mask was produced for each 
region.  The final mosaic was produced from a second reduction using 
this mask to define areas of emission, with individual scans cropped to exclude noisy edge regions. 

Detection of emission structure smaller in size than the spatial filter are robust both inside and outside of the masked regions. 
A spatial filter of 600\arcsec\ is used in the reduction, and tests with simulated data show 
that flux recovery is robust for sources with a Gaussian FWHM less than 
2.5\arcmin, provided that the source peak brightness is several times the noise. Sources between 2.5\arcmin\ and 7.5\arcmin\ will be 
detected, but both the flux and the size are underestimated because 
Fourier components with scales greater than 5\arcmin\ are removed by the 
filtering process. Detection of sources larger than 7.5\arcmin\ is 
dependent on the mask used for reduction. In this analysis, we focus on compact disk sources, which, even with their remnant envelopes, have size scales $<$1\arcmin\ (0.04\,pc at a distance of 140\,pc), where we expect flux recovery to be robust at both 850\,$\mu$m and 450\,$\mu$m.

The data are calibrated in mJy\,pixel$^{-1}$, using aperture FCFs of 2.34\,Jy/pW/arcsec$^{2}$ and 
4.71\,Jy/pW/arcsec$^{2}$ at 850\,$\mu$m and 450\,$\mu$m, respectively, 
derived from average values of JCMT calibrators \citep{dempsey2013}, 
and correcting for the pixel area. The PONG scan pattern leads to lower 
noise in the map centre and overlap regions, while data reduction and 
emission artefacts can lead to small variations in the noise over the 
whole map.

\begin{figure*}
\includegraphics[width=8cm]{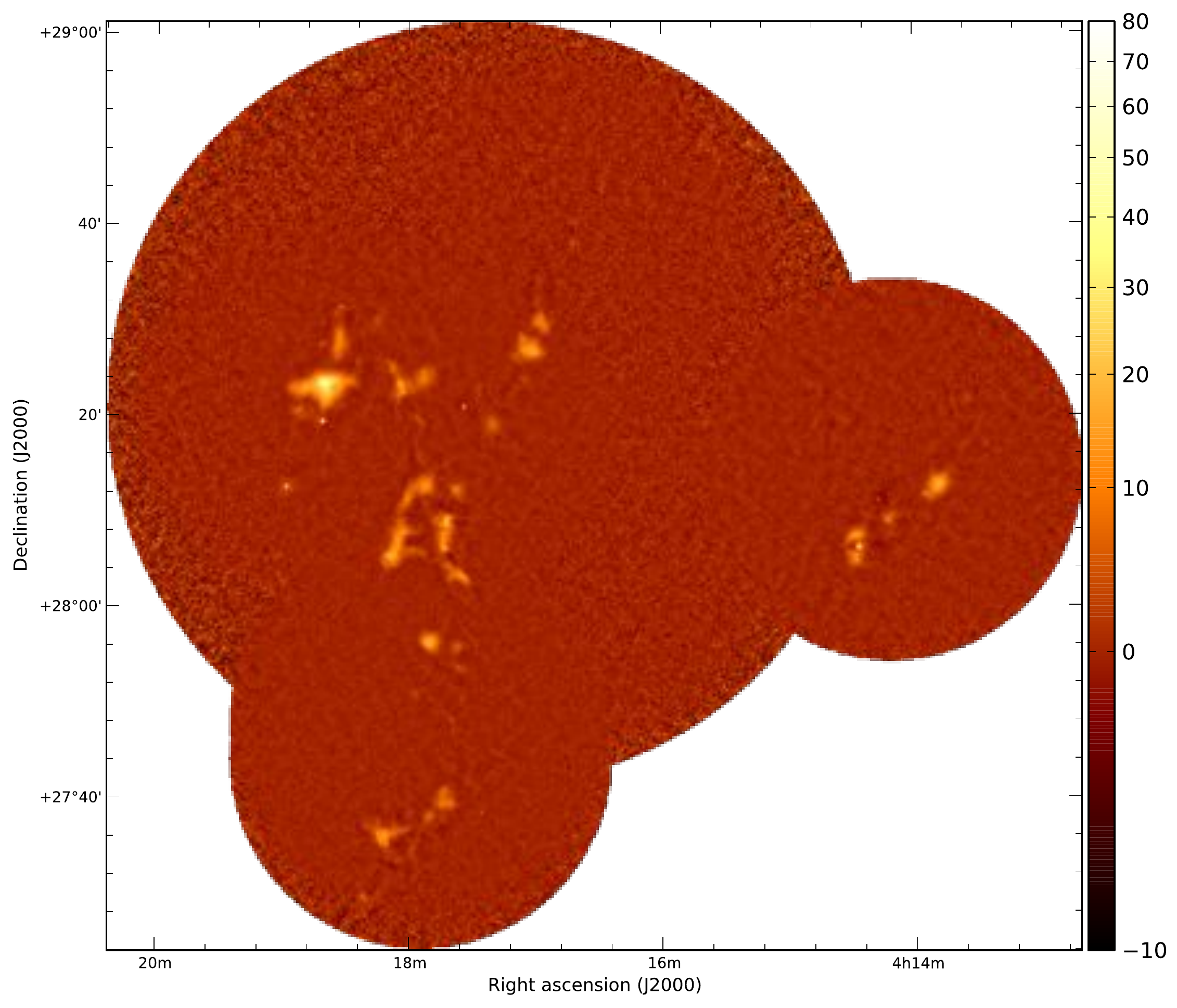}
\includegraphics[width=8cm]{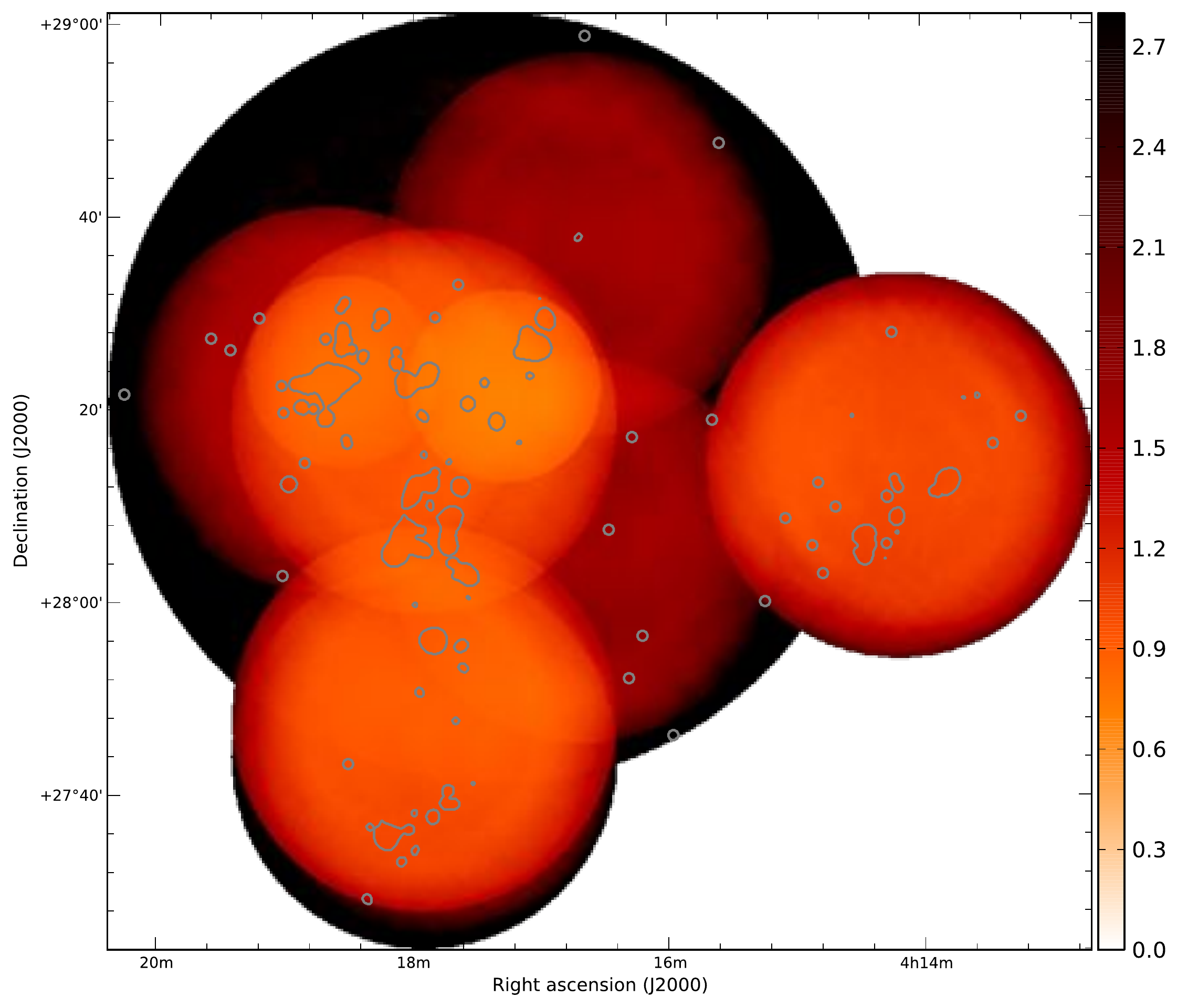}
\includegraphics[width=8cm]{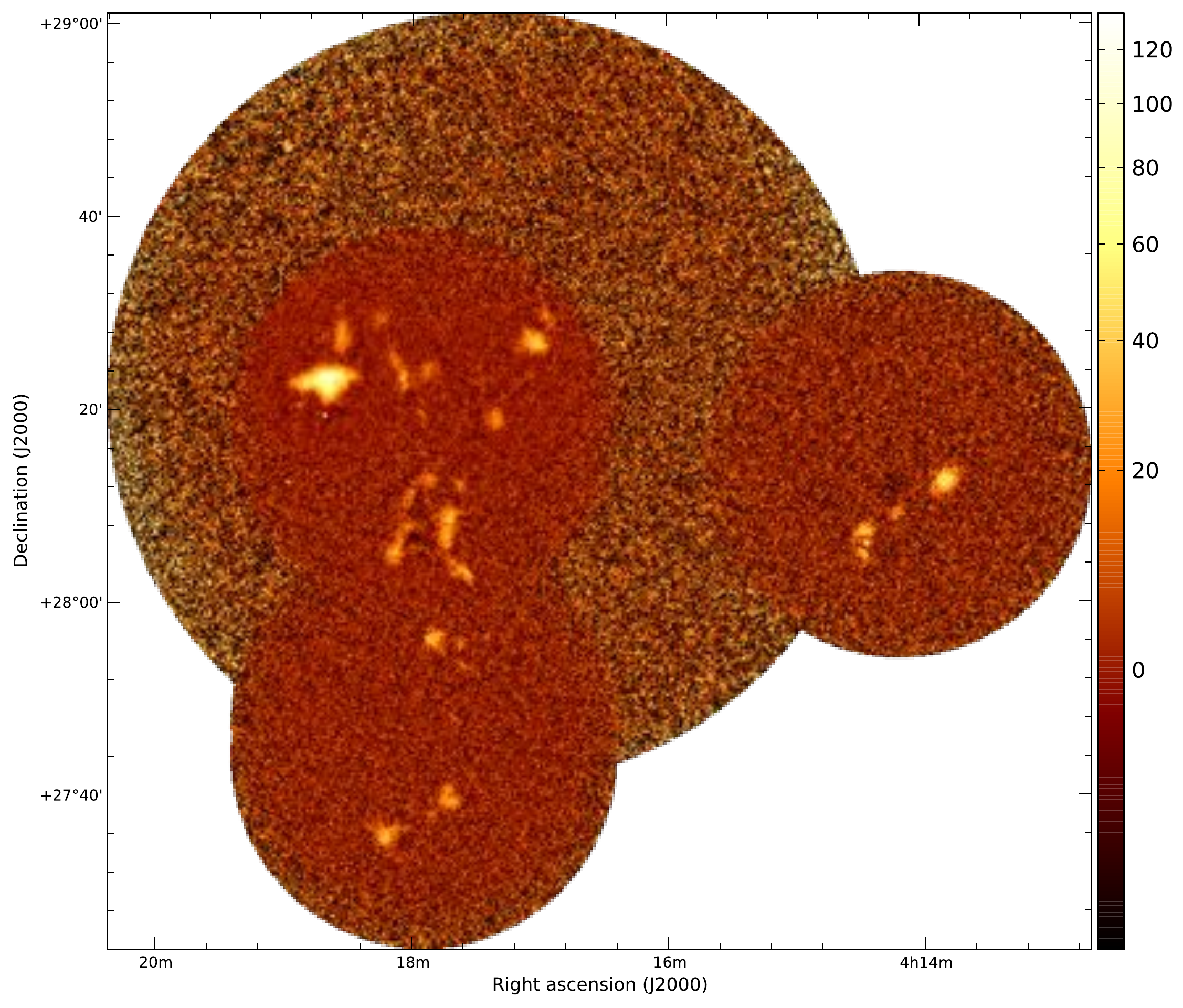}
\includegraphics[width=8cm]{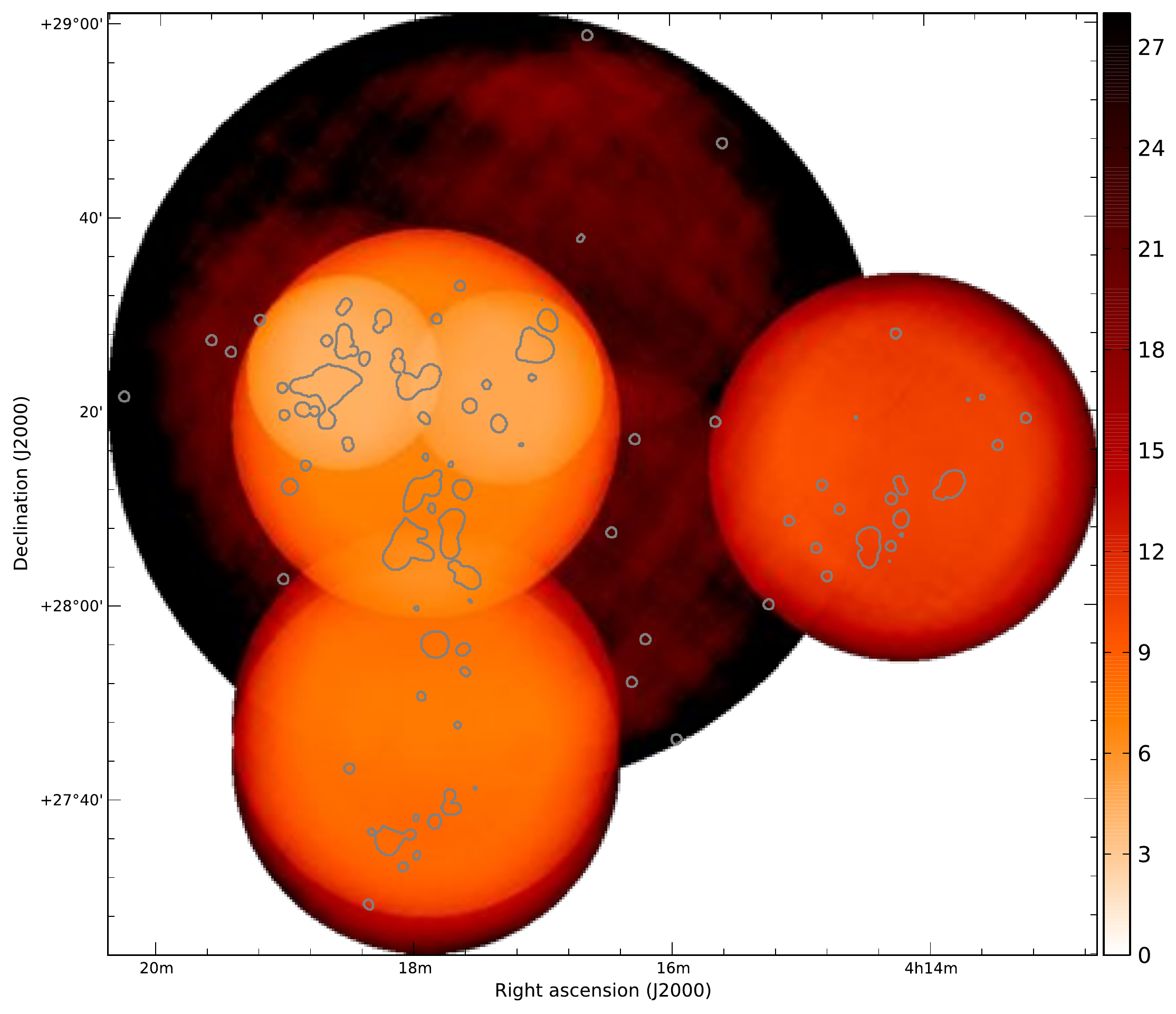}
\caption{SCUBA-2 850\,$\mu$m (top) and 450\,$\mu$m (bottom) continuum images (left) and error maps (right) of L1495. Fluxes are in mJy\,pixel$^{-1}$. Contours on the error maps show the mask used in the data reduction (Sec.\,\ref{sec-dr}). Note that the data maps are on a logarithmic colour scale, while the error maps are on a linear scale, with the colour-coding reversed .\label{fig-cont}}
\end{figure*}

Fig.\,\ref{fig-cont} shows the calibrated emission maps at 850\,$\mu$m and 450\,$\mu$m on the left, and the error maps on the right, which are also contoured with the mask used for data reduction.  A mean 1$\sigma$ uncertainty of 1.7\,mJy\,pixel$^{-1}$ at 850\,$\mu$m was achieved in the final L\,1495 mosaic. The inclusion of science verification data leads to a variation in noise across the final mosaic, due to the differences in spatial coverage. We achieve a better, more consistent sensitivity within the masked regions, however. Within the masked region, which is the focus of the science presented here, a mean 1$\sigma$ rms of 0.98\,mJy\,pixel$^{-1}$, equivalent to 2.5\,mJy\,beam$^{-1}$, was achieved. For the 450\,$\mu$m data, a mean 1$\sigma$ rms of 5.1\,mJy\,pixel$^{-1}$, equivalent to 13.3\,mJy\,beam$^{-1}$, was achieved inside the masked region.  

\subsection{Disk sources in L\,1495}
We use the most comprehensive disk catalogues of Taurus \citep{luhman2010,esplin2014} to identify Class I, Class II, and Class III sources in L\,1495, which are listed in Table\,\ref{tabflux}. We have applied updated spectral types published by \citet{herczeg2014} and \citet{pecaut2013}, as well as details of multiplicity of PMS stars in Taurus \cite[e.g.,][]{akeson2014} and the classification of sources as transition disks \citep[e.g.,][]{najita2007,furlan2011}. Note that there are no known Class 0 sources in this region. Figures \ref{diskfig0}--\ref{diskfig55} in Appendix\,\ref{Aplots} show thumbnail continuum images of all the disks in our sample. 
\subsection{CO contamination of dust continuum}
Dust continuum measurements in the submillimetre, observed using wide-band filters, can be contaminated by flux from molecular lines. Lines of $^{12}$CO are particularly problematic, as they are the brightest lines in the submillimetre wavelength range. For SCUBA-2, the $^{12}$CO $J=3\rightarrow2$ transition is near the central transmission peak of the 850\,$\mu$m bandpass filter, and can dominate dust continuum observations in regions of energetic molecular outflow \citep{drabek2012}. The JCMT GBS datasets include large-scale molecular line maps in $^{12}$CO $J=3\rightarrow2$, which we utilise, following the techniques described in \citet{drabek2012} to quantify the contribution of $^{12}$CO emission to our dust continuum measurements.

We use the JCMT GBS HARP observations of Taurus \citep{davis2010}, which comprise CO $J = 3\rightarrow2$ maps along the north-west region of Taurus known as L\,1495. The region contains a diffuse cloud \citep[named as a bowl or hub,][]{davis2010} which is the focus of the dust continuum observations presented here.  The CO data were processed through the same iterative mapmaker as the 850\,$\mu$m data, to apply the same spatial filtering \citep{hatchell2013,sadavoy2013}.

In the quiescent Taurus region, emission from CO is relatively diffuse, with only a few locations of strong emission, associated with high velocity protostellar outflows \citep{davis2010}.  The level of CO contamination does not exceed 15\%, the 850\,$\mu$m systematic flux uncertainty, and is generally below 5\%, as shown in Table \ref{tabflux}  for the disk sources in our sample. Sources not detected at 850\,$\mu$m are marked as `-', while sources which were not observed in $^{12}$CO are marked as `N/O' in the Table.
The results are consistent with results from other clouds  \citep[e.g.][]{drabek2012}, where high levels of CO contamination in the 850\,$\mu$m dust continuum emission are associated with energetic molecular outflows. Such outflows are rarer in the L\,1495 region observed here, which is more evolved than other parts of Taurus \citep{davis2010}. Of the two Class I and three Class II disk sources with CO contamination levels $>$5\%, IRAS\,04111+2800\,G (Class I) drives a high velocity outflow, while CoKu\,Tau\,1 (Class I), MHO3 (Class II)  and V410\,X-ray\,6 (Class II) are associated with high velocity red-shifted CO emission \citep{davis2010}.
The remaining Class II source with contamination $>$5\%, LR1, is a compact dust source embedded in weaker extended emission (both dust emission and CO integrated intensity) with local CO spectra having multiple velocity components.   

Given the low levels of CO contamination in the 850\,$\mu$m band, we do not expect any measurable contamination in the 450\,$\mu$m band, where the $^{12}$CO $J=6\rightarrow5$ transition lies at the edge of the SCUBA-2 bandpass filter \citep{drabek2012}.
\section{Analysis}
\label{sec-an}
Background subtracted mean and integrated flux densities for the disk sample were measured using 20\arcsec\ diameter apertures, with the background level estimated from nearby pixels in annuli between 30\arcsec\ and 40\arcsec. For a few crowded regions, nearby background regions with 20\arcsec\ diameter apertures were chosen manually.  Table \ref{tabflux} shows the integrated flux values with 1\,$\sigma$ uncertainties for each disk source, as well as information on the Class and spectral type. 
\begin{table*}
 \centering
 \begin{minipage}{180mm}
  \caption{Background-subtracted continuum flux measurements, from aperture photometry, with 1\,$\sigma$ uncertainties on the  mean and integrated fluxes, which do not include the calibration errors. Sources are ordered by Class, then Right Ascension, with names, Class and spectral types as given in the catalogues \citep{luhman2010,esplin2014} \label{tabflux}}
  \begin{tabular}
  {lllrrrrrr} 
  \#$^a$  & ID &   Class &   Sp.Type &mean F$_{\rm 850}$& total F$_{\rm 850}$&mean F$_{\rm 450}$ &total F$_{\rm 450}$ & \% CO \\
       &     &              &                  &mJy\,pixel$^{-1}$                &mJy                        & mJy\,pixel$^{-1}$                   &mJy                         &   \\
 \hline  
 1 &    J04153566+2847417 	&   I&    -- &   	 2.85$\pm$1.17 &   25.7$\pm$10.6 &    4.25$\pm$7.16 &   82.9$\pm$140 & N/O\\ 
{\bf \underline 2} & SSTTau041831p2+28261 &   I &    -- &    3.40$\pm$0.36 &   30.7$\pm$3.27 &    6.20$\pm$1.32 &  120$\pm$26 & $<$5\\ 
{\bf  \underline 3} &      IRAS04108+2803B &   I &    K0 &    5.91$\pm$0.44 &   53.3$\pm$4.01 &   14.8$\pm$3.14 &  288$\pm$61.6 & $<$5\\ 
{\bf  \underline 4} &      IRAS04111+2800G &   I &    K0 &    4.60$\pm$0.50 &   41.5$\pm$4.54 &   11.0$\pm$2.94 &  215$\pm$57.8 &  13\\ 
{\bf  \underline 5} &      IRAS04108+2803A &   I & M4.75 &    4.63$\pm$0.44 &   41.8$\pm$3.99 &   14.7$\pm$3.14 &  287$\pm$61.6 & $<$5\\ 
{\bf  \underline 6} &       IRAS04154+2823 &   I&  M2.5 &    7.20$\pm$0.43 &   65.0$\pm$3.89 &   11.3$\pm$1.37 &  221$\pm$26.9 & N/O\\ 
{\bf   \underline 7} &            CoKuTau/1 &   I &    M0 &    1.91$\pm$0.40 &   17.3$\pm$3.64 &    4.58$\pm$1.25 &   89.3$\pm$24.6 &   8\\ 
{\bf  \underline 8} &       IRAS04158p2805 &   I & M5.25 &   22.3$\pm$0.46 &  201$\pm$4.15 &   56.7$\pm$2.24 & 1110$\pm$44.0 & $<$5\\ 
{\bf  9} &           V410X-ray7 &  II & M0.75 &    1.38$\pm$0.36 &   12.5$\pm$3.27 &    2.61$\pm$1.37 &   50.9$\pm$27.0 & $<$5\\ 
10 &    J04141188+2811535 &  II & M6.25 &    0.94$\pm$0.44 &    8.50$\pm$3.96 &   -3.84$\pm$3.13 &  -74.9$\pm$61.6 & $<$5\\ 
11 &           V773\,Tau\,A+B &  II$^{\rm T}$ &    K3 &    1.23$\pm$0.44 &   11.1$\pm$3.96 &   -6.02$\pm$3.14 & -117$\pm$61.7 & $<$5\\ 
12 &                FM\,Tau &  II &  M4.5 &    1.19$\pm$0.48 &   10.8$\pm$4.37 &   -2.17$\pm$2.88 &  -42.3$\pm$56.5 & $<$5\\ 
{\bf 13} &                FN\,Tau &  II&  M3.5 &    1.62$\pm$0.53 &   14.7$\pm$4.78 &    8.07$\pm$3.11 &  158$\pm$61.0 & N/O\\ 
{\bf  \underline{14}} &                CW\,Tau &  II &  K3.0 &    6.55$\pm$0.49 &   59.2$\pm$4.41 &   16.9$\pm$2.90 &  330$\pm$57.0 & $<$5\\ 
15 &                CIDA1 &  II &  M4.5 &    0.39$\pm$0.50 &    3.56$\pm$4.58 &    4.59$\pm$2.99 &   89.6$\pm$58.6 & $<$5\\ 
{\bf  \underline{16}} &               MHO1/2 &  II &  M2.5 &   48.9$\pm$0.46 &  442$\pm$4.20 &   85.7$\pm$3.03 & 1670$\pm$59.5 & $<$5\\ 
{\bf 17} &                 MHO3 &  II &  M2.2 &    2.02$\pm$0.51 &   18.3$\pm$4.62 &    2.30$\pm$3.03 &   44.8$\pm$59.6 &  10\\ 
{\bf 18} &             FO\,Tau\,A+B &  II$^{\rm T}$&  M3.9 &    1.80$\pm$0.46 &   16.2$\pm$4.20 &    7.82$\pm$2.73 &  153$\pm$53.5 & N/O\\ 
19 &    J04153916+2818586 &  II &  M4.0 &    1.64$\pm$0.78 &   14.8$\pm$7.04 &    1.36$\pm$5.59 &   26.5$\pm$110 & $<$5\\ 
20 &    J04155799+2746175 &  II&  M5.2 &    2.31$\pm$2.02 &   20.8$\pm$18.3 &   13.9$\pm$11.1 &  271$\pm$217 & N/O\\ 
21 &    J04161210+2756385 &  II& M4.75 &   -0.28$\pm$0.84 &   -2.54$\pm$7.66 &   -6.06$\pm$6.04 & -118$\pm$119 & N/O\\ 
22 &    J04163911+2858491 &  II&  M5.5 &   -0.33$\pm$1.90 &   -3.02$\pm$17.20 &    3.82$\pm$7.01 &   74.6$\pm$137 & N/O\\ 
{\bf  \underline{23}} &                CY\,Tau &  II &  M2.3 &   13.7$\pm$0.30 &  124$\pm$2.68 &   10.0$\pm$1.38 &  196$\pm$27.0 & $<$5\\ 
{\bf 24} &               KPNO10 &  II &    M5 &    2.21$\pm$0.44 &   20.0$\pm$4.03 &   -1.35$\pm$1.94 &  -26.3$\pm$38.1 & $<$5\\ 
25 &           V410X-ray1 &  II&  M3.7 &   -0.12$\pm$0.45 &   -1.06$\pm$4.09 &   -0.03$\pm$1.99 &   -0.60$\pm$39.0 & N/O\\ 
26 &           V410Anon13 &  II& M5.75 &   -0.68$\pm$0.41 &   -6.16$\pm$3.75 &   -0.01$\pm$1.31 &   -0.28$\pm$25.7 & N/O\\ 
{\bf  \underline{27}} &             DD\,Tau\,A+B &  II &  M4.8 &    2.10$\pm$0.38 &   19.0$\pm$3.41 &    6.17$\pm$1.43 &  121$\pm$28.1 & $<$5\\ 
28 &             CZ\,Tau\,A+B &  II &  M4.2 &   -0.06$\pm$0.37 &   -0.56$\pm$3.38 &   -0.77$\pm$1.43 &  -15.0$\pm$28.0 & $<$5\\ 
29 &           V410\,X-Ray\,2 &  II&    M0 &    1.25$\pm$0.42 &   11.3$\pm$3.84 &    0.59$\pm$1.31 &   11.6$\pm$25.8 & N/O\\ 
{\bf  \underline{30} }&              V892\,Tau &  II &    A0 &   43.3$\pm$0.34 &  391$\pm$3.06 &   74.3$\pm$1.24 & 1450$\pm$24.3 & $<$5\\ 
{\bf  \underline{31}} &                  LR\,1 &  II &  K4.5 &    3.05$\pm$0.40 &   27.5$\pm$3.65 &   11.7$\pm$1.23 &  229$\pm$24.1 &  13\\ 
32 &           V410\,X-Ray\,6 &  II$^{\rm T}$ &  M5.9 &    1.10$\pm$0.43 &    9.90$\pm$3.94 &    3.53$\pm$1.44 &   68.9$\pm$28.3 &   9\\ 
33 &               KPNO12 &  II&    M9 &    0.33$\pm$0.97 &    2.96$\pm$8.77 &    4.25$\pm$6.18 &   82.9$\pm$121 & N/O\\ 
34 &             FQ\,Tau\,A+B &  II$^{\rm T}$&  M4.3 &   -0.31$\pm$0.74 &   -2.77$\pm$6.73 &    0.70$\pm$5.60 &   13.6$\pm$110 & N/O\\ 
35 &              V819\,Tau &  II$^{\rm T}$&  K8.0 &   -0.33$\pm$0.75 &   -2.94$\pm$6.83 &   10.4$\pm$5.68 &  203$\pm$112 & N/O\\ 
36 &                FR\,Tau &  II&  M5.3 &    0.25$\pm$0.77 &    2.26$\pm$6.98 &    4.22$\pm$5.59 &   82.4$\pm$110 & N/O\\ 
37 &    J04201611+2821325 &  II&  M6.5 &    3.83$\pm$2.33 &   34.6$\pm$21.1 &    2.39$\pm$12.1 &   46.6$\pm$238 & N/O\\ 
38 &                LkCa1 & III &    M4 &   -0.26$\pm$0.51 &   -2.38$\pm$4.64 &   -2.70$\pm$3.10 &  -52.7$\pm$60.8 & $<$5\\ 
39 &                Anon1 & III &    M0 &   -0.64$\pm$0.48 &   -5.74$\pm$4.37 &    0.90$\pm$2.93 &   17.5$\pm$57.5 & $<$5\\ 
40 &           XEST20-066 & III &  M5.2 &    0.42$\pm$0.54 &    3.77$\pm$4.87 &   -4.96$\pm$3.01 &  -96.8$\pm$59.0 & $<$5\\ 
41 &           XEST20-071 & III &  M3.1 &    1.34$\pm$0.48 &   12.1$\pm$4.38 &    0.02$\pm$2.81 &    0.42$\pm$55.1 & $<$5\\ 
42 &                CIDA2 & III &  M5.5 &    0.63$\pm$0.48 &    5.71$\pm$4.34 &   -0.33$\pm$2.79 &   -6.35$\pm$54.7 & $<$5\\ 
43 &                KPNO1 & III&  M8.5 &    0.10$\pm$0.91 &    0.93$\pm$8.27 &    0.53$\pm$5.37 &   10.3$\pm$106 & N/O\\ 
44 &    J04161885+2752155 & III& M6.25 &   -0.79$\pm$0.82 &   -7.17$\pm$7.43 &    3.89$\pm$5.93 &   75.9$\pm$116.3 & N/O\\ 
45 &                LkCa4 & III&  M1.3 &    1.03$\pm$0.81 &    9.32$\pm$7.37 &   16.0$\pm$6.21 &  312$\pm$122 & N/O\\ 
46 &                LkCa5 & III&  M2.2 &   -0.54$\pm$0.48 &   -4.91$\pm$4.32 &   -3.85$\pm$2.10 &  -75.2$\pm$41.3 & N/O\\ 
47 &            V410Xray3 & III &  M6.5 &   -0.87$\pm$0.41 &   -7.88$\pm$3.69 &   -1.74$\pm$1.30 &  -33.9$\pm$25.6 & $<$5\\ 
{\bf 48 }&           V410Anon25 & III &    M1 &    1.68$\pm$0.36 &   15.18$\pm$3.26 &    1.81$\pm$1.32 &   35.4$\pm$25.9 & $<$5\\ 
49 &               KPNO11 & III&  M5.9 &   -0.08$\pm$0.47 &   -0.68$\pm$4.28 &    0.54$\pm$2.23 &   10.5$\pm$43.8 & N/O\\ 
50 &         V410\,Tau\,A+B+C & III &    K7 &    0.43$\pm$0.40 &    3.91$\pm$3.59 &   -0.24$\pm$1.24 &   -4.76$\pm$24.3 & $<$5\\ 
{\bf 51} &           V410X-ray4 & III &    M4 &    1.94$\pm$0.39 &   17.50$\pm$3.58 &   -0.98$\pm$1.21 &  -19.2$\pm$23.8 & $<$5\\ 
52 &              Hubble4 & III &  K8.5 &   -0.17$\pm$0.40 &   -1.52$\pm$3.59 &    0.61$\pm$1.24 &   12.0$\pm$24.3 & $<$5\\ 
53 &                KPNO2 & III &  M7.5 &    0.49$\pm$0.46 &    4.46$\pm$4.21 &   -4.92$\pm$1.94 &  -96.0$\pm$38.1 & $<$5\\ 
54 &          V410X-ray5a & III&  M5.5 &   -0.17$\pm$0.43 &   -1.50$\pm$3.95 &   -0.93$\pm$1.41 &  -18.1$\pm$27.7 & N/O\\ 
55 &    J04144158+2809583 & III &    -- &    0.94$\pm$0.48 &    8.45$\pm$4.33 &    0.56$\pm$2.81 &   10.9$\pm$55.1 & $<$5\\ 
56 &    J04161726+2817128 & III &    -- &    1.26$\pm$0.80 &   11.4$\pm$7.28 &    0.78$\pm$6.03 &   15.1$\pm$118 & $<$5\\ 
\hline
\multicolumn{5}{l}{$^{\rm a}$ 850\,$\mu$m detected sources are indicated in boldface, and underlined if also detected at 450\,$\mu$m}\\
\multicolumn{5}{l}{$^{\rm T}$ identifies a transition disk}\\
 \end{tabular}
\end{minipage}
\end{table*}

Using a detection criteria of 3\,$\sigma$ or better in total flux, we detect 88\% (7/8) of the Class I sources at each wavelength. For Class II sources, we detect 38\% (11/29) at 850\,$\mu$m and 21\%  (6/29) at 450\,$\mu$m.  Only 11\% (2/19) of the Class III sources are detected 850\,$\mu$m, and none at 450\,$\mu$m. Table \ref{tabsum} summarises detections at the 3\,$\sigma$ level for all the sources, by wavelength, Class, spectral type and multiplicity. 
As discussed in Sec.\,\ref{sec-dr}, our reduction algorithm filters out emission on scales of a few arcminutes, and so these observations are not sensitive to emission from the  surrounding cloud. The presence of large remnant envelopes can increase the submillimetre continuum around disk sources, particularly for more embedded, younger sources. 
Removing the local background using close annuli should compensate in part for such contributions, although the results are then more likely to be affected by contamination from the error beam. 
We investigate the contribution of emission from surrounding envelopes and loud material  in Sec.\,\ref{sec:class}. 

The sources in our sample are dominated by disks with stellar hosts of spectral type M, which have lower detection rates than for disks with hosts of spectral type K, as previously found in studies focussing on late spectral types \citep{andrews2013,schaefer2009}. Several of the sources are known binary or multiple sources, and these have a slightly higher detection rates (29\% and 22\% for multiple and single sources, respectively, at both wavelengths) than sources thought to be single systems, although there is, again, a large difference in the numbers of sources in each category. There are no known multiple Class I sources in our sample, and we do not detect the single Class III multiple source. We analyse these trends further in Sec.\,\ref{sec-evolution}.
\begin{table}
 \centering
  \caption{3$\sigma$ detection statistics in percentages for the disk sources.   \label{tabsum}}
  \begin{tabular}
  {lrrr} 
  Category &\# sources&\multicolumn{2}{c}{\% Detections}\\
  &  & 850\,$\mu$m&      450\,$\mu$m\\
  \hline
       All &56&  36&  23\\
       \hline
         I & 8 & 88&  88\\
        II &29&  38&  21\\
       III &19 & 11&   0\\
       \hline
 Sp Type K & 8&  50&  50\\
 Sp Type M &43 & 33&  16\\
 \hline
  Multiple&  7&  43&  29\\
    Single &49&  35 & 22\\
\hline
 \end{tabular}
\end{table}
For sources that are not detected, a flux limit can be obtained from the mean and standard error on the mean of the associated integrated flux values, listed in Table \ref{tabflux}. For Class II and III sources at 850\,$\mu$m, the mean flux and standard error on the mean are 6.24\,$\pm$\,2.38\,mJy and 1.55\,$\pm$\,1.57\,mJy, respectively. At 450\,$\mu$m, the mean flux and standard error on the mean are 42.7\,$\pm$\,19.7\,mJy and 5.1\,$\pm$\,19.8\,mJy for Class II and III respectively.
\subsection{Physical characteristics of dust in the disk sample}
A measure of the planet-forming capabilities of circumstellar disks is determining if they contain sufficient mass ($>$\,20\,M$_{\rm JUP}$ of dust and gas) to create the equivalent of a primitive Solar System \citep{davis2005}. Accurately measuring disk masses is a vital step in understanding disk evolution and the diversity of exoplanets. The opacity of dust is known to be one of the largest uncertainties in accurately assessing disk mass \citep[e.g.][]{beckwith1990}. With observations at two wavelengths for our complete sample of disks we investigate the spectral slope, and through it the dust opacity and temperature, before calculating disk masses. The analyses are described in the following sections, and the results presented in Table \ref{tabmass}. Results and calculations are all for detections at $>$3\,$\sigma$ unless otherwise stated.
\subsubsection{Continuum slopes}
\label{sec-alpha}
\begin{figure}
\includegraphics[width=8cm]{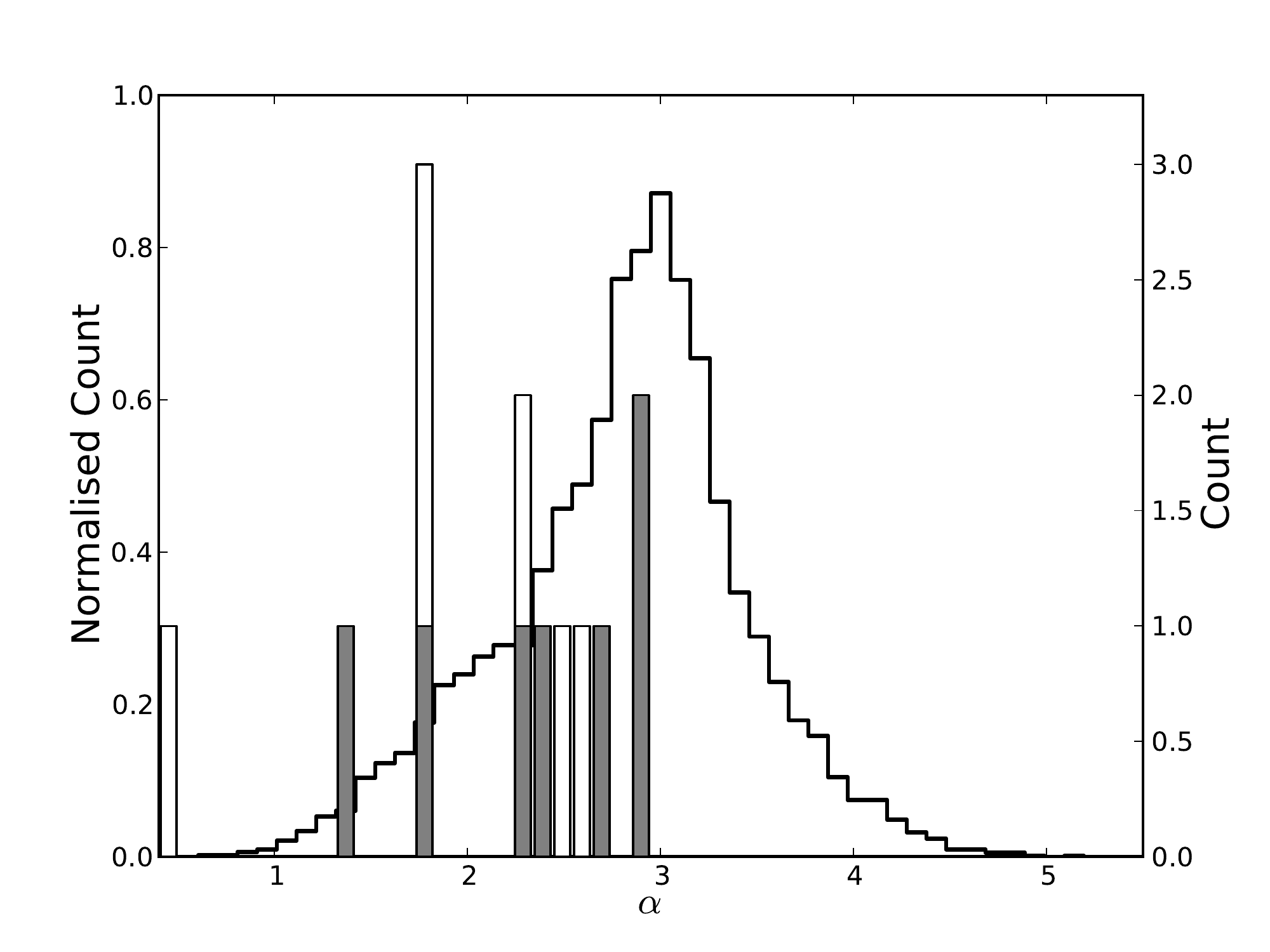}
\caption{Normalised distribution of $\alpha$ values (left axis, stepped histogram) calculated from the simple spectral index (Eq.\,\ref{eqalpha}) for all pixels detected with $>$3\,$\sigma$ in the L\,1495 region. Individual counts (right axis) for Class I (grey) and Class II (white) disk sources are also shown.\label{fig-alphapdf}}
\end{figure}

The slope ($\alpha$) of the continuum submillimetre emission can be calculated from the simple spectral index:
\begin{equation}
\alpha =\frac{\ln(F_{450})-\ln(F_{850} )}{\ln(850)-\ln(450)}\label{eqalpha}
\end{equation}
for sources with emission detected at 3\,$\sigma$ or greater in both wavelengths. For the disk sources, we use the integrated fluxes reported in Table \ref{tabflux} to calculate individual $\alpha$ values. For optically thin and isothermal emission, where the Rayleigh-Jeans criteria applies (e.g., dust temperatures $>$20\,K for $h\nu\gg\,kT$ at 850\,$\mu$m, and $>$35\,K at 450\,$\mu$m), then $\alpha$=2+$\beta$, where $\beta$ is the power law index of the dust opacity ($\kappa$) at submillimetre wavelengths:
\begin{equation}
\kappa_{\nu_1}=\kappa_{\nu_2}\left(\frac{\nu_1}{\nu_2}\right)^\beta . \label{eqkv}
\end{equation}
In this case, we expect $\alpha$ values of 3 if $\beta$=1.0, as appropriate for disk sources \citep[e.g.][]{beckwith1990}, or tending to higher values if $\beta>$1.0. For optically thick emission, or for dust grains much larger than the wavelength, values of $\alpha$\,$\approx$2 are expected, regardless of $\beta$ values.  To ensure the same scales are sampled at both wavelengths, we  use cross-convolved datasets for this, by convolving the 850\,$\mu$m map with 2-dimensional Gaussians representing the primary and secondary beams at 450\,$\mu$m, and similarly for the 450\,$\mu$m map to produce resolution-matched maps. The FWHM of the primary/secondary beams are 7.9\arcsec/13.0\arcsec\ and 25.0\arcsec/48.0\arcsec\ at 450\,$\mu$m and 850\,$\mu$m, respectively. The error beam at 450\,$\mu$m contributes significantly to flux in a large area \citep{dempsey2013}, where the relative volumes of the primary and secondary beams are 0.78/0.22 at 850\,$\mu$m and 0.61/0.39 at 450\,$\mu$m. 
Fig.\,\ref{fig-alphapdf}, (left axis, stepped histogram), shows the distribution of $\alpha$ across L\,1495, obtained from all pixels in Fig.\,\ref{fig-cont} with a detection $>$3\,$\sigma$. The distribution has a mean and standard deviation of 2.8$\pm$0.6. Since regions with significant detections at 450\,$\mu$m are compact, and generally towards star-forming cores in the L\,1495 region, this distribution suggests we are tracing little of the extended cloud material, where $\beta$ values of 1.5 (and $\alpha\geq$3.5 for $\alpha$=2+$\beta$) or higher may be expected.  

Fig.\,\ref{fig-alphapdf} also shows the distributions of $\alpha$ values in the disk sources, which are similar for Class I and Class II sources. CY\,Tau has an anomalously low value of 0.42, although this is consistent with previously published submillimetre values \citep[e.g.,][]{andrews2005}.
There are four sources with $\alpha>2.5$, implying $\beta\sim\,$1.0, and eight sources with $\alpha<2.5$, implying either grain growth, or optically thick emission.  Although we are not able to distinguish between these scenarios, the similar distribution we find between Class I and Class II disks provides evidence that disk evolution must be rapid. Under the assumption of optically thin, isothermal and warm ($>$20\,K) emission from dust in the disks, these results suggest that a value of $\beta$=1.0 is appropriate to use in the rest of the analyses presented here.

\subsubsection{Temperature}
\begin{figure}
\includegraphics[width=8cm]{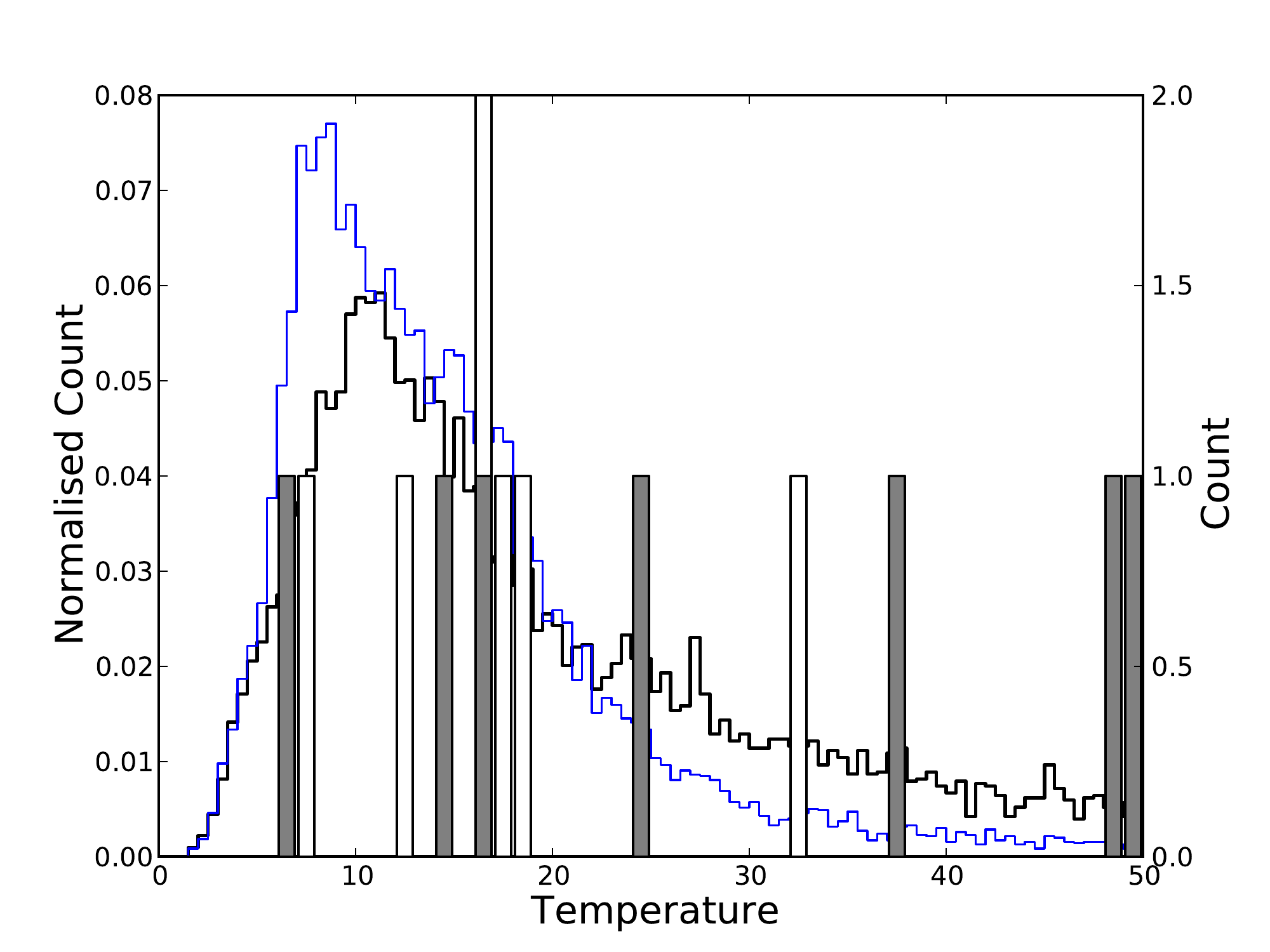}
\caption{Normalised distribution of dust temperatures (left axis) calculated from the flux ratio map (Eq.\,\ref{eqratio}), assuming $\beta$=1.8 (blue) and  $\beta$=1.0 (black). Temperature counts for Class I (grey) and Class II (white) disk sources are shown on the right axis, assuming $\beta$=1.0. \label{fig-tdust}}
\end{figure}

Alternatively, the dust temperature can be inferred from the flux ratio (Eq.\,\ref{eqratio}), given an assumed value of $\beta$:
\begin{equation}
\frac{F_{\rm 450}}{F_{\rm 850}}= \left({\frac{850}{450}}\right)^{3+\beta}\left(\frac{\exp({\rm hc}/\lambda_{850}kT_{\rm dust})-1}{\exp({\rm hc}/\lambda_{450}kT_{\rm dust})-1}\right) \label{eqratio}
\end{equation}
A flux ratio map was created using the cross-convolved datasets, as described in Sec.\,\ref{sec-alpha}
The dust temperature map was created from the ratio map for $\beta$=1.0, as determined in Sec.\,\ref{sec-alpha}, with a high temperature cap of 50\,K, above which the flux ratio does not constrain the dust temperature \citep[e.g.,][]{salji2014}.  Fig.\,\ref{fig-tdust} (left axis) shows the normalised temperature distribution calculated from the flux ratio (Eq.\,\ref{eqratio}) of the L1495 region for $\beta$=1.0 and $\beta$=1.8. A value of $\beta$=1.8 is more appropriate for emission from the ISM, while the disk sources are more accurately described with $\beta$=1.0.  Also shown is the temperature distribution for Class I and Class II disk sources, calculated assuming $\beta$=1.0.  Although both Classes have similar distributions of temperatures, the sources with the highest temperatures are all Class I. The median flux ratio for all the sources is 4.3, implying a dust temperature of 20\,K if $\beta$=1.0. The Class I sources' flux ratios extend to higher values than those of the Class II sources, with median flux ratios of 4.6/4.0 leading to dust temperatures of 23/18\,K, respectively.  We are removing the local background emission in which the disk sources are embedded. However, it is possible that the envelopes may be still contributing to flux from the Class I sources. If so, this will affect the flux ratios to a larger extent than for the Class II sources. Higher 450/850 flux ratios can be due to higher dust temperatures and/or a higher opacity index $\beta$ \citep{hatchell2013}, and requires flux measurements at additional wavelengths to disentangle.

Two disk sources have 5\,K$\leq$T$_{\rm dust}$$\leq$10\,K, while three have 30\,K$\leq$T$_{\rm dust}$$<$50\,K, and one has a flux ratio too high to estimate the temperature reliably. Note, however,  that the majority of the disk sources are in the regime where the Rayleigh-Jeans criteria fails at one or both wavelengths, and so the continuum slope is not well constrained. These sources may have contributions from surrounding material with different dust emission properties, requiring different $\beta$ values, and for these sources interferometric observations will be required to isolate emission from the compact disk. The similar range of dust temperatures calculated for both Classes suggests that adopting a single constant dust temperature for all sources not detected at both wavelengths will provide consistent results. A dust temperature of 20\,K is consistent with the median values derived for both Class I and Class II sources, and also in line with previous work \citep[e.g.,][]{mohanty2013,andrews2005,williams2013}. We therefore adopt a constant disk temperature of 20\,K in calculating disk masses for the rest of the analyses presented here.  

\subsubsection{Masses}
To determine the disk mass for all of the sources detected in at least one wavelength, we use values of T$_{\rm dust}$=20\,K and $\beta$=1.0, consistent with the above analyses. We assume that the emission is optically thin, the opacity law does not change from source to source, and can be described with a single characteristic temperature across the unresolved disk, so that the disk mass (dust + gas) scales with the observed flux density ($F_{\nu}$):
\begin{equation}
M_d= \frac{F_{\nu} D^2}{\kappa_{\nu} B_{\nu}(T_{\rm dust})} \label{eqmass}
\end{equation}
where $D$ is the distance to Taurus \cite[we adopt the canonical value of 140\,pc,][]{elias78}. $B_{\nu}$(T$_{\rm dust}$) is the Planck function for the given dust temperature, and $\kappa_{\nu}$ is the wavelength-dependent dust opacity.  We adopt a dust opacity of $\kappa_{850}$\,=\,0.0035\,m$^2$\,kg$^{-1}$, which assumes a gas--to--dust ratio of 100:1, and follows from the commonly adopted value for disks of $\kappa_{1.3{\rm mm}}$\,=\,0.0023\,m$^2$\,kg$^{-1}$ \citep[e.g.,][]{beckwith1990}, with $\beta$\,=\,1 (Eq.\,\ref{eqkv}). Where sources were detected at both wavelengths, masses have also been calculated from the inferred dust temperature (Eq. \ref{eqratio}), and these are both listed in Table \ref{tabmass}.

\begin{figure}
\includegraphics[width=8cm]{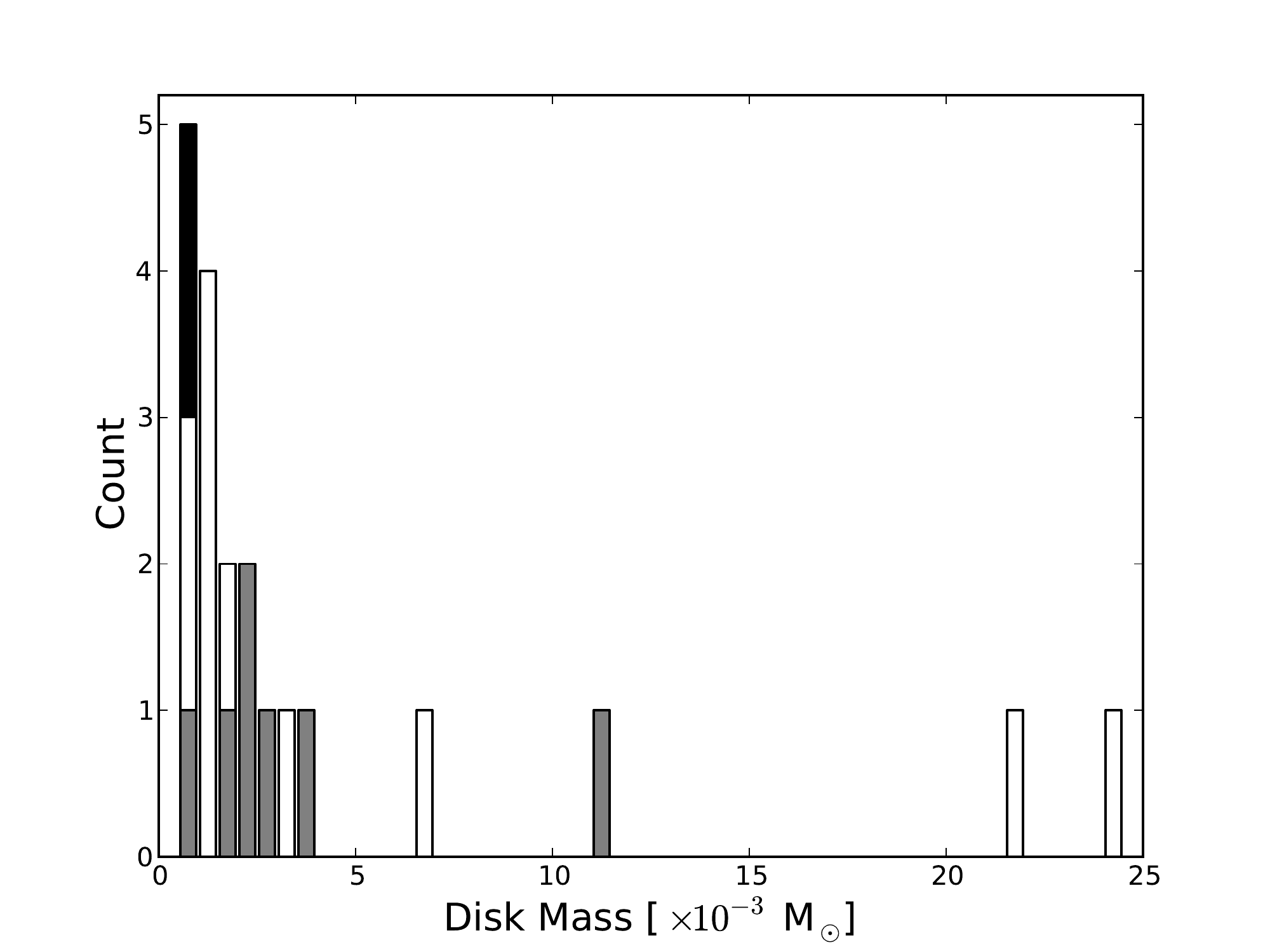}
\caption{Mass distribution of disk sources, assuming T$_{\rm dust}$=20\,K. Class I, II and III are shown in grey, white and black respectively. One Jupiter mass, M$_{\rm JUP}$\,=\,$1.05\times10^{-3}$\msol . \label{fig-mass}}
\end{figure}

Fig.\,\ref{fig-mass} shows the mass distribution of the disk sources, assuming T$_{\rm dust}$=20\,K, which has a mean disk mass of 4.5\,$\times 10^{-3}$\,\msol\ ($\sim$4\,M$_{\rm JUP}$), and a median disk mass of 1.6\,$\times 10^{-3}$\,\msol\ ($\sim$1\,M$_{\rm JUP}$). The Class II sources have higher mass, on average, than the Class I sources, dominated by the bright A0 source, V892\,Tau, and the multiple  source MHO\,1/2, both with disk masses $>$\,20\,M$_{\rm JUP}$. 

For sources where we are able to derive a dust colour temperature, the two cold sources, with  T$_{\rm dust}<$10\,K, have M(T$_{\rm dust})>$\,20\,M$_{\rm JUP}$ (assuming $\beta$\,=\,1). One of these, CY\,Tau, has been shown in recent high resolution molecular line observations to have a compact disk, at a low inclination \citep{guilloteau2014}. If the disk is compact, then the large mass in a small disk indicates high column densities, and so the assumption of optically thin dust emission may not be valid, and the disk mass underestimated. 

\begin{figure}
\includegraphics[width=4.2cm]{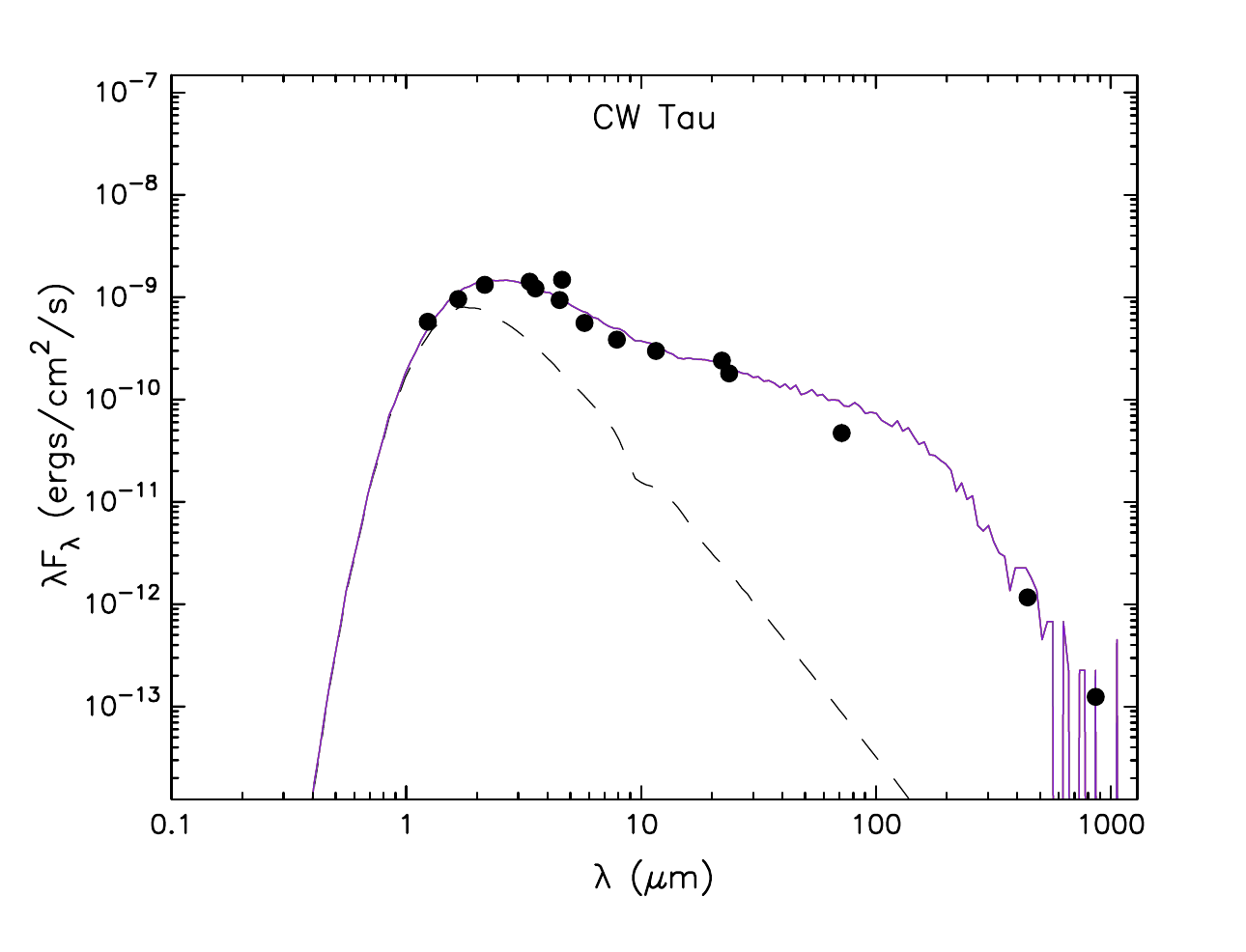}
\includegraphics[width=4.2cm]{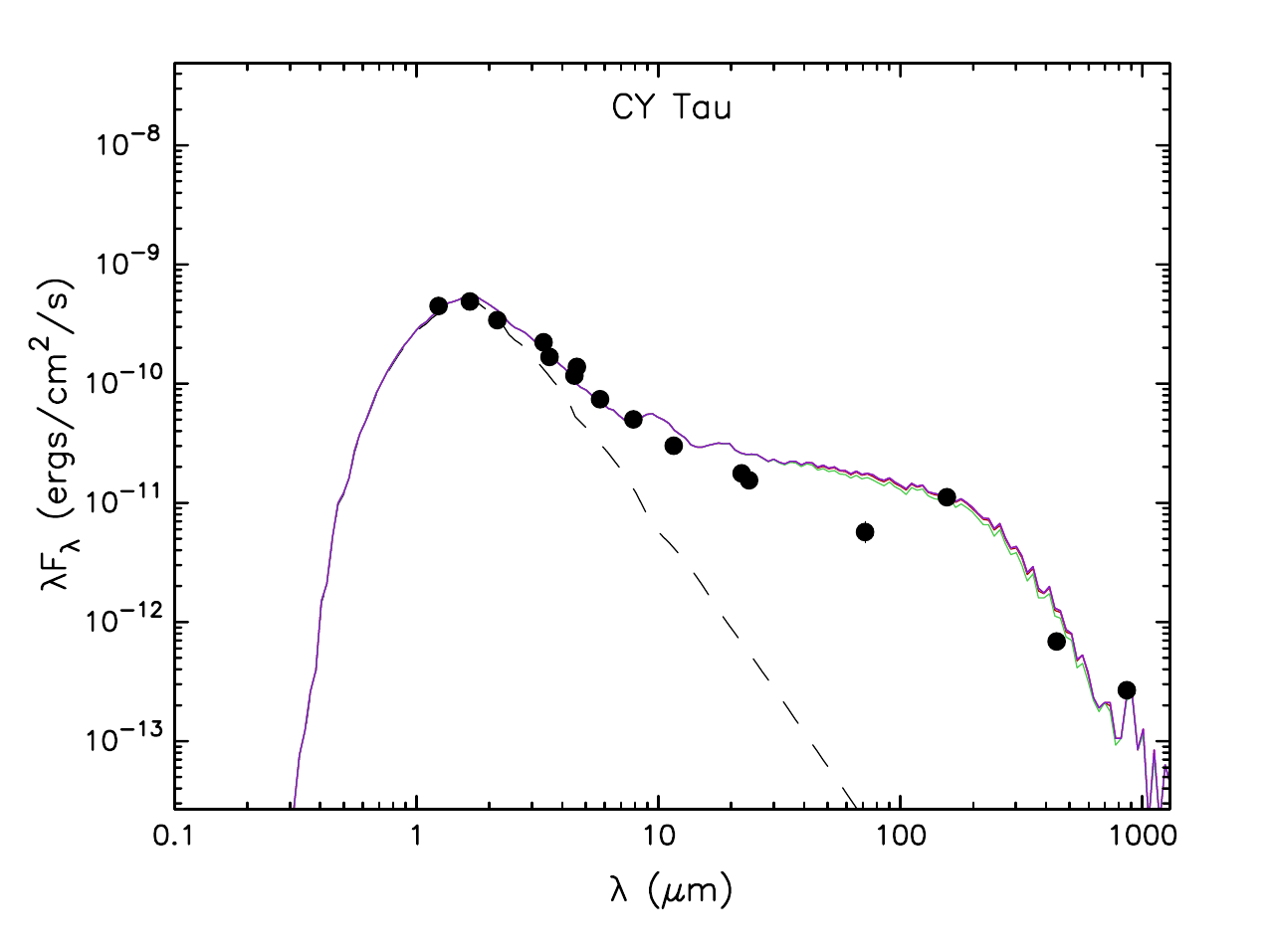}
\caption{The filled circles show the input fluxes and the coloured lines show the best fits to the SED \citep{rob2007}. The dashed line shows the stellar photosphere corresponding to the central source of the best fitting model, as it would look in the absence of circumstellar dust (but including interstellar extinction). Disk masses derived from the best-fit models of the SED are 7.34$\times 10^{-3}$\msol\  (CW\,Tau) and 27.9 $\times 10^{-3}$\msol\  (CY\,Tau). \label{sedfig}}
\end{figure}

The above analysis demonstrates that the simultaneous dual wavelength wide-area mapping observations attainable with SCUBA-2 provide valuable constraints on the physical characteristics of the dusty circumstellar disks.  To further demonstrate the validity of these results, we can compare disk masses derived above with those derived using SED models of young stellar objects. We use the Web-based tools of  \citet{rob2007}, obtaining shorter wavelength photometric data from 2MASS, IRAC and MIPS using published catalogues \citep{cutri2013,redbull2010,roeser2008,cutri2003}. The SED-fitting tool \citep{rob2007}  utilises a large grid of pre-computed models that span a wide range of parameter space in order to fit the SED, and provides the resulting model fits as datafiles and as SED plots, shown in Fig.~\ref{sedfig} for two of the sources detected at both wavelengths, CY\,Tau and CW\,Tau. The SCUBA-2 observations suggest that CY Tau has a more massive disk than CW Tau, which results from SED fitting support. The masses derived from SCUBA-2 observations, and those derived from SED fitting, agree to within a factor of 2 for both sources.  

We are also able to calculate masses from the 450\,$\mu$m flux, which should be similar to the masses derived from 850\,$\mu$m fluxes, if our assumptions on the dust opacity index (Eq.\,\ref{eqkv}) are correct and extend to 450\,$\mu$m. For a constant dust temperature of 20\,K, the mean and standard error of the M$_{450}$/M$_{850}$ ratio is 1.16$\pm$0.11. The lower mass disks have the largest variation in the mass ratio, but differences in derived dust temperature with respect to the characteristic temperature of 20\,K do not account for this difference. The mass ratios with the largest deviations from 1.0 are for those sources with $\alpha$$\ll$2 or $\alpha$$>$3, and indicate that the largest uncertainties in the mass estimates are from variations in the dust opacity. 
We investigate variations in the mass ratio with respect to source and stellar host characteristics in Sec.\,\ref{sec-evolution}.

\begin{table}
 \centering
  \caption{Physical parameters derived from continuum measurements \label{tabmass}}
  \begin{tabular}
  {llrrrr} 
 \#   ID  &Class  &  Mass\,(20K) &Tdust&Mass\,(T$_{\rm dust}$) &$\alpha$\\
              &         & $\times 10^{-3}$\msol\ &K&$\times 10^{-3}$\msol\ &\\
\hline
 2&  I& 1.69&24.5& 1.27& 2.69\\ 
 3&  I& 2.94&48.9& 0.91& 2.36\\ 
 4&  I& 2.29&37.6& 0.98& 1.80\\ 
 5&  I& 2.31&50.0& 0.70& 2.91\\ 
 6&  I&3.59& 6.96&28.0& 1.38\\ 
 7&  I& 0.95&16.8& 1.24& 2.92\\ 
 8&  I&11.1&14.6&18.3& 2.32\\ 
 9& II& 0.69& 5.47& --& --\\ 
13& II&1.05&16.1& --& --\\ 
14& II& 3.26&18.2& 3.78& 2.47\\ 
16& II&24.4&17.3&30.3& 1.73\\ 
17& II& 1.01&--& --& --\\ 
18& II&0.86&--& --& --\\ 
23& II& 6.84& 7.55&43.3& 0.43\\ 
24& II& 1.10&--& --& --\\ 
27& II& 1.05&16.7& 1.38& 2.32\\ 
30& II&21.6&12.8&44.9& 1.80\\ 
31& II& 1.52&32.2& 0.79& 2.59\\ 
48&III& 0.84&--& --& --\\ 
51&III& 0.97& --& --& --\\ 
\hline
 \end{tabular}
\end{table}

 Almost two-thirds (63\%)  of the sources were not detected at the 3$\sigma$ level or above. The mean and standard error of mass for these sources, calculated using the mean of fluxes for sources not detected at the 3$\sigma$ level, reported in Table\,\ref{tabflux}, and using a dust temperature of 20\,K, is 1.07$\pm$0.12\,$\times10^{-3}$\,\msol. We have a similar number of non-detections for the Class II and Class III sources, which have mean and standard errors of 1.26$\pm$0.21\,$\times10^{-3}$\,\msol\ and 0.83$\pm$0.06\,$\times10^{-3}$\,\msol.  We investigate further the mass limits and structure of non-detections in Sec.\,\ref{sec:class}. Of the five Class II sources characterised as probable transition disks, we detect one, the multiple source FO\,Tau A \& B, and calculate a disk mass of 0.86\,$\times 10^{-3}$\,\msol. 
Multiplicity in disk sources hampers efforts to characterise their physical characteristics \citep{najita2007}, which we investigate in Sec.\,\ref{sec-evolution}.
\subsection{Source characteristics by Class}
\label{sec:class}
The detection rate for our disk sample decreases with evolutionary Class. For example, at 850\,$\mu$m, there is only one non-detection for the eight Class I sources, but only two detections for the nineteen Class III sources. The wide-field mapping available with the SCUBA-2 datasets means that we can investigate in a statistical manner the material surrounding each of the disk sources. By stacking the data at the nominal disk positions, we can  compare average properties by Class.

\begin{figure} 
\includegraphics[width=8cm]{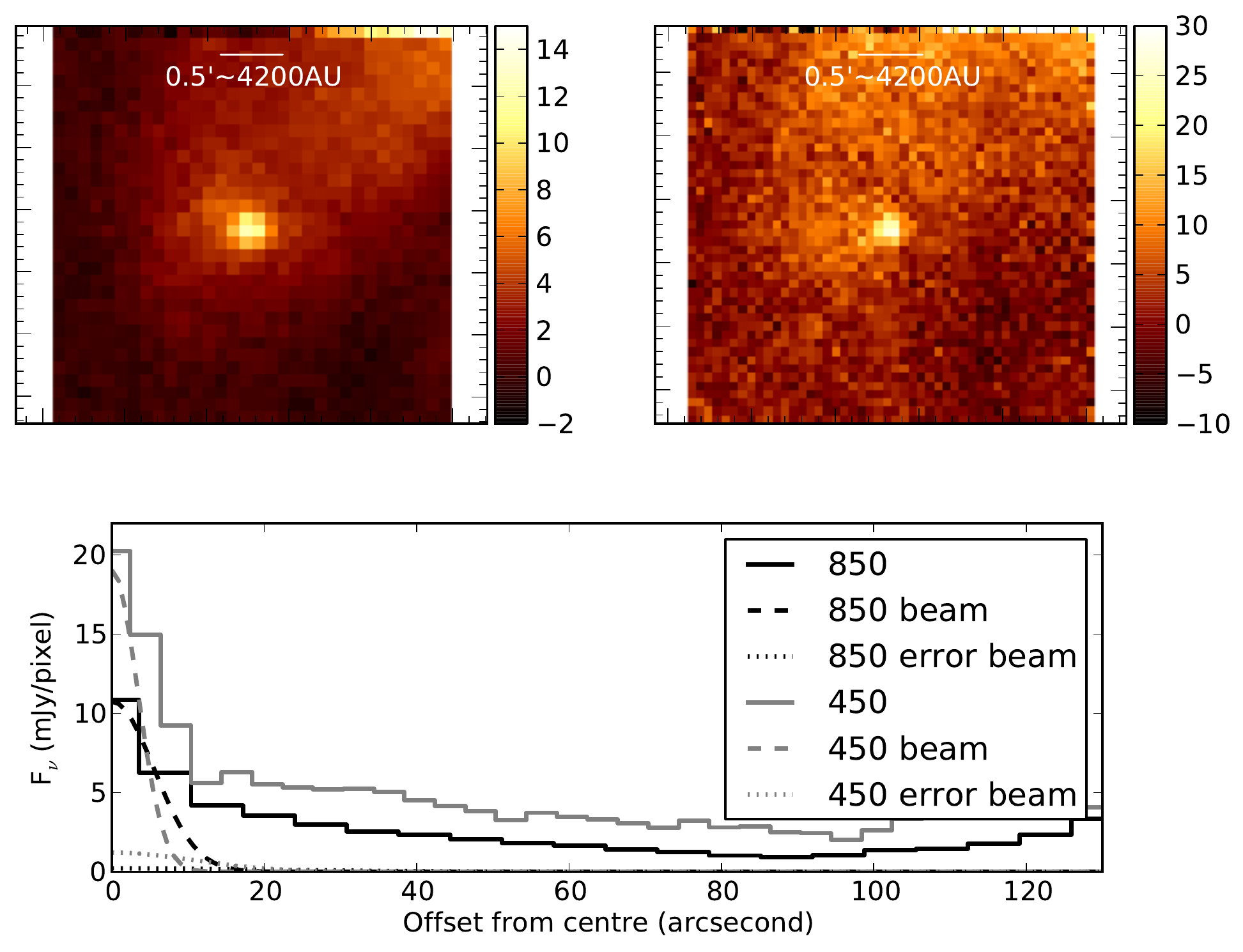}
\caption{{\it Top:} Stacked image of detected Class I sources at 850\,$\mu$m (left) and 450\,$\mu$m (right), with flux shown in mJy\,pixel$^{-1}$.  {\it Bottom:} Radially averaged emission profiles for the stacked images, showing the low-level extended emission in which the compact disk is embedded. Also shown are the relative contributions of the primary and secondary (error) beams at each wavelength.
Using a 2-component Gaussian fit to model the radial profiles, the extended emission will contribute 30\% to the observed flux if not removed through background subtracted aperture photometry.
\label{fig-stacked}}
\end{figure}

\begin{figure}
\includegraphics[width=8cm]{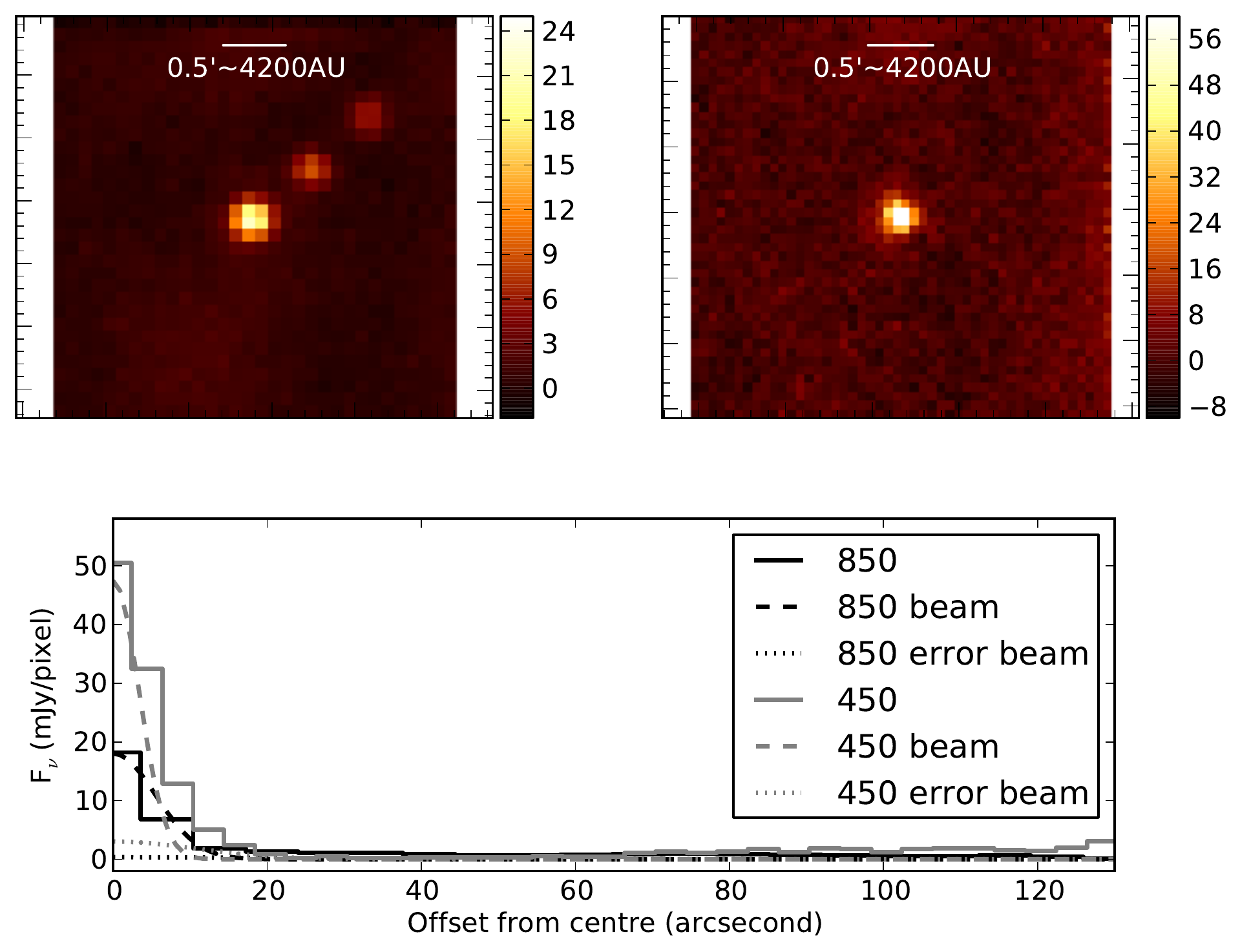}
\includegraphics[width=8cm]{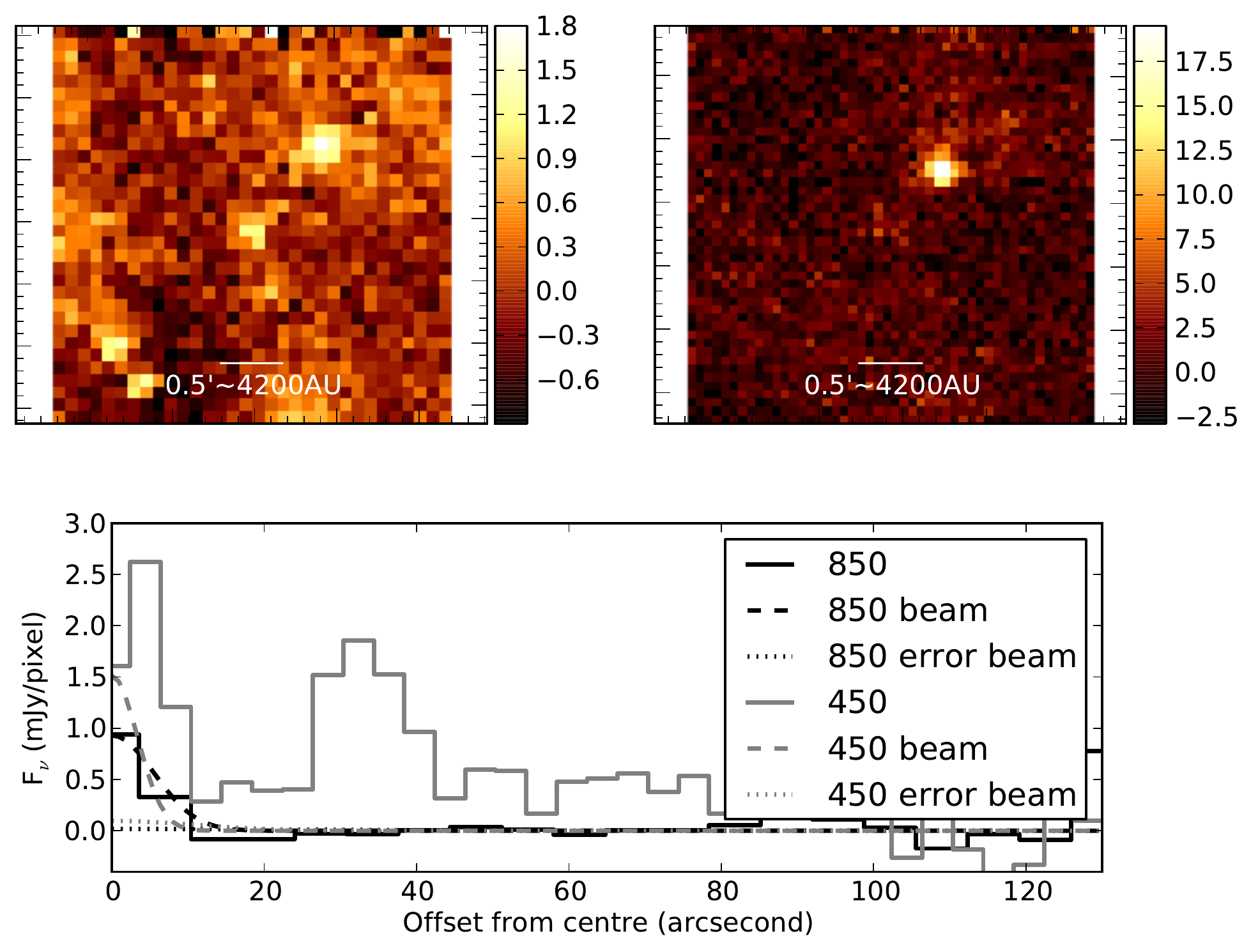}
\caption{As Fig. \ref{fig-stacked} for detected (top) and non-detected (bottom) Class II sources.  In addition to the disk source at the central position, nearby disk sources, bright enough to appear in the stacked image, are also seen, but are clearly separated from the nominal disk position. \label{fig-stackedII}}
\end{figure}

\begin{figure}
\includegraphics[width=8cm]{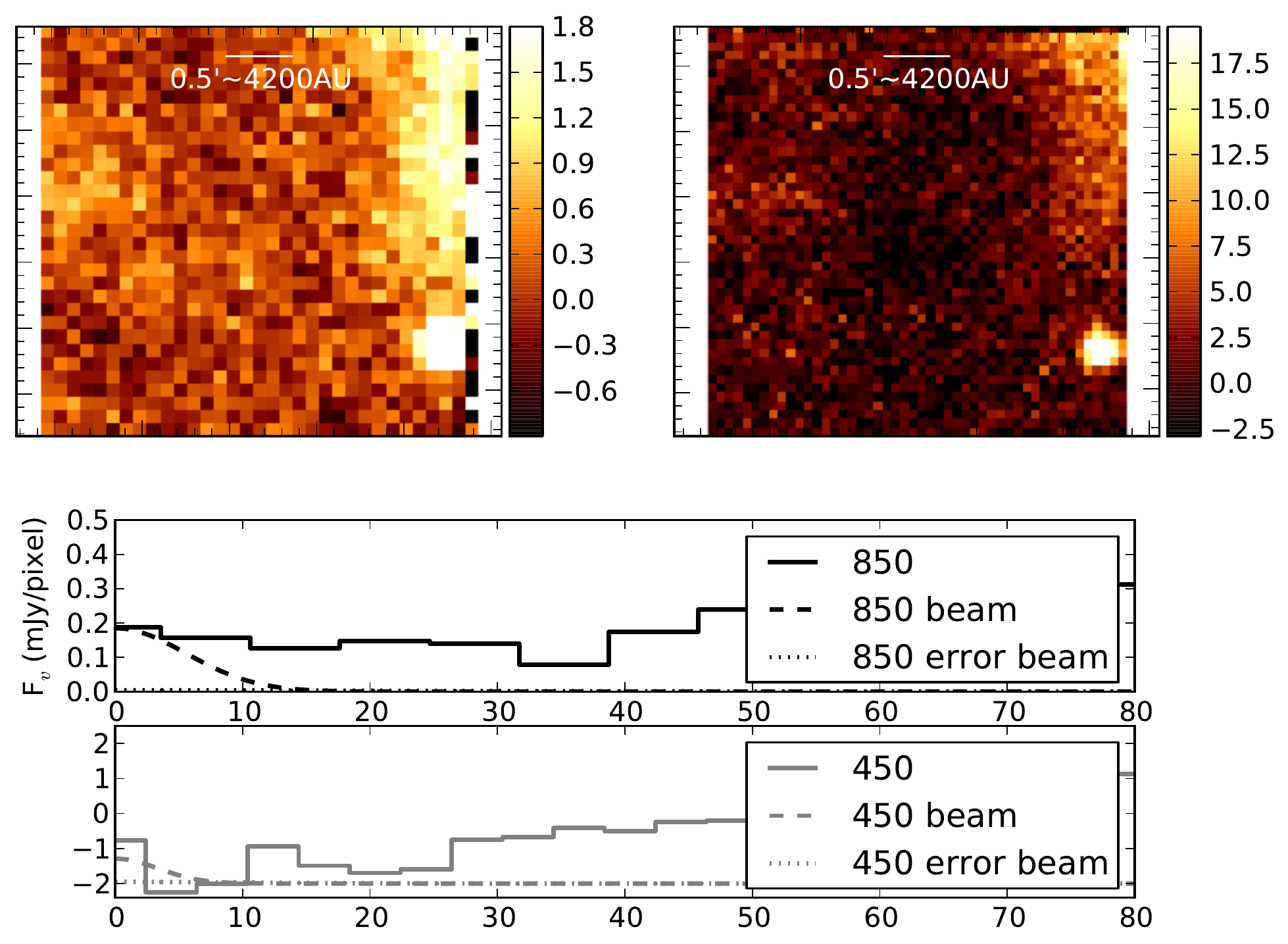}
\caption{As Fig. \ref{fig-stacked} for non-detected Class III sources, with the radial profiles shown separately for each wavelength. Bright nearby sources have been removed from the stacked data (see text). \label{fig-stackedIII}}
\end{figure}

Figs.\,\ref{fig-stacked}--\ref{fig-stackedIII} show the stacked images at 850\,$\mu$m (left) and 450\,$\mu$m (right), with no local background subtracted. Also shown are the radially averaged emission profiles over the central 2\arcmin\ of the stacked image and the relative contributions of the primary and secondary (error) beams at each wavelength.

Fig.\,\ref{fig-stacked} shows the stacked images and radially-averaged emission profiles for the detected Class I sources. At both wavelengths,  we see compact emission peaks within extended low-level emission. The apertures used to measure background-subtracted fluxes for these disk sources, with a diameter of 20\arcsec\ and local background measured in annuli of 30\arcsec\--40\arcsec,  probe size scales of 2800\,AU--5600\,AU, and should reduce contamination in the observed disk flux due to emission from the envelope, which could extend to scales $\sim$20000\,AU. A 2-component Gaussian fit to the radial profiles indicates that the low-level wide component -- associated with the envelope -- will contribute 30\% to the fluxes at both wavelengths if not removed.  The compact peak can be associated with an unresolved disk, with widths of 6.3\arcsec\--7.7\arcsec\ ($\sim$1000\,AU).  At SCUBA-2's resolutions, we are not able to resolve any accretion disc that might be present.

For the Class II sources, we can compare the structure of detected and non-detected sources using stacked data, shown in Fig. \ref{fig-stackedII}. Nearby compact sources are bright enough to be seen in the Class II stacked data for detections and non-detections, but are clearly separated from the nominal disk positions. For the detected and non-detected sources at both wavelengths, the compact peak has a width of 6.3\arcsec\--7.0\arcsec\ ($\sim$1000\,AU). At 450\,$\mu$m, the stacked data for non-detections show a weak central peak, which is offset by just a few arcseconds from the nominal central position, making it difficult to determine the contribution to the measured disk flux. This offset may indicate dust asymmetry in the disk, suggesting that the non-detected Class II disks are more evolved, rather than just less luminous. At 850\,$\mu$m, and for detected 450\,$\mu$m sources, the narrow component contributes at least 95\% of the flux. 

The stacked data for the Class III sources are dominated by extended arcs of emission at offsets of $\sim$1\arcmin\ from the nominal central position, which arise in a small number of sources  located near the strongest emission in the map, the bright reflection nebula illuminated by the Class II Ae star V892 Tau.  These sources (V410\,Tau and V410\,X-Ray\,3 undetected at both wavelengths,  V410\,X-Ray\,4 and V410\,Anon\,25 undetected at 450\,$\mu$m) have been removed from the final stacked data, shown in Fig.\,\ref{fig-stackedIII}. Extended emission at distances $>$1\arcmin\ from the disk position remains in the stacked data, however. The emission decreases radially towards the disk position, and at offsets $<$0.5\arcmin\ from the disk position, the flux is more consistent with a flat profile, particularly at 850\,$\mu$m. There may be emission from the central 6\arcsec\--10\arcsec\ region, but higher sensitivity data are required to investigate whether or not the dust `hole' for these sources is real.

With the disks in both Class I and Class II sources unresolved at this resolution, we are unable to determine differences in observable disk size at submillimetre wavelengths between the Classes. The differences in emission structure that we do observe are instead dominated by  envelope and/or cloud material.  
The derived physical characteristics show only small differences on average between Class I and Class II objects, with those in each phase covering similar ranges in temperature, $\alpha$, and disk mass. In the next Section, we investigate alternative indicators of disk evolution.
\section{Disk Evolution}
\label{sec-evolution}
Disk masses are known to decline with age, and to be intrinsically lower for late-type stellar hosts. \citet{andrews2013} have shown that the disk luminosity and mass are correlated with stellar host mass for Class II sources, with more massive stars more likely to have more massive disks. In this section, we investigate this correlation for M-- and K-- spectral type Class I and Class II sources in L\,1495.  

A wide variety of stellar masses have been computed from spectral types of young, pre-main sequence stars, and to date it is been difficult to determine dynamical stellar masses \citep{schaefer2009}. We therefore use stellar types as a proxy for stellar mass, to minimise the assumptions made about stellar hosts. For Class I sources with a range of possible spectral types, the median value has been used. 

\begin{figure}
\includegraphics[width=8cm]{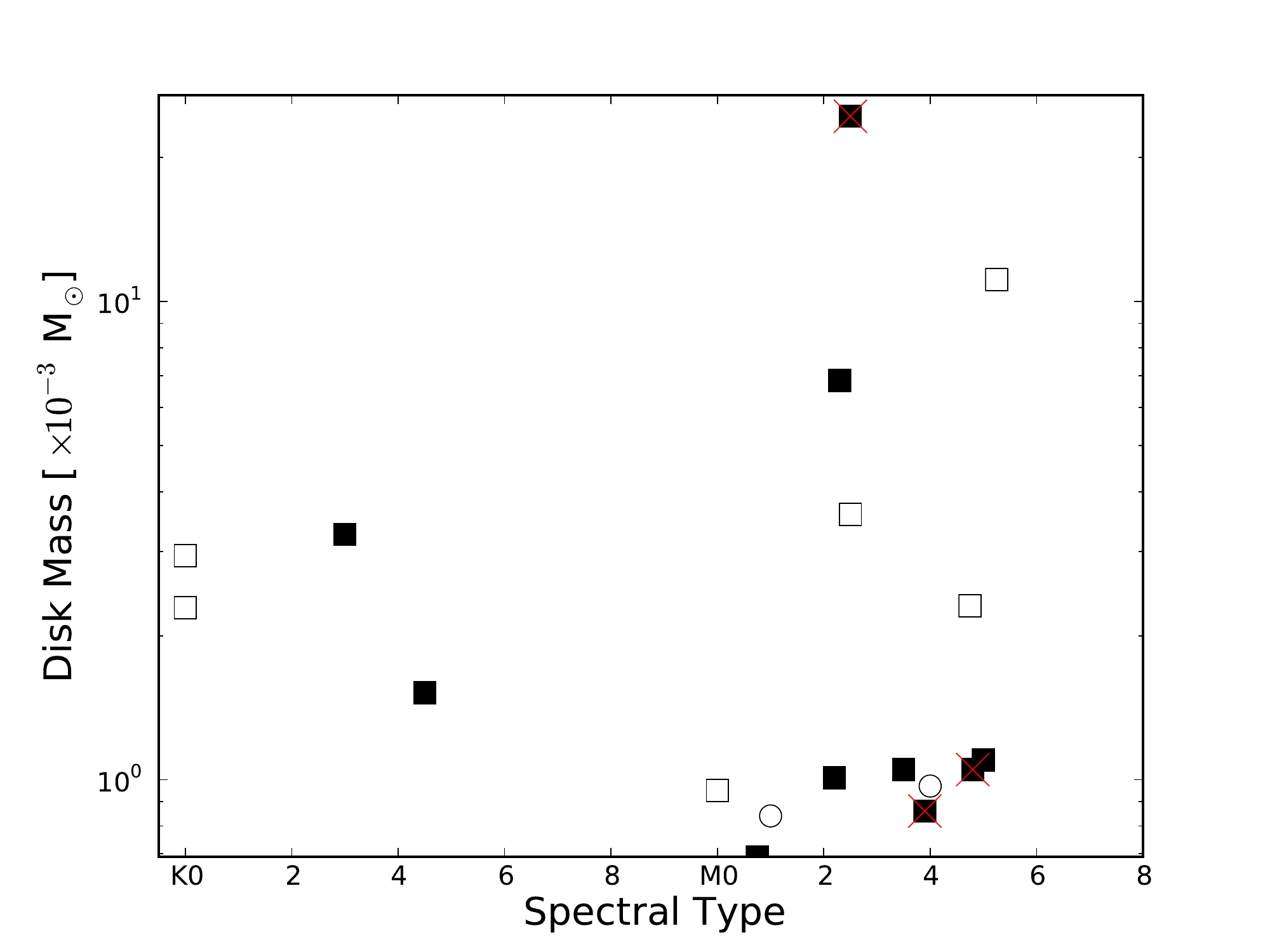}
\caption{Disk mass (at 20\,K) as a function of spectral type for Class I (open squares), Class II (closed squares) and Class III (open circles). Red crosses mark known multiple sources. \label{fig-masssp}}
\end{figure}

Fig. \,\ref{fig-masssp} shows the disk mass, calculated assuming a dust temperature of 20\,K, as a function of spectral type. Since binary or higher multiple disk sources may have multiple and/or circumbinary disks, we identify known binaries in the plot. The multiple sources all have late spectral types, and with the exception of MHO1/2,  are at the lower end of the mass distribution.  For both Classes, M-type stellar hosts have a wider distribution of disk mass than K-type stellar hosts, with stellar hosts of spectral types M2 to M6 having the largest range of mass. For the Class II sources, with the exception of the anomalous CY\,Tau and multiple MHO1/2 sources, M-type stellar hosts have lower masses than K-type stellar hosts. These data  support results from 
studies of late spectral type PMS stars in Taurus at high resolution, which  have similarly found evidence that disk masses decrease over the limited spectral range of K--M stars, with a difference in flux observed at the M2--M3 spectral type \citep{schaefer2009} and $M_d~\propto~M_*$ in a linear scaling \citep{andrews2013}.  

Fig. \,\ref{fig-masssp} shows a slight trend in the opposite direction for the Class I disk sources, with later spectral types extending to larger masses, which may be associated with a difference in evolutionary timescales for more massive stellar hosts. Given the difficulty in determining spectral types for Class I sources, however, this trend is much less clearly defined than for the Class II sources.

\begin{figure}
\includegraphics[width=8cm]{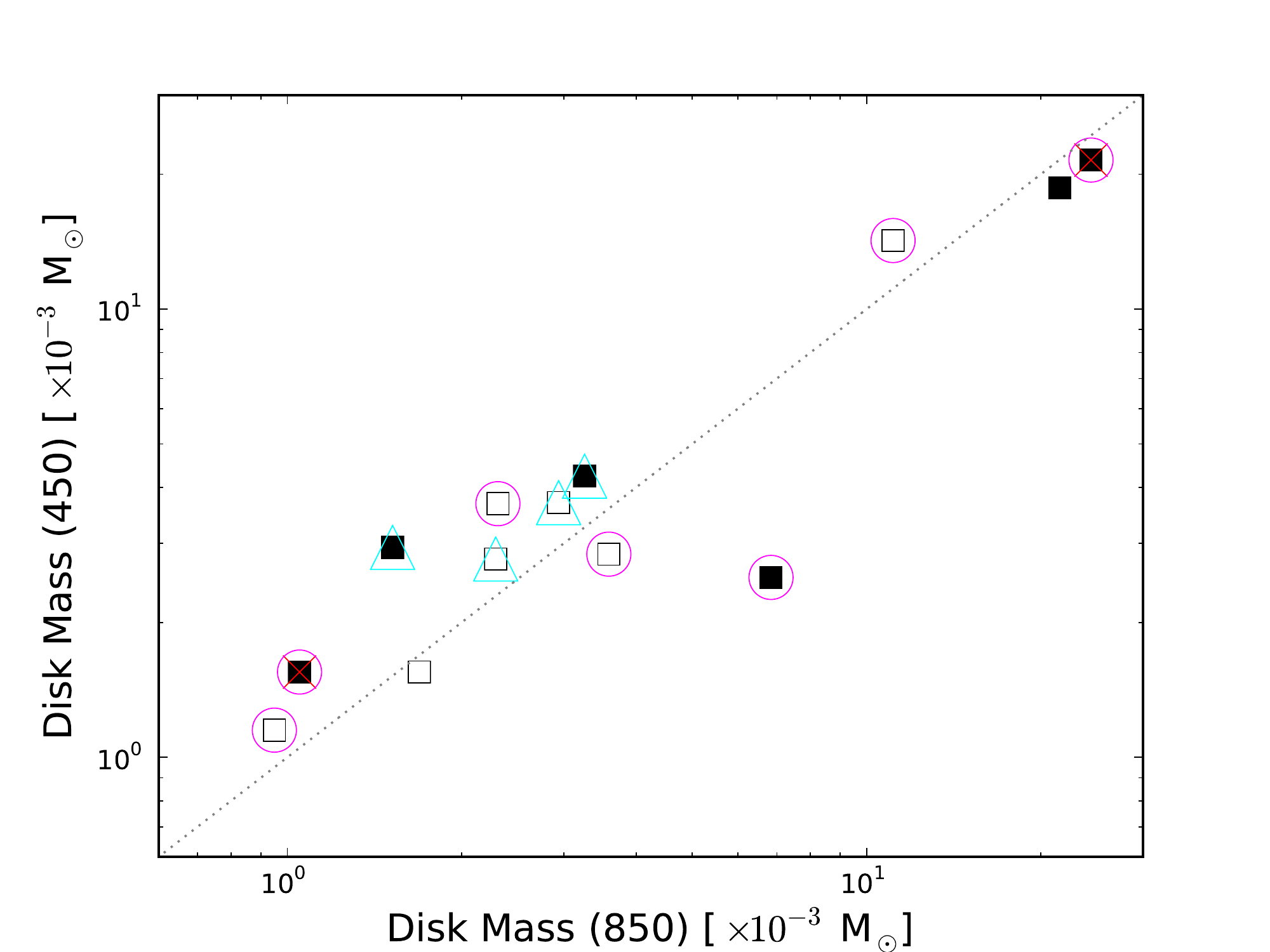}
\caption{Disk mass (at 20\,K) calculated from 850\,$\mu$m flux against that calculated from 450\,$\mu$m flux. Disk class and multiple sources as in Fig.\,\ref{fig-masssp}, with magenta circles and cyan triangles marking disks of spectral type M and K, respectively. The dotted line shows the 1:1 ratio. \label{fig-massr}}
\end{figure}

Fig. \,\ref{fig-massr} shows the disk mass derived at each wavelength, calculated assuming T$_{\rm dust}$=20\,K, with source Class and multiple sources marked as in Fig.\,\ref{fig-masssp}. Additionally, magenta circles and cyan triangles mark disks with stellar hosts of spectral type M and K, respectively.  Disks with stellar hosts of spectral type K consistently have higher 450\,$\mu$m disk masses than 850\,$\mu$m disk masses. For disks with stellar hosts of spectral type M, and for the different disk classes, there are no clear trends in the mass ratio. Although variations in dust opacity are the largest source of uncertainty in disk mass calculations, these results suggest that there could be an intrinsic difference in the dust opacity, or the dust temperature, or a combination of both, for disks surrounding K-type stellar hosts compared to those surrounding M-type stellar hosts.  Earlier type stars have higher luminosities, which generally leads to higher dust temperatures in the disk \citep{andrews2013}. Once complete, the full GBS dataset will provide a much larger sample with which to investigate this difference.

\section{Conclusions}
\label{sec-con}
\begin{itemize}
\item CO contamination in the continuum at 850\,$\mu$m is negligible in L\,1495, with only five sources showing contamination levels over 5\%, all of which are are below 15\% 
\item Across the L\,1495 region, the spectral slope $\alpha$\,=\,2.8\,$\pm$\,0.6, suggesting that much of the dust detected by SCUBA-2 has undergone grain growth. 
\item Class I and Class II disks have similar distributions of $\alpha$ values, with four sources having 2.5\,$<\alpha<3$, and eight sources with 1.4\,$<\alpha<2.5$. These results imply that $\beta\simeq$1 for the majority of disk sources, but that both Classes show evidence either for small, optically thick disks, or rapid grain growth.
\item Class I sources have slightly higher dust colour temperatures than Class II sources, although both Classes have similar ranges of dust temperature, 7--40\,K. In two Class I sources (IRAS 04108+2803 A and IRAS 04108+2803 B) the 450/850\,$\mu$m flux ratio is approaching the upper limit of 50\,K, where the dust colour temperature is unconstrained, particularly for low $\beta$ values.  
\item In L\,1495, 2/29 Class II sources have disk masses $M_d>$20\,M$_{\rm JUP}$. If masses are calculated from the dust colour temperature, then the two coldest sources have much larger masses i.e., 3/29 Class II sources and 1/8 Class I sources have $M_d>$20\,M$_{\rm JUP}$.
\item We detect a higher percentage of disks with stellar hosts of spectral type K than those of spectral type M, with detection rates of 50\% and 33\% respectively at 850\,$\mu$m. At 450$\mu$m, we detect the same 50\% of objects with stellar hosts of spectral type K, but only 16\% of those of spectral type M. Variations in the dust opacity lead to the largest uncertainties in disk mass calculations, but our results suggest there is an intrinsic difference in the dust opacity, and/or the dust opacity between disks with stellar hosts of spectral type M and K. The latter all have higher disk masses calculated using 450\,$\mu$m fluxes than calculated using 850\,$\mu$m data.
\end{itemize}

The Legacy datasets forming the JCMT Gould Belt Survey will be used to extend this analysis across a range of star-forming regions, so that comparisons between regions and star-forming environments can be made.

The largest uncertainty in calculating disk masses is in determining the dust opacity, particularly for observations where the disks are not resolved. Research in this field will make significant progress when the full capabilities of ALMA become available, and nearby disks can be imaged at millimetre/submillimetre wavelengths on scales of a few AU.

\section*{Acknowledgments}
The James Clerk Maxwell Telescope is operated by the Joint Astronomy Centre on behalf of the Science and Technology Facilities Council of the United Kingdom, the National Research Council of Canada, and (until 31 March 2013) the Netherlands Organisation for Scientific Research. Additional funds for the construction of SCUBA-2 were provided by the Canada Foundation for Innovation. DJ is supported by the National Research Council of Canada and by a Natural Sciences and Engineering Research Council of Canada (NSERC) Discovery Grant.

\label{lastpage}

\clearpage

\appendix
\section{Disk continuum images}
\label{Aplots}
\begin{figure}
\includegraphics[width=4cm]{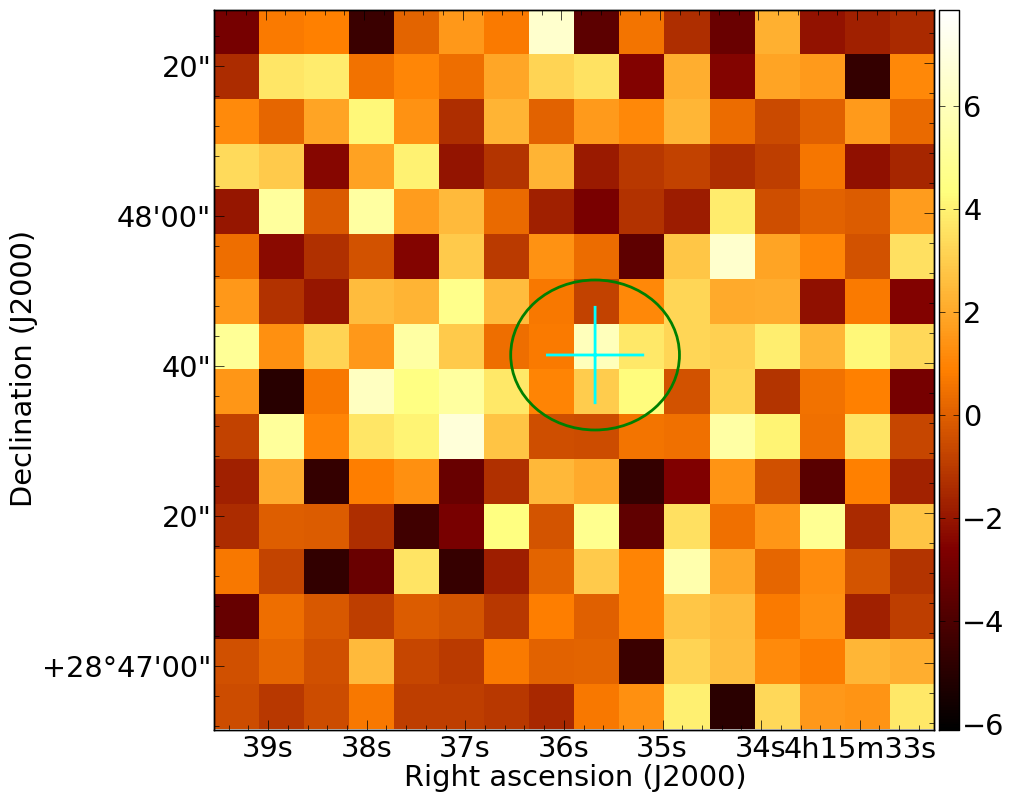}
\includegraphics[width=4cm]{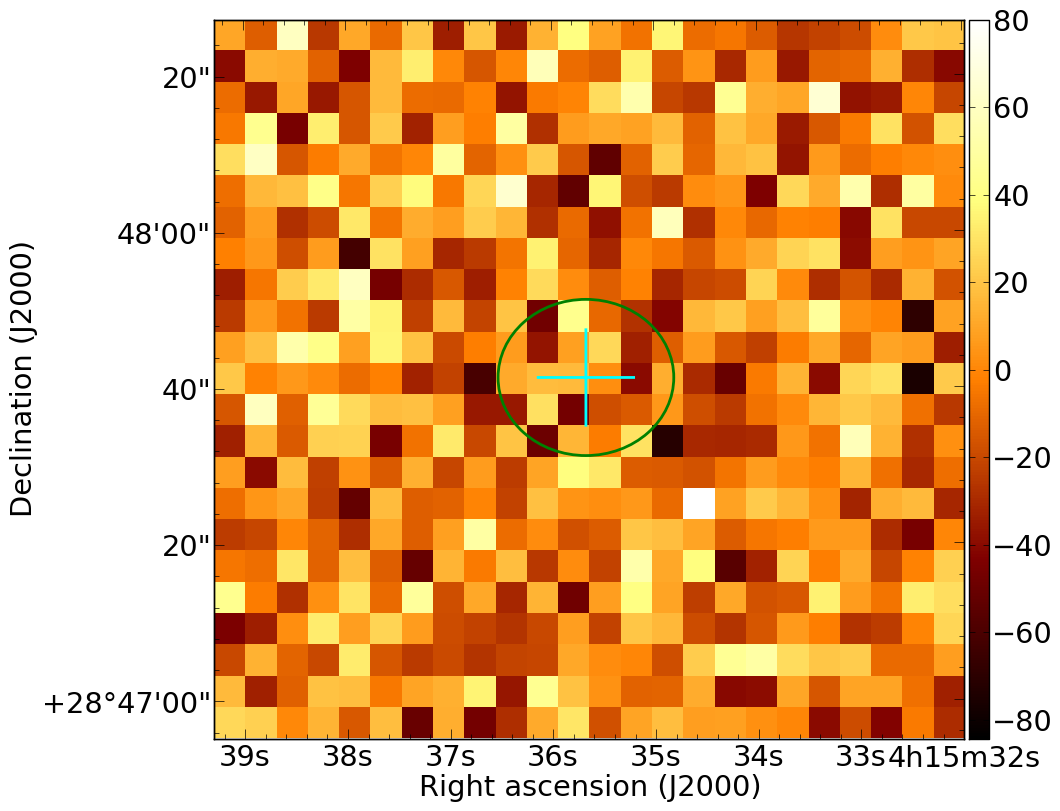}
\caption{850\,$\mu$m (left) and 450\,$\mu$m (right) dust continuum towards J04153566+2847417 Class I \label{diskfig0}}
 \end{figure}

\begin{figure}
\includegraphics[width=4cm]{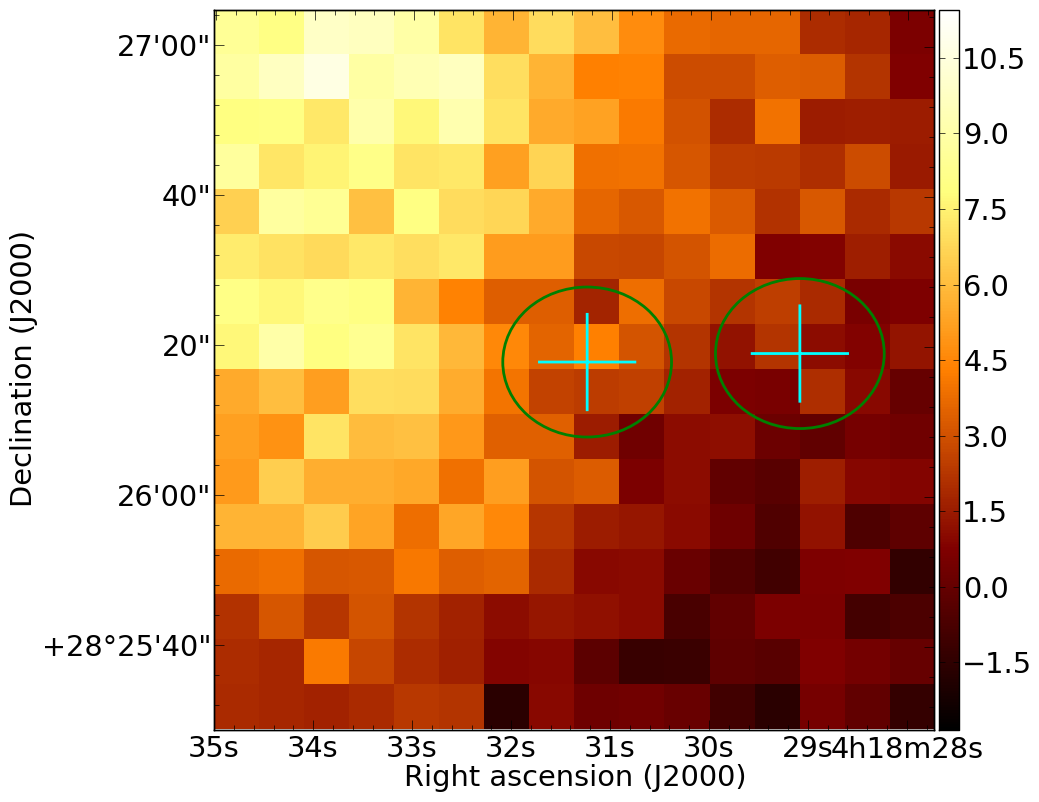}
\includegraphics[width=4cm]{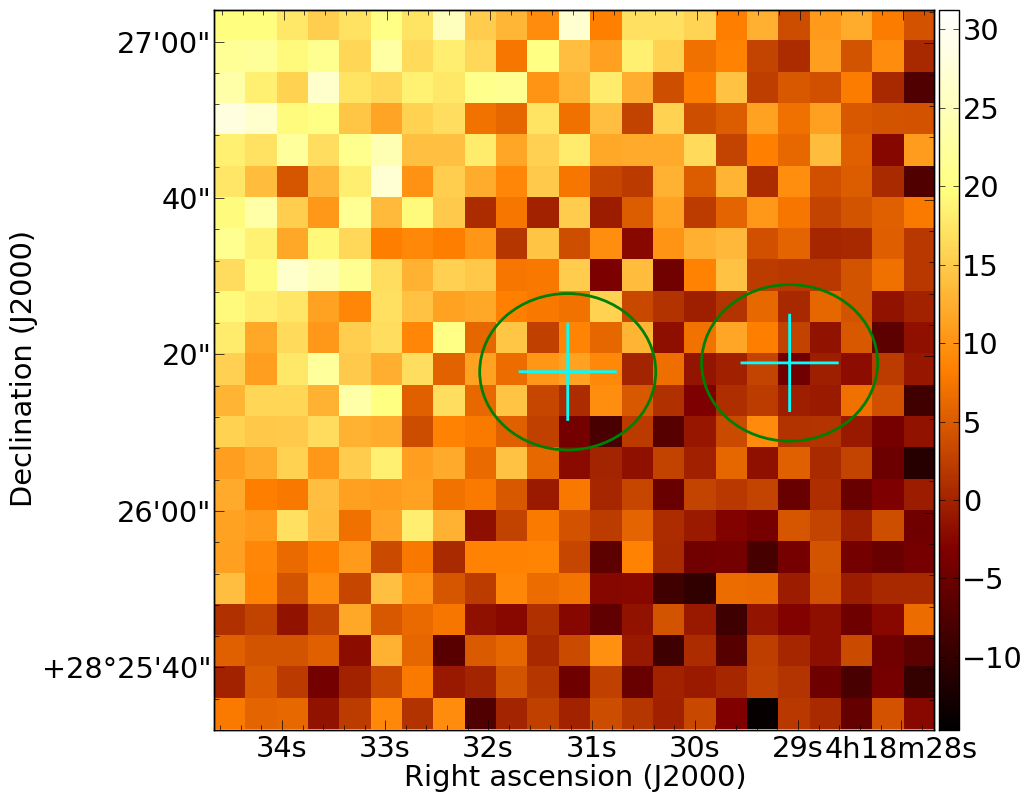}
\caption{as Fig.~\ref{diskfig0} SSTTau041831.2+28261 Class I \label{diskfig1}}
 \end{figure}

\begin{figure}
\includegraphics[width=4cm]{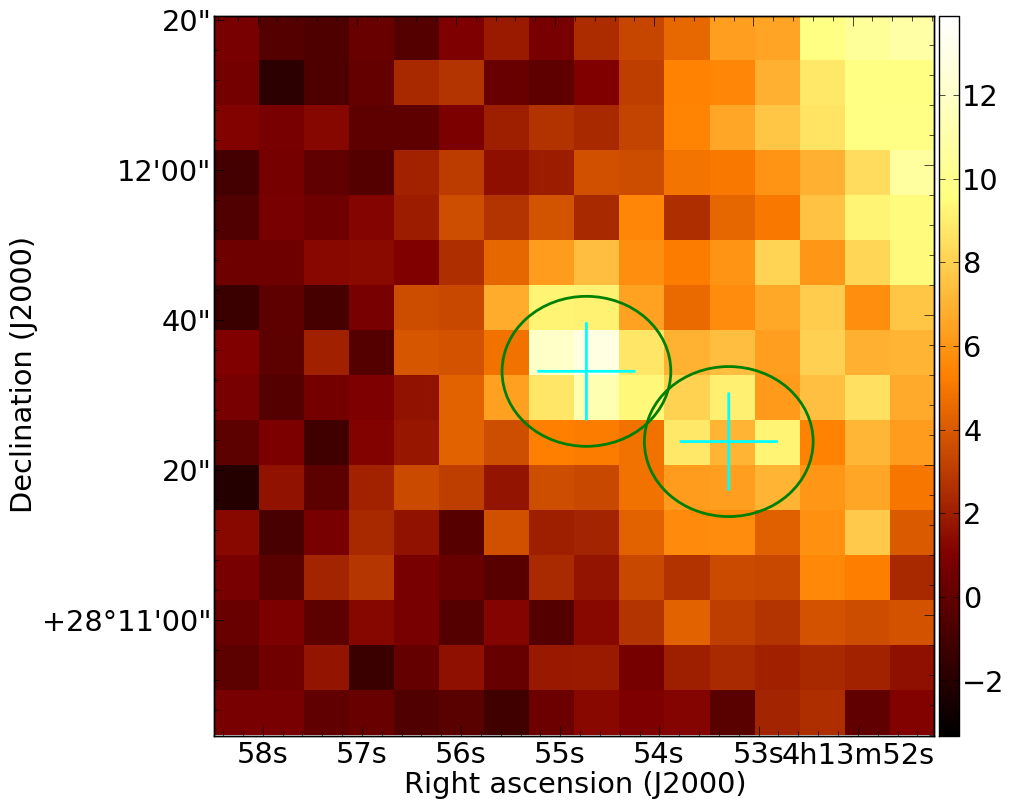}
\includegraphics[width=4cm]{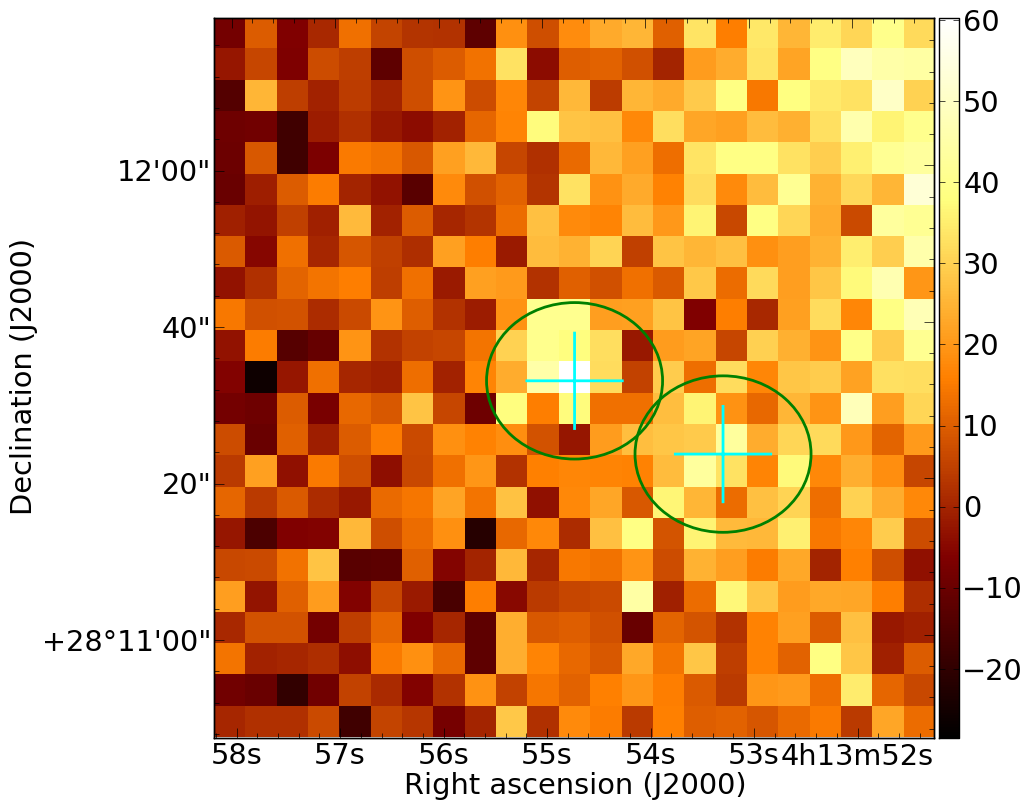}
\caption{as Fig.~\ref{diskfig0} IRAS04108+2803B Class I \label{diskfig2}}
 \end{figure}

\begin{figure}
\includegraphics[width=4cm]{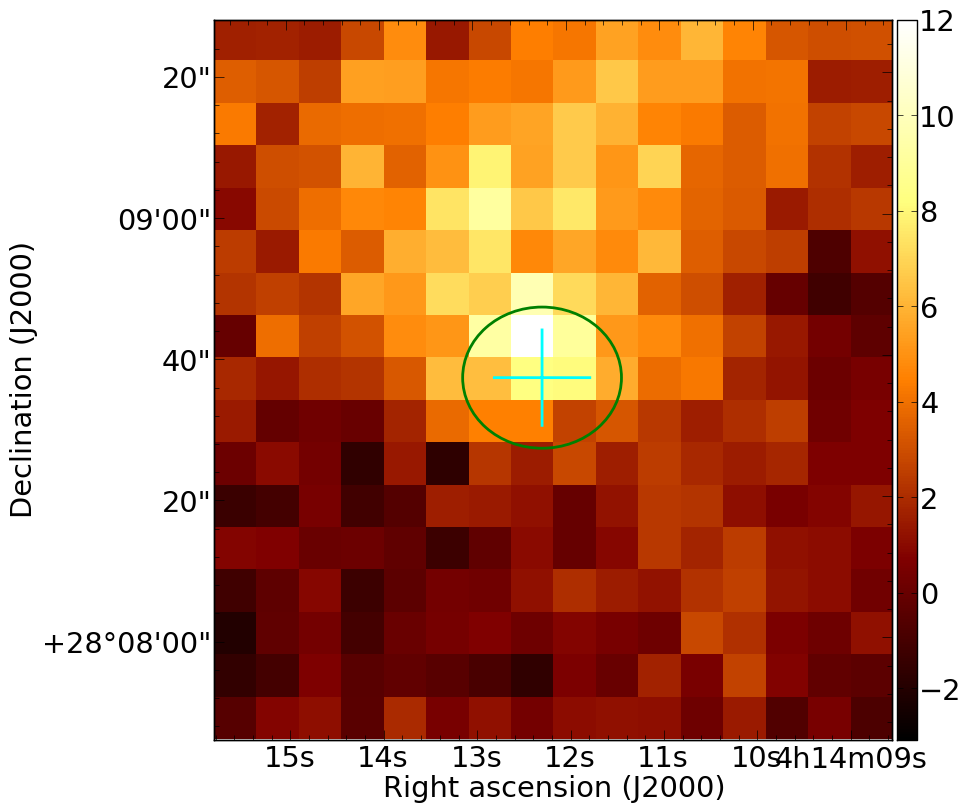}
\includegraphics[width=4cm]{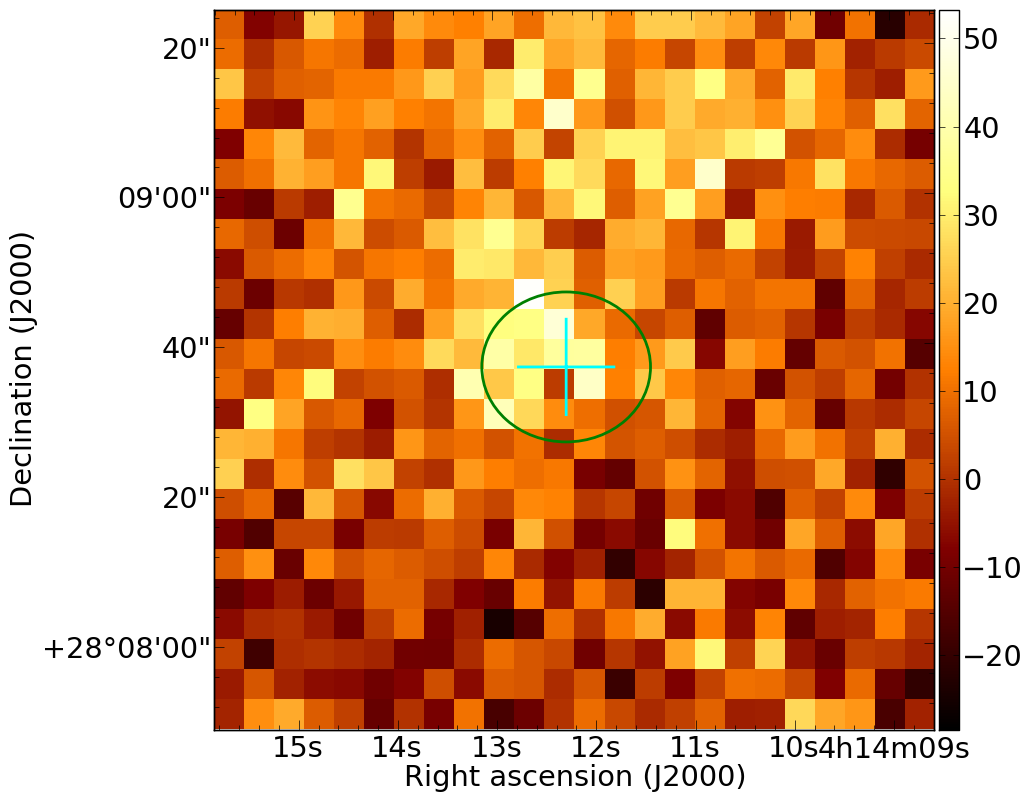}
\caption{as Fig.~\ref{diskfig0} IRAS04111+2800G Class I \label{diskfig3}}
 \end{figure}

\begin{figure}
\includegraphics[width=4cm]{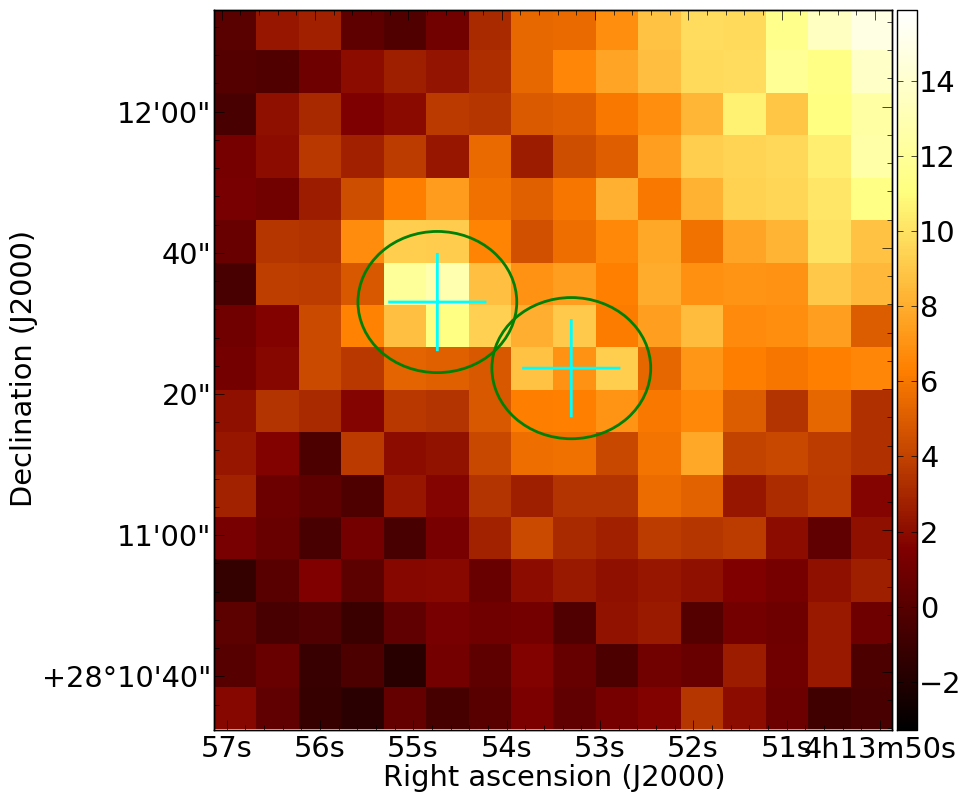}
\includegraphics[width=4cm]{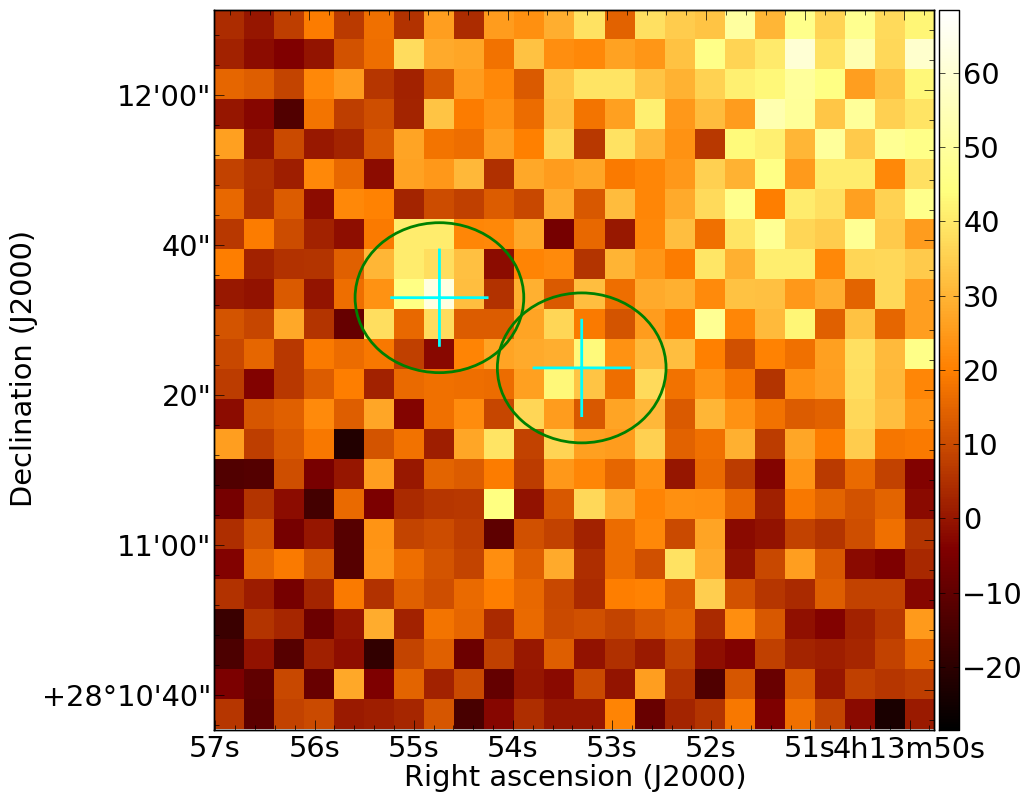}
\caption{as Fig.~\ref{diskfig0} IRAS04108+2803A Class I \label{diskfig4}}
 \end{figure}

\begin{figure}
\includegraphics[width=4cm]{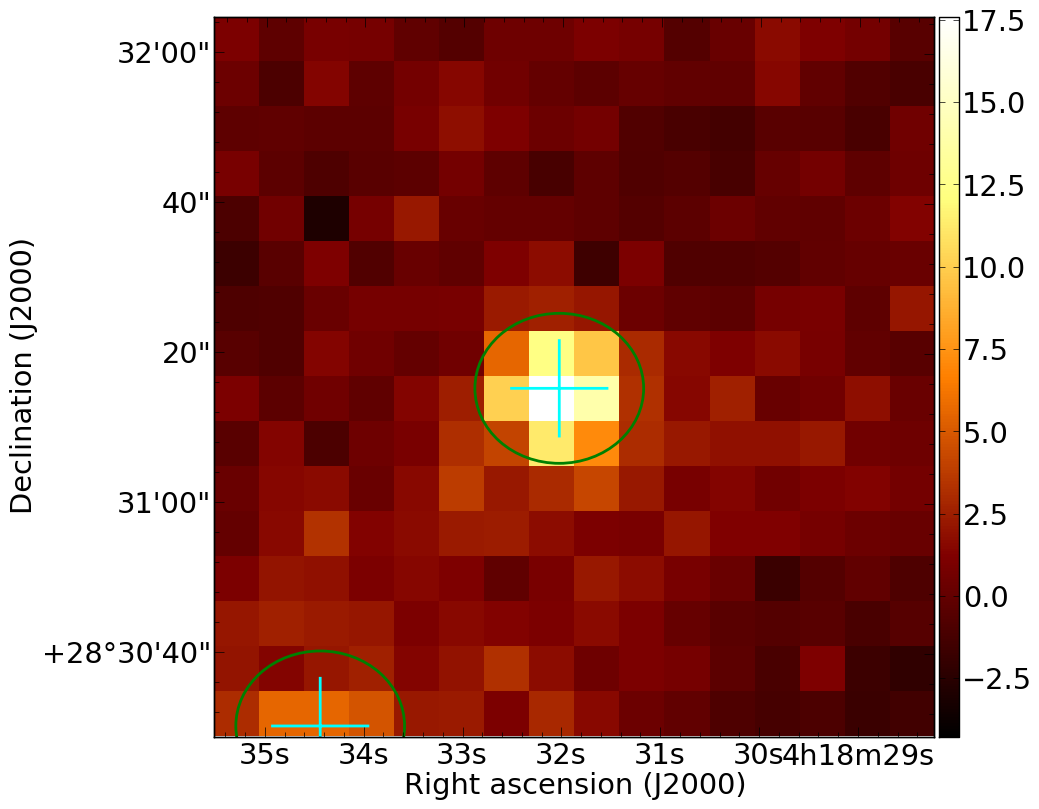}
\includegraphics[width=4cm]{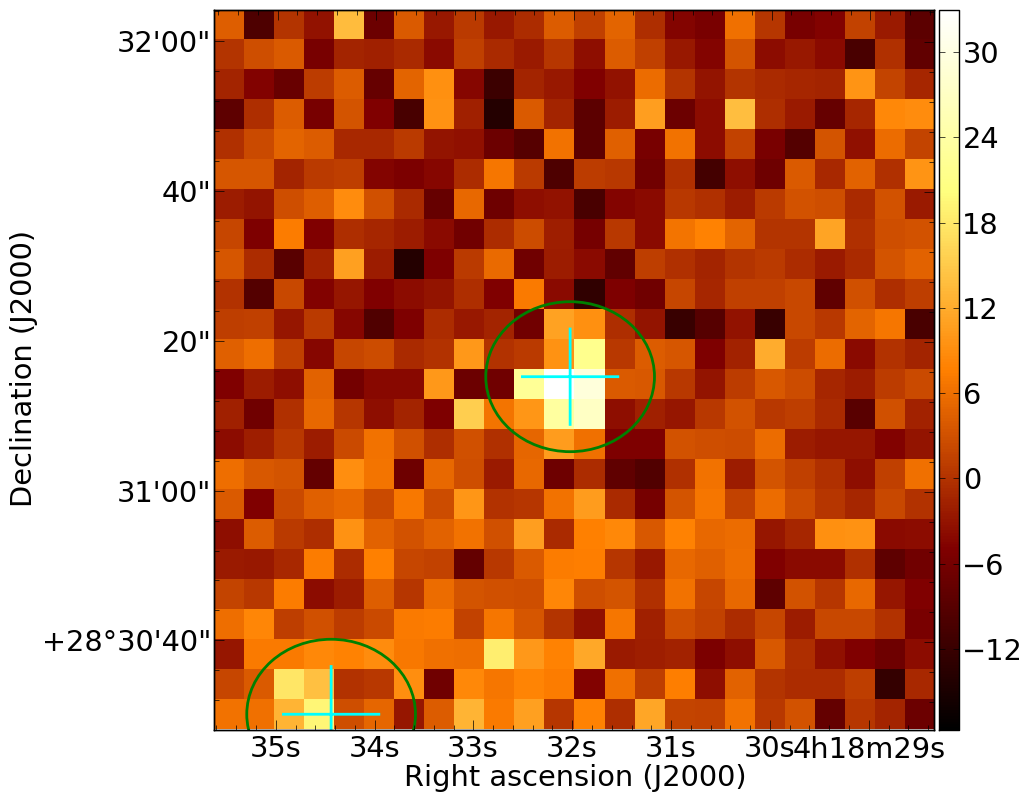}
\caption{as Fig.~\ref{diskfig0} IRAS04154+2823 Class I \label{diskfig5}}
 \end{figure}

\begin{figure}
\includegraphics[width=4cm]{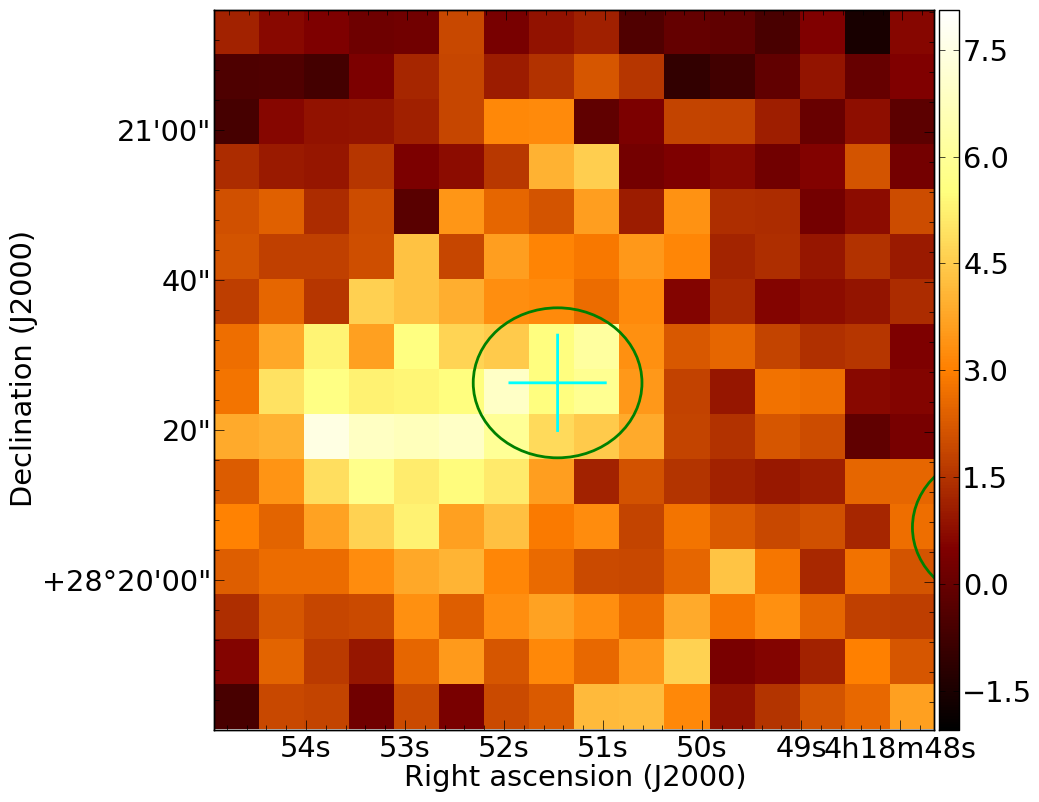}
\includegraphics[width=4cm]{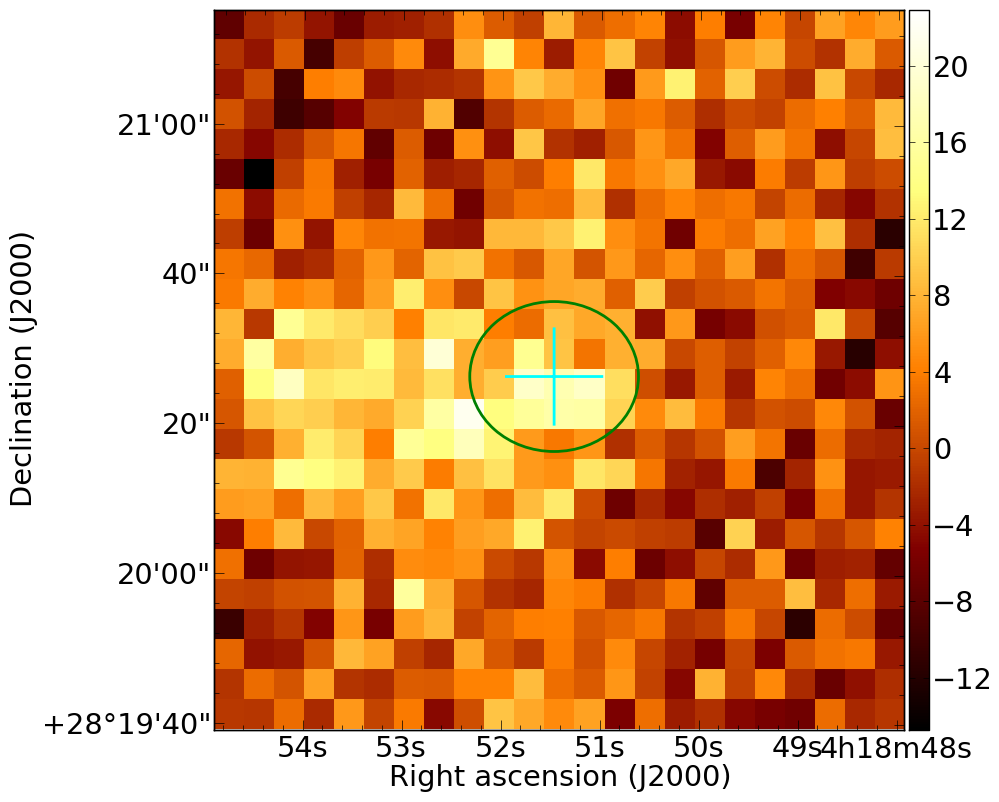}
\caption{as Fig.~\ref{diskfig0} CoKuTau/1 Class I \label{diskfig6}}
 \end{figure}

\begin{figure}
\includegraphics[width=4cm]{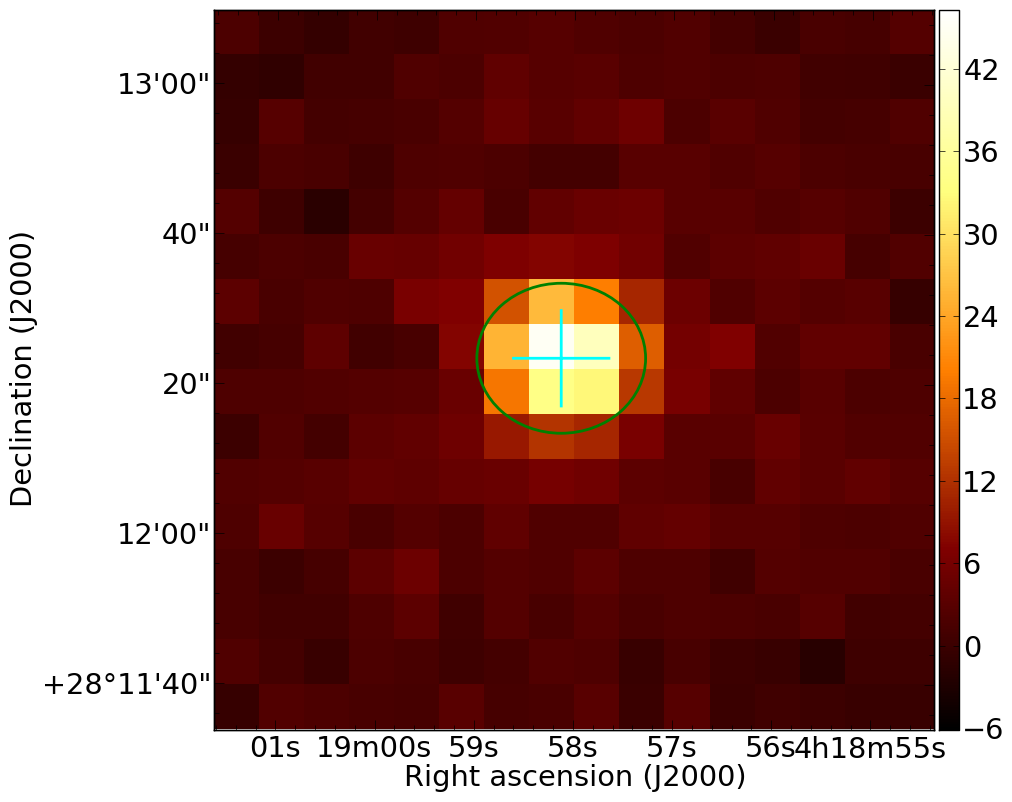}
\includegraphics[width=4cm]{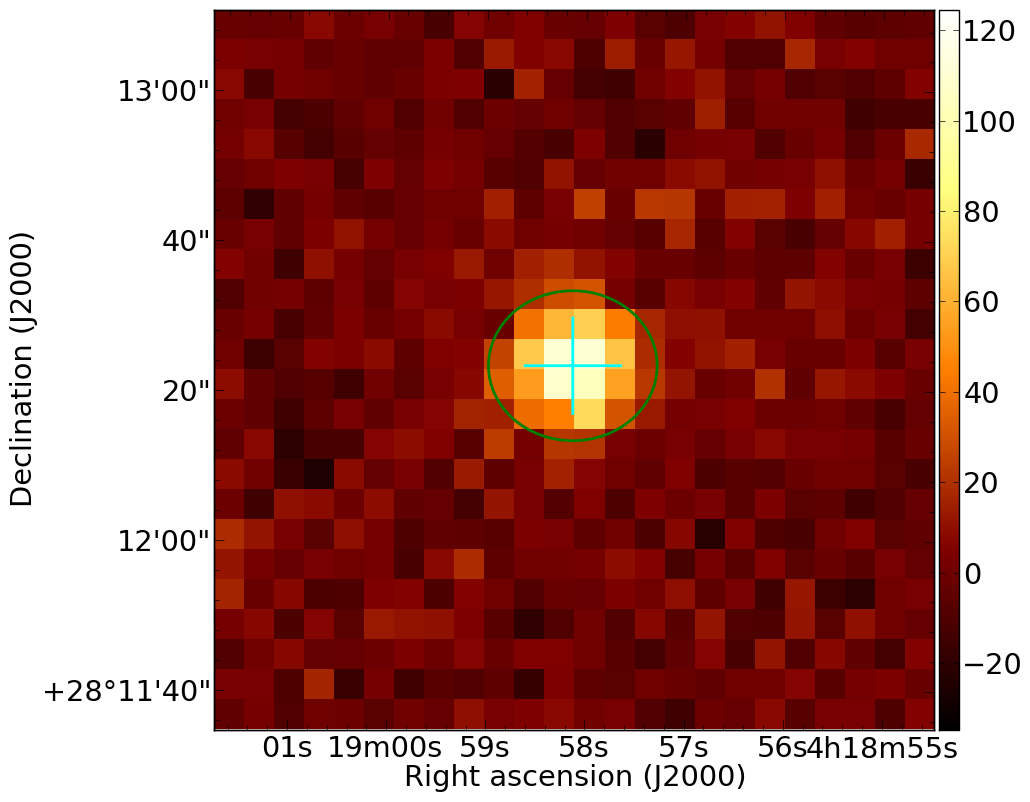}
\caption{as Fig.~\ref{diskfig0} IRAS04158+2805 Class I \label{diskfig7}}
 \end{figure}

\begin{figure}
\includegraphics[width=4cm]{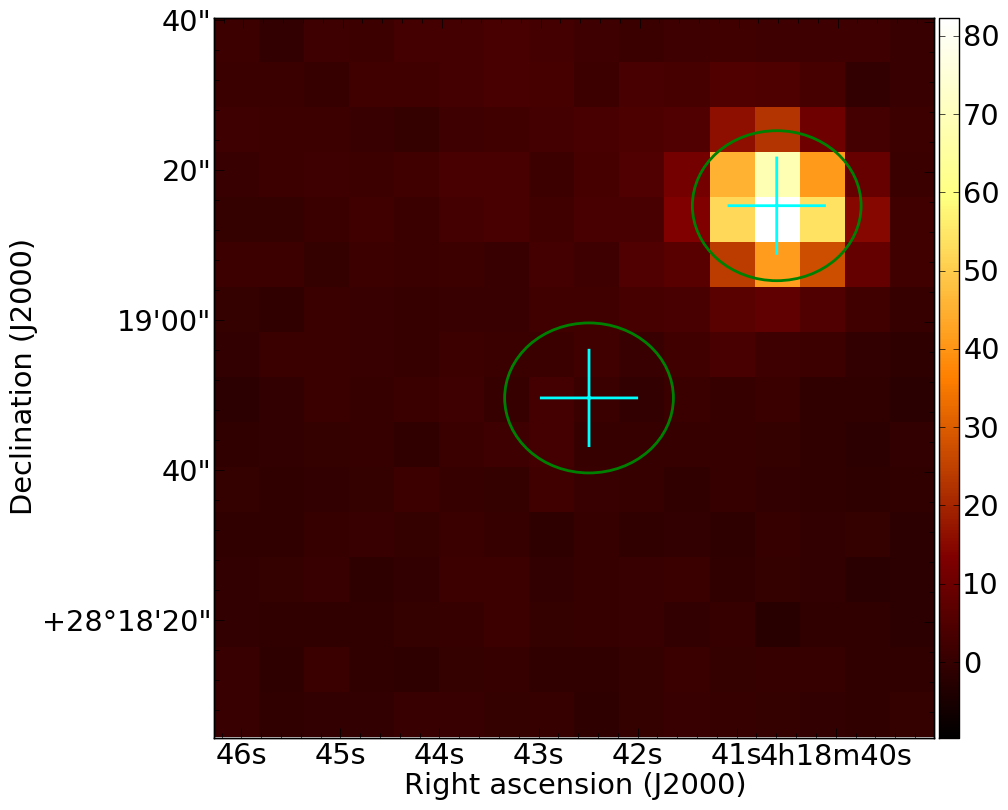}
\includegraphics[width=4cm]{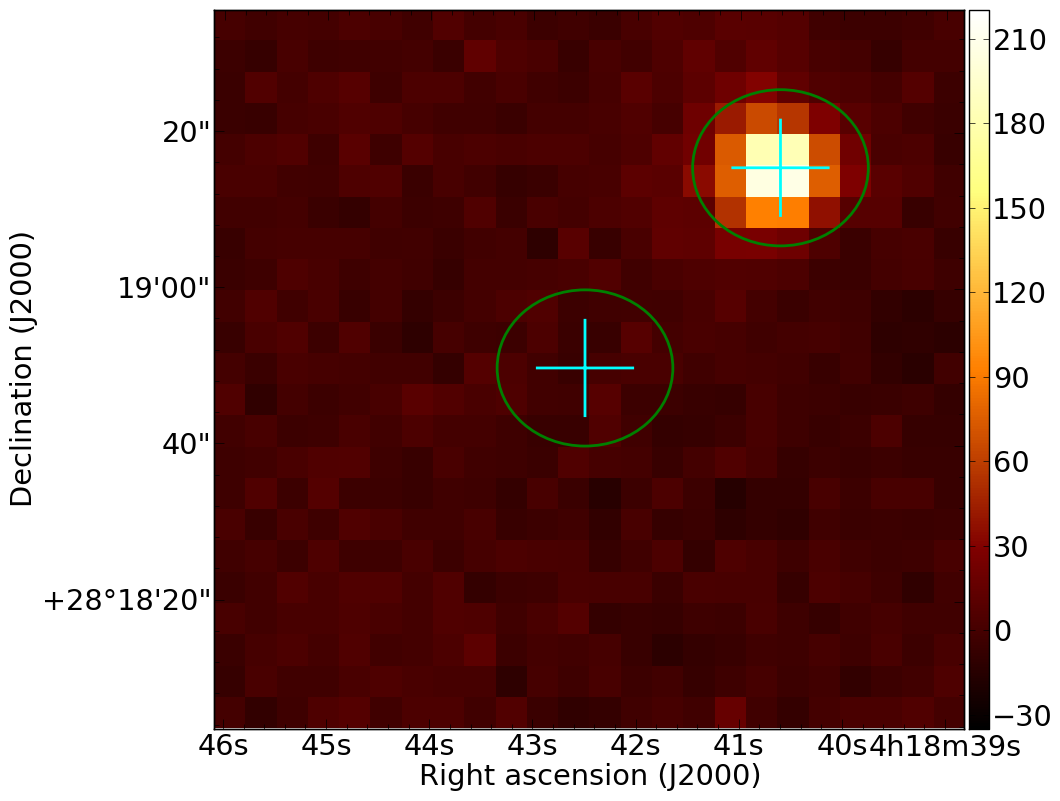}
\caption{as Fig.~\ref{diskfig0} V410X-ray7 Class II \label{diskfig8}}
 \end{figure}

\begin{figure}
\includegraphics[width=4cm]{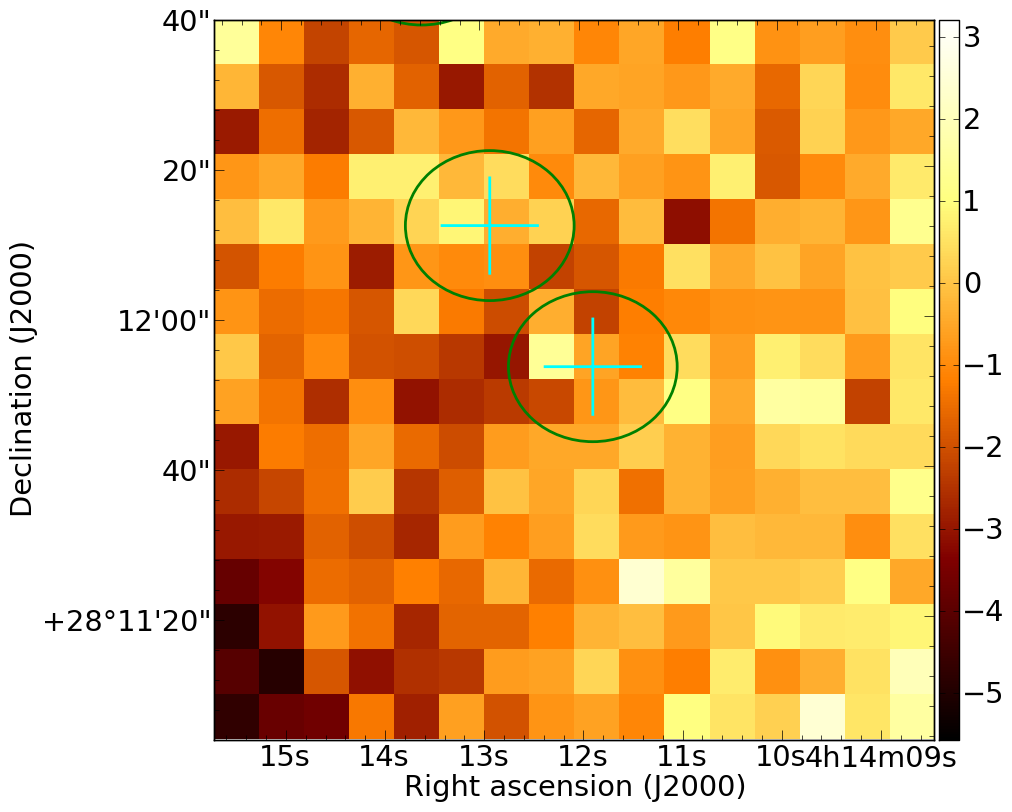}
\includegraphics[width=4cm]{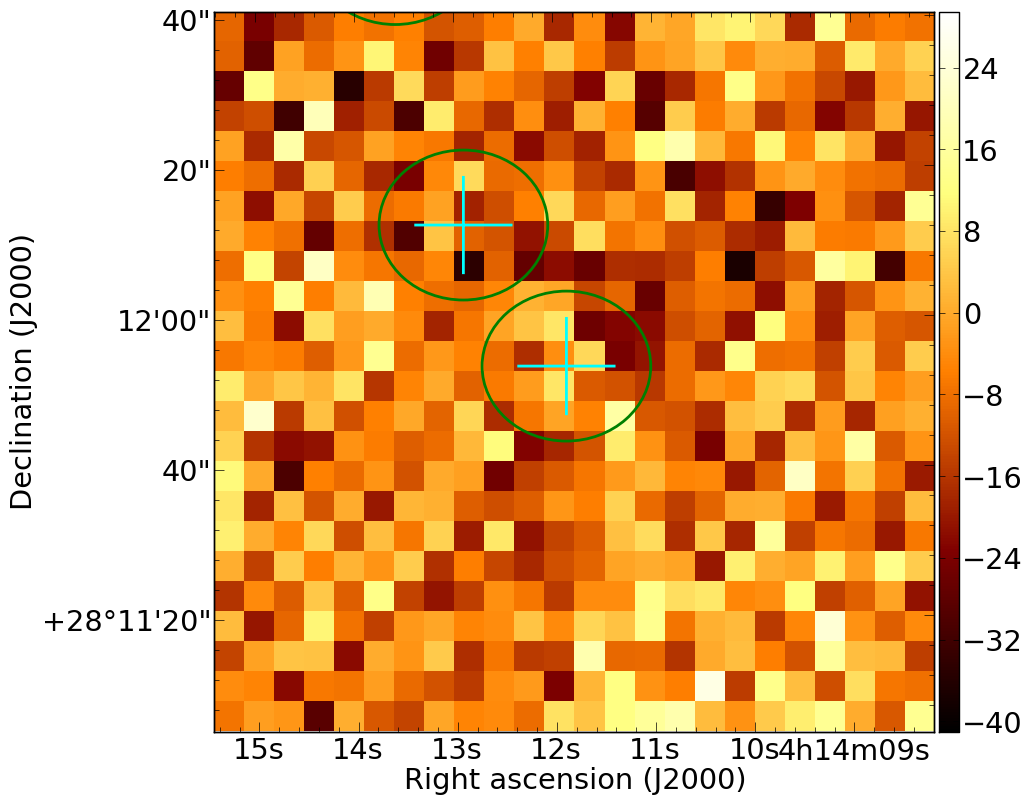}
\caption{as Fig.~\ref{diskfig0} J04141188+2811535 Class II \label{diskfig9}}
 \end{figure}

\begin{figure}
\includegraphics[width=4cm]{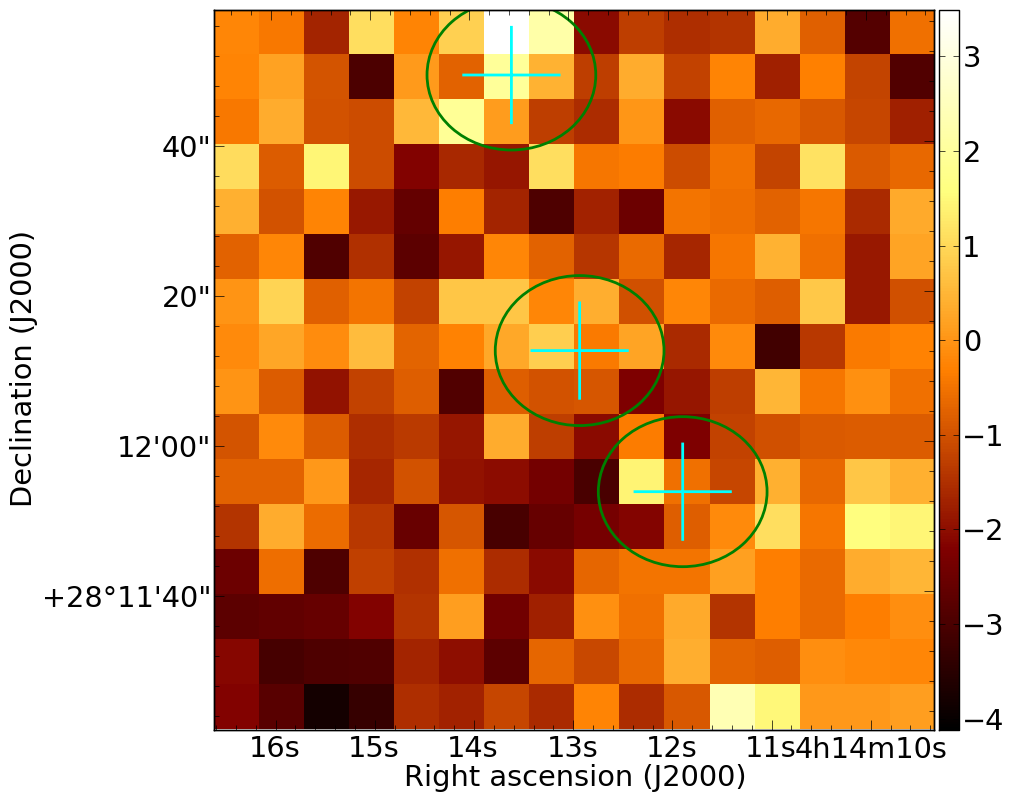}
\includegraphics[width=4cm]{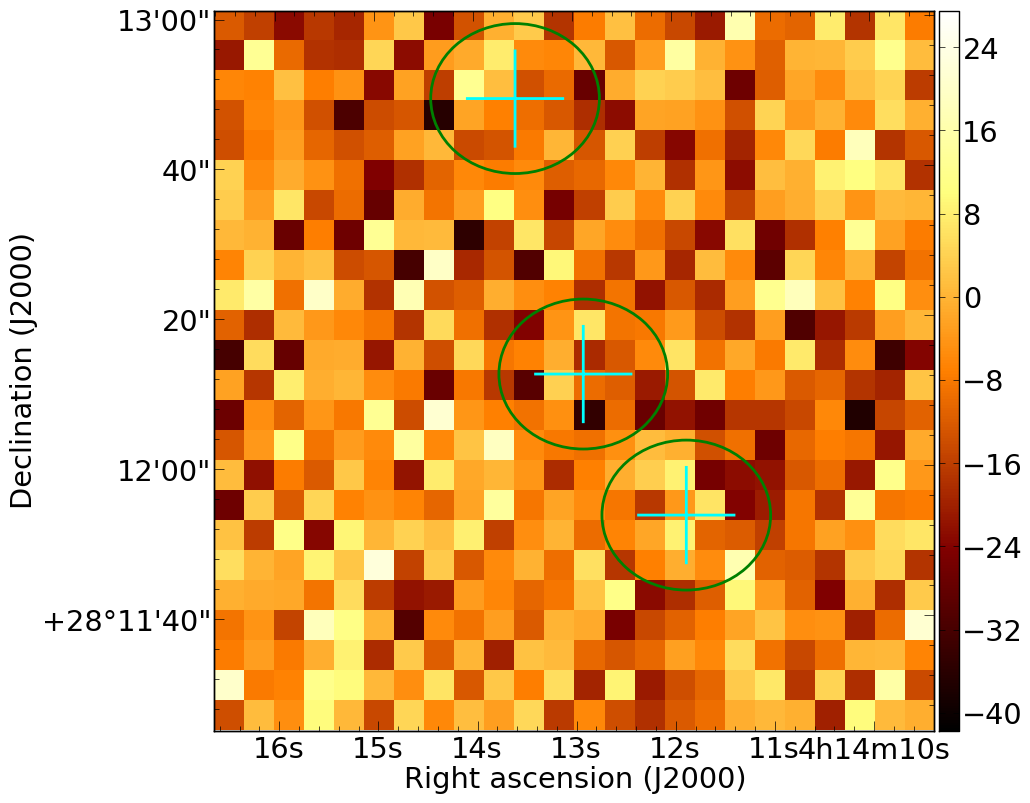}
\caption{as Fig.~\ref{diskfig0} V773TauA+B Class II \label{diskfig10}}
 \end{figure}

\begin{figure}
\includegraphics[width=4cm]{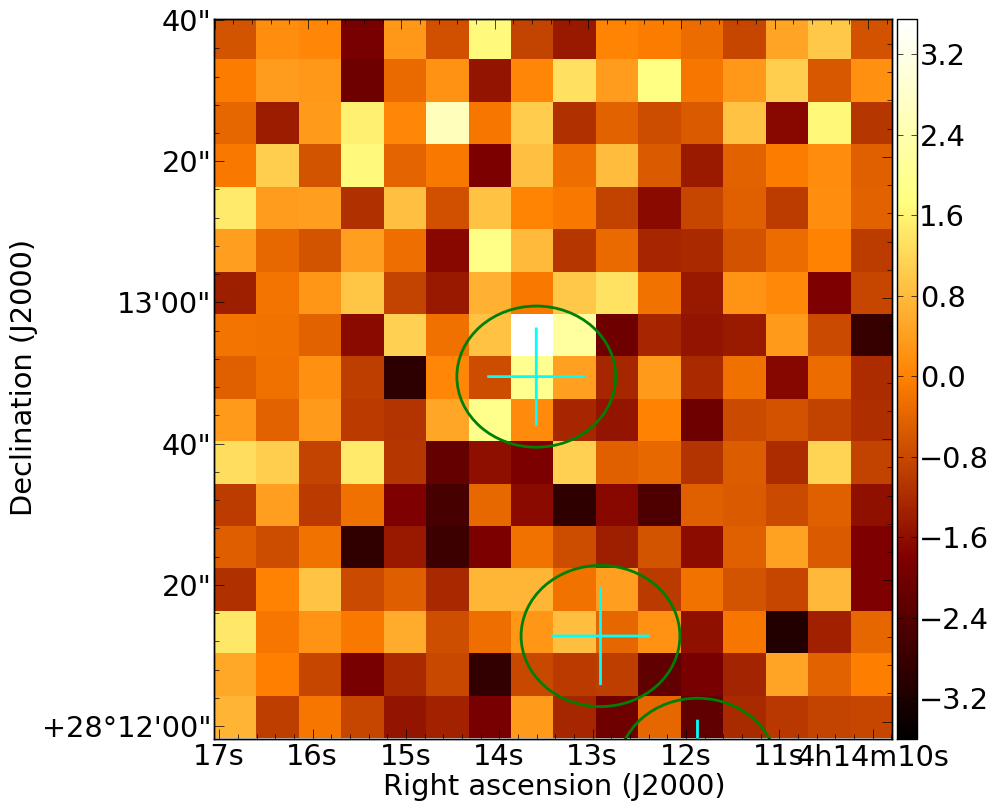}
\includegraphics[width=4cm]{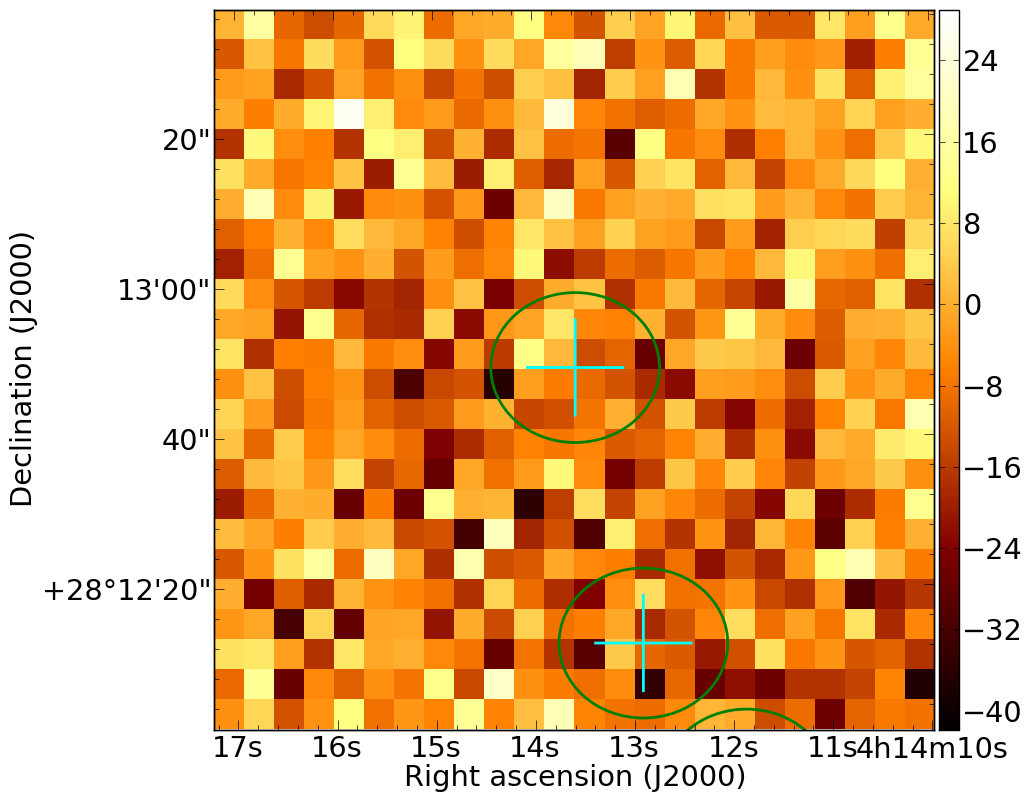}
\caption{as Fig.~\ref{diskfig0} FMTau Class II \label{diskfig11}}
 \end{figure}

\begin{figure}
\includegraphics[width=4cm]{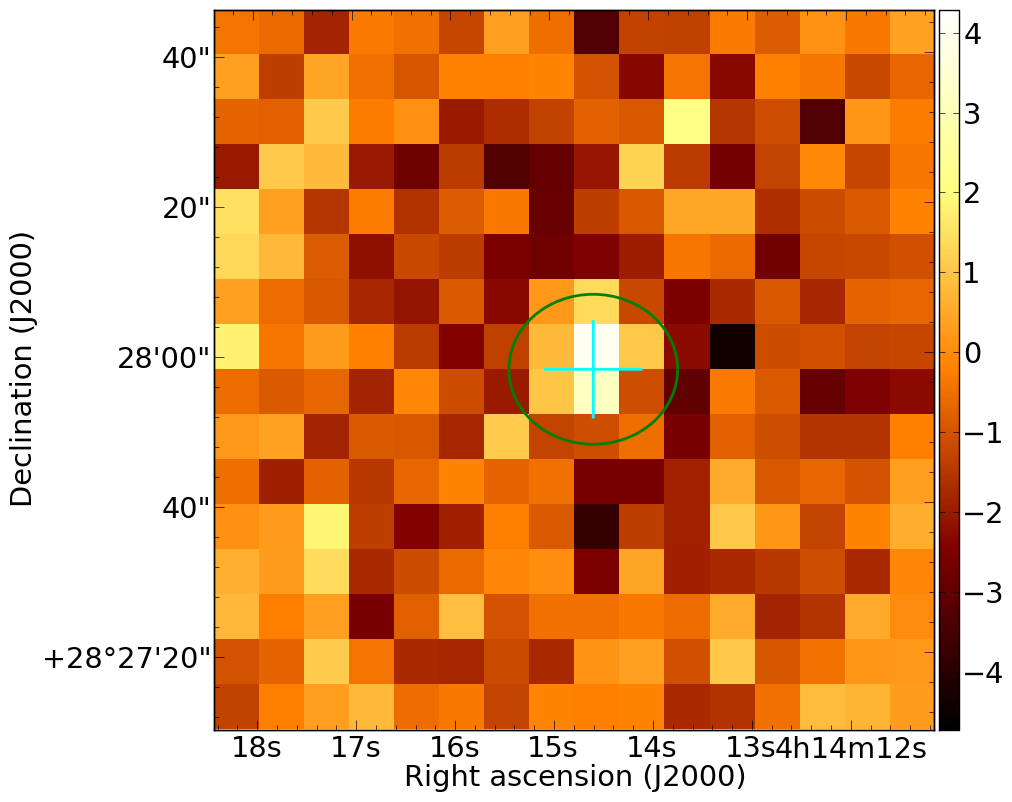}
\includegraphics[width=4cm]{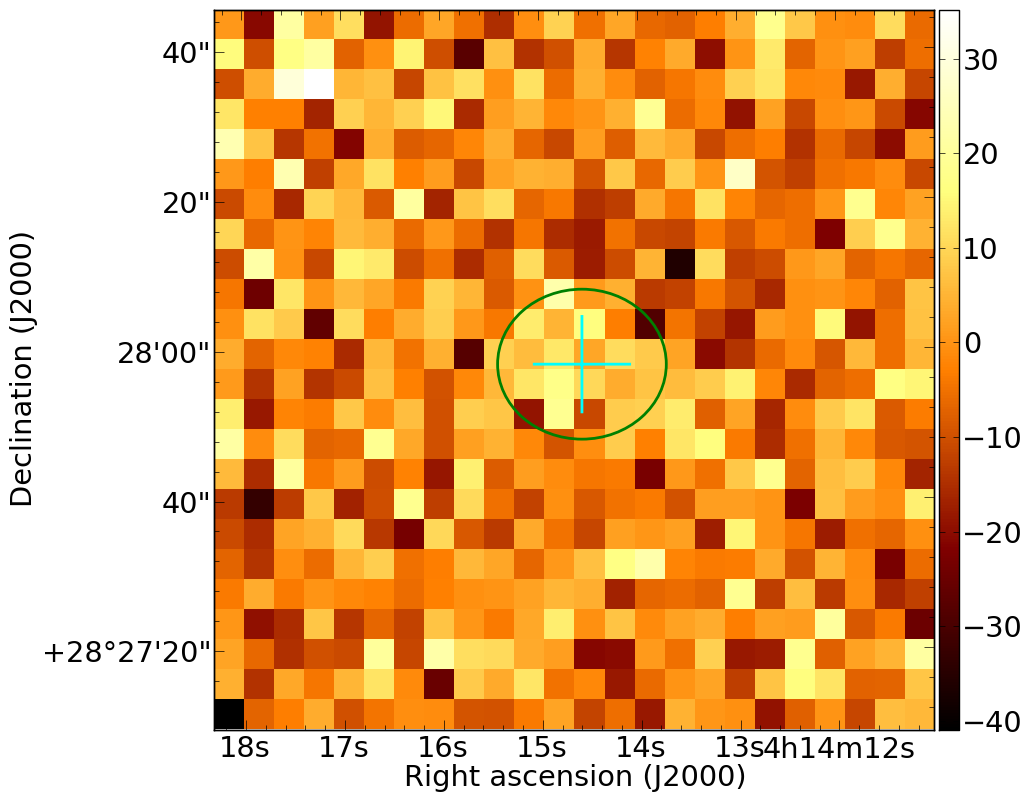}
\caption{as Fig.~\ref{diskfig0} FNTau Class II \label{diskfig12}}
 \end{figure}

\begin{figure}
\includegraphics[width=4cm]{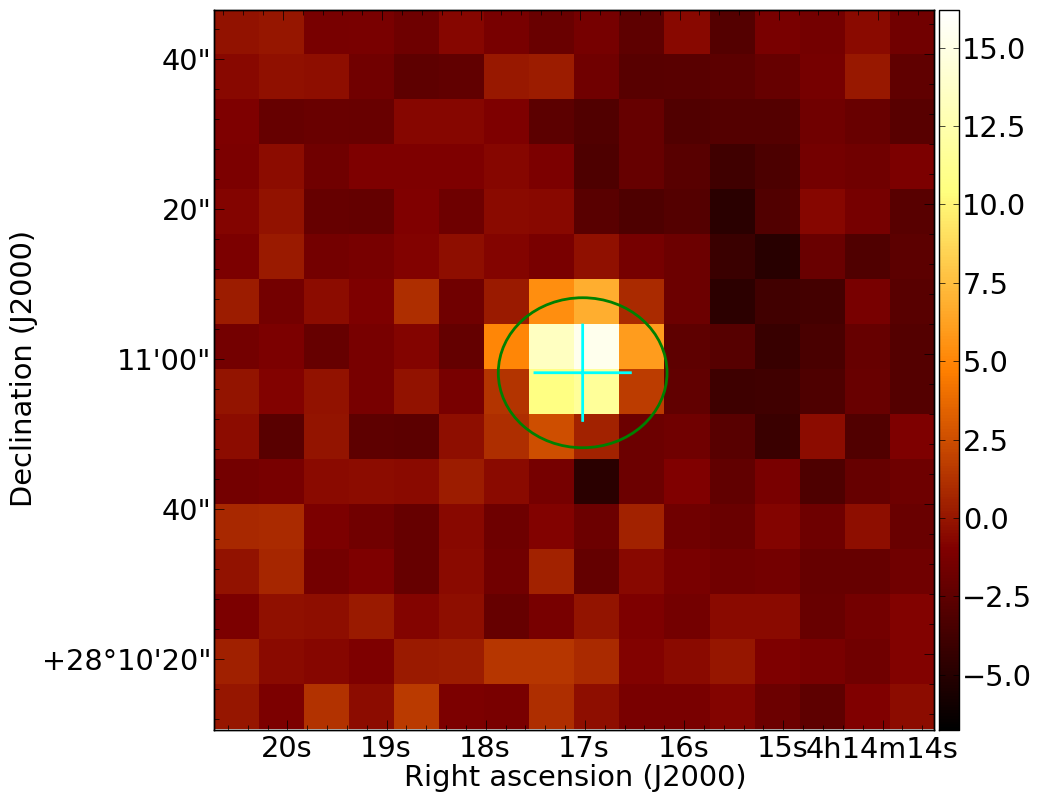}
\includegraphics[width=4cm]{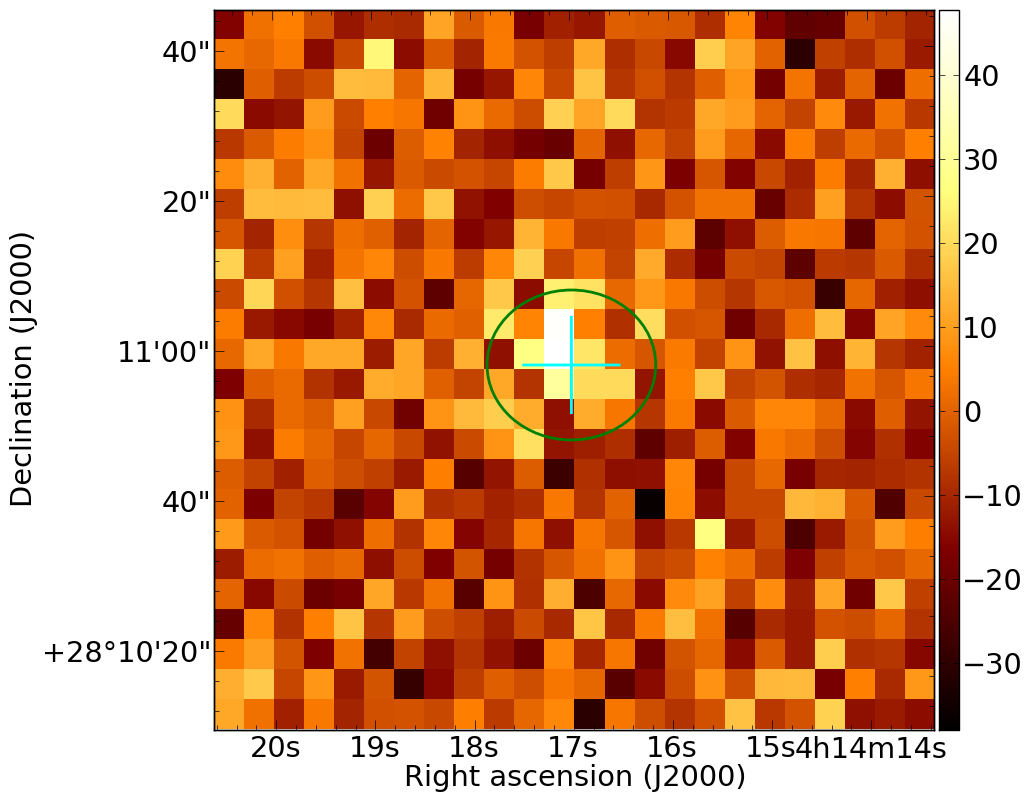}
\caption{as Fig.~\ref{diskfig0} CWTau Class II \label{diskfig13}}
 \end{figure}

\begin{figure}
\includegraphics[width=4cm]{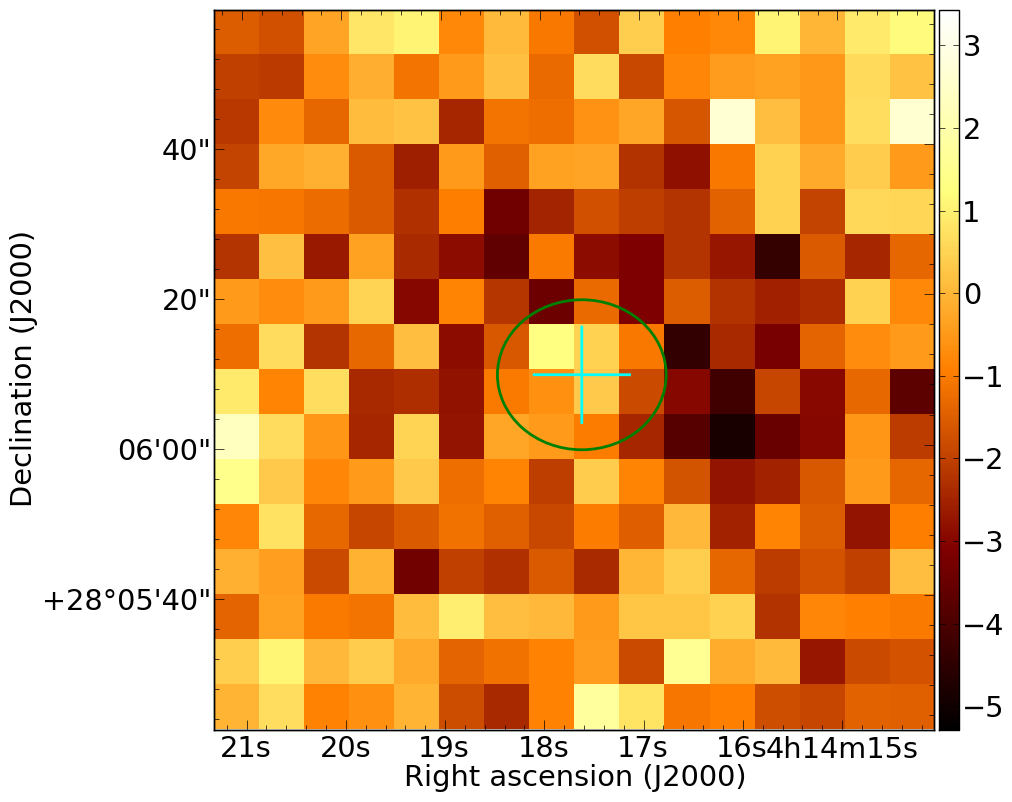}
\includegraphics[width=4cm]{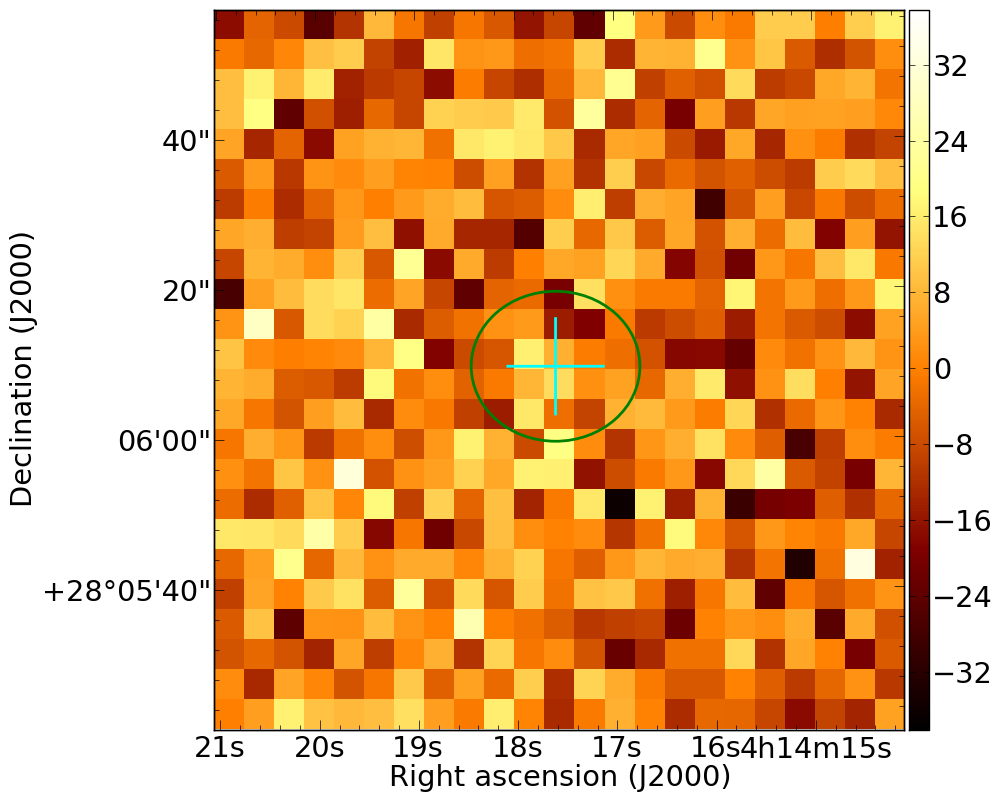}
\caption{as Fig.~\ref{diskfig0} CIDA1 Class II \label{diskfig14}}
 \end{figure}

\begin{figure}
\includegraphics[width=4cm]{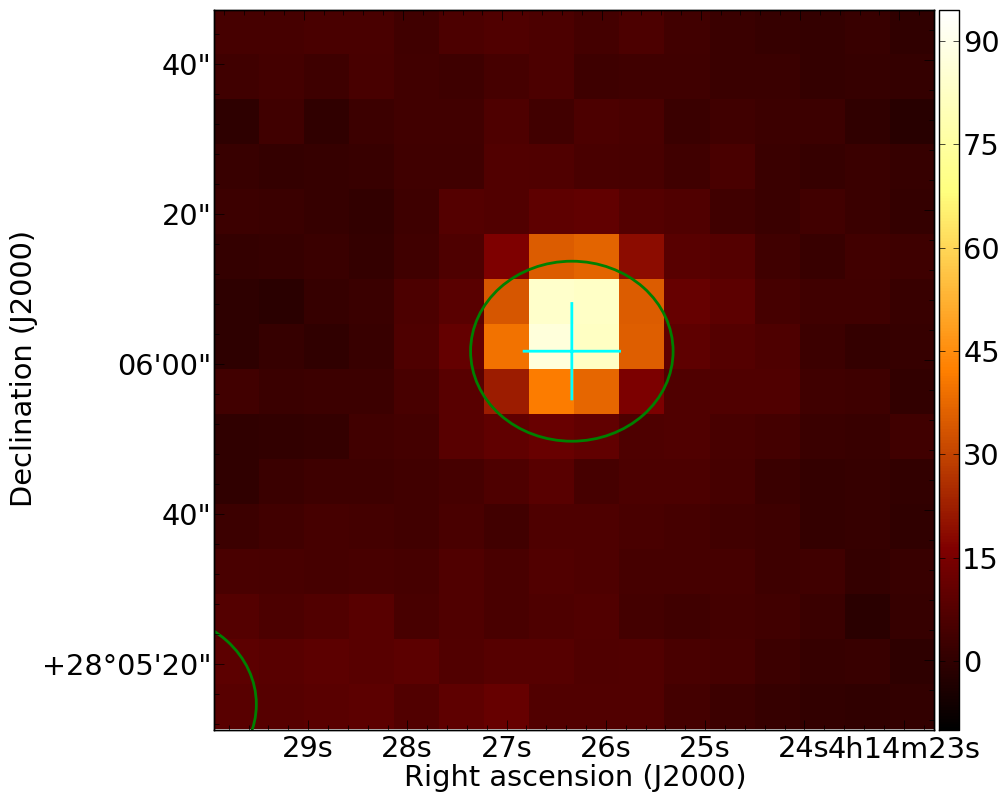}
\includegraphics[width=4cm]{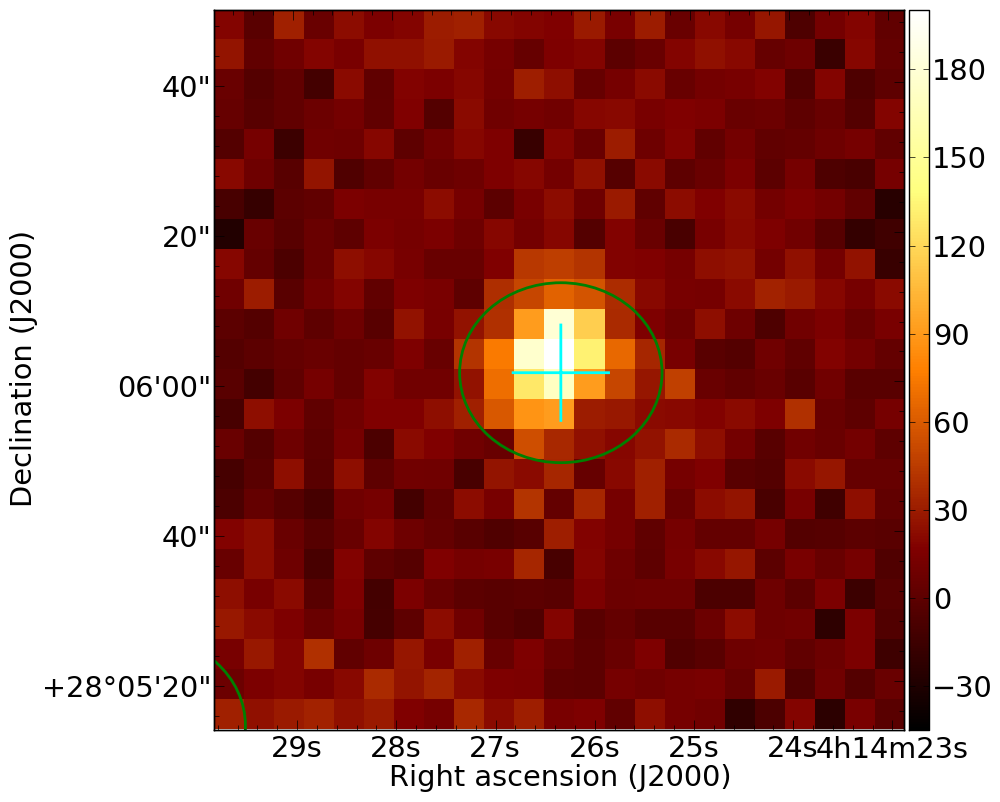}
\caption{as Fig.~\ref{diskfig0} MHO1/2 Class II. A larger 24\arcsec\ aperture was used for this source. \label{diskfig15}}
 \end{figure}

\begin{figure}
\includegraphics[width=4cm]{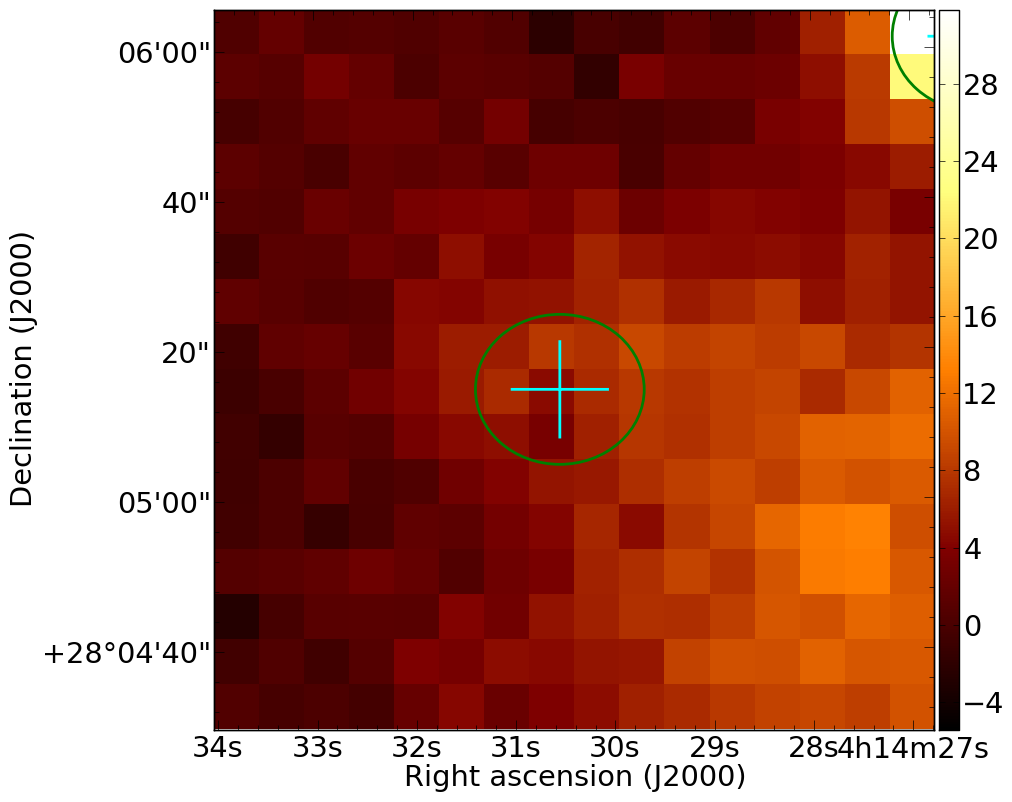}
\includegraphics[width=4cm]{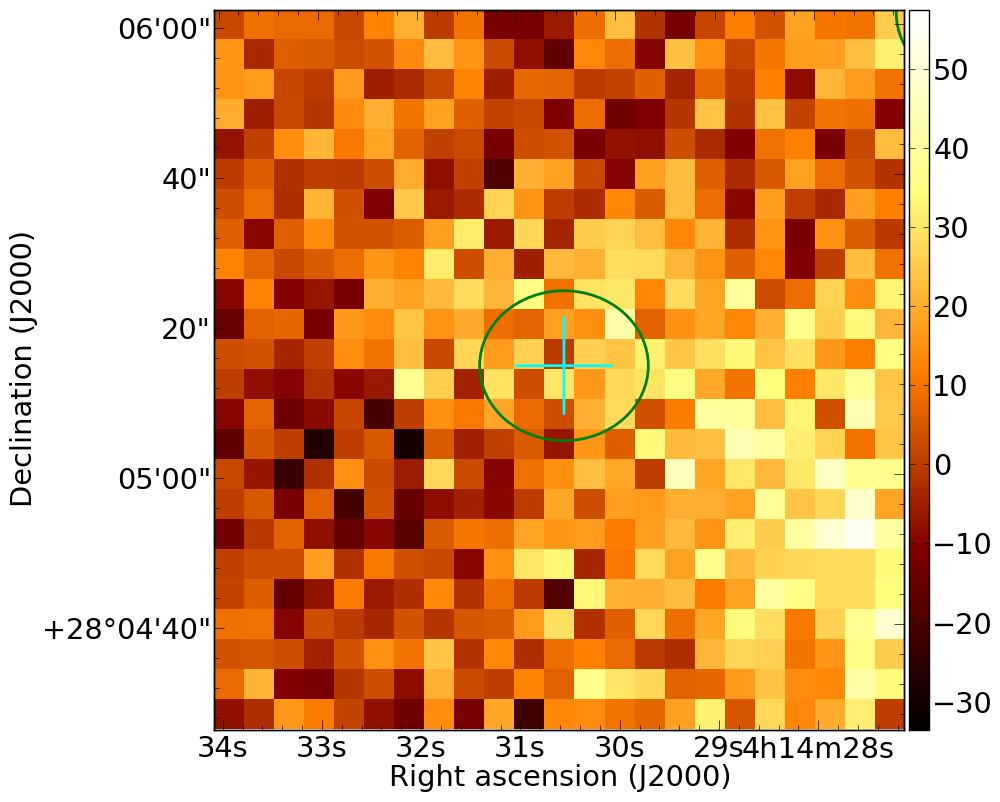}
\caption{as Fig.~\ref{diskfig0} MHO3 Class II \label{diskfig16}}
 \end{figure}

\begin{figure}
\includegraphics[width=4cm]{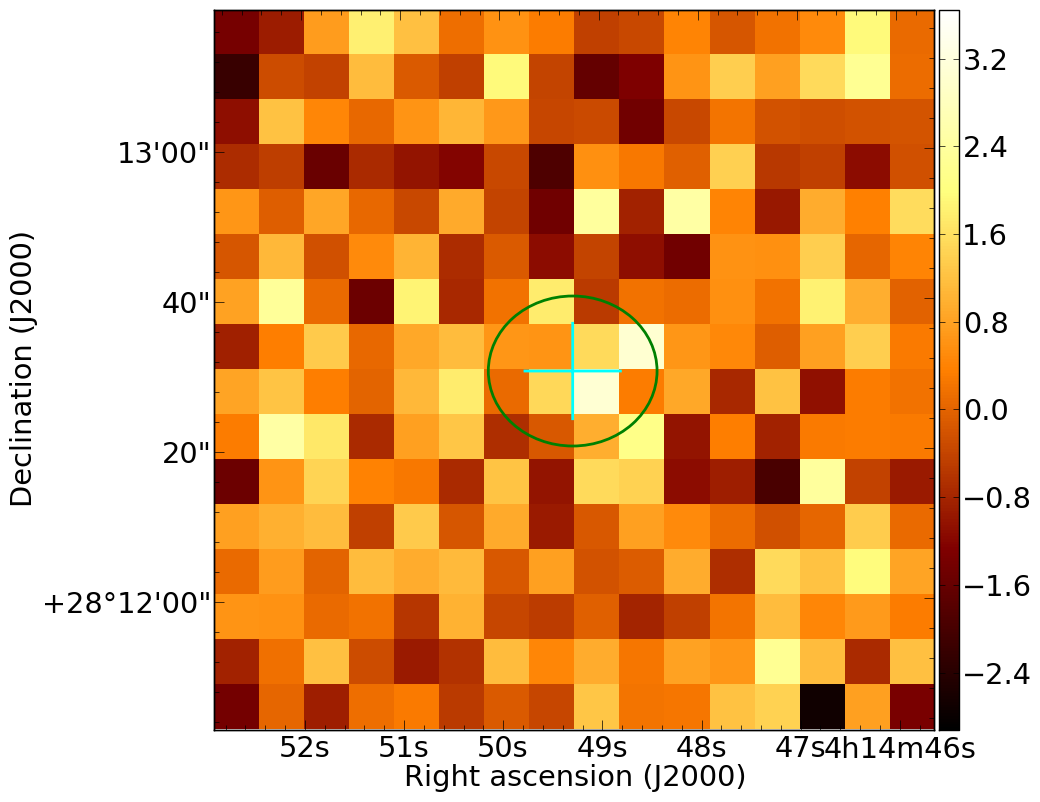}
\includegraphics[width=4cm]{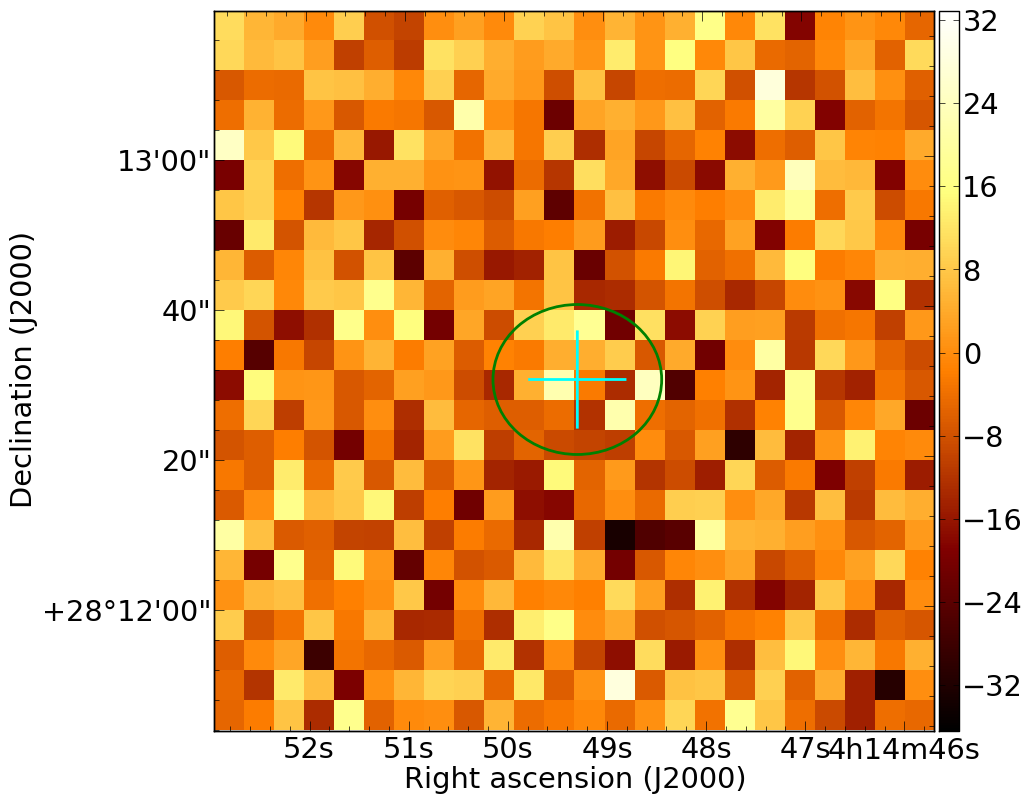}
\caption{as Fig.~\ref{diskfig0} FOTauA+B Class II \label{diskfig17}}
 \end{figure}

\begin{figure}
\includegraphics[width=4cm]{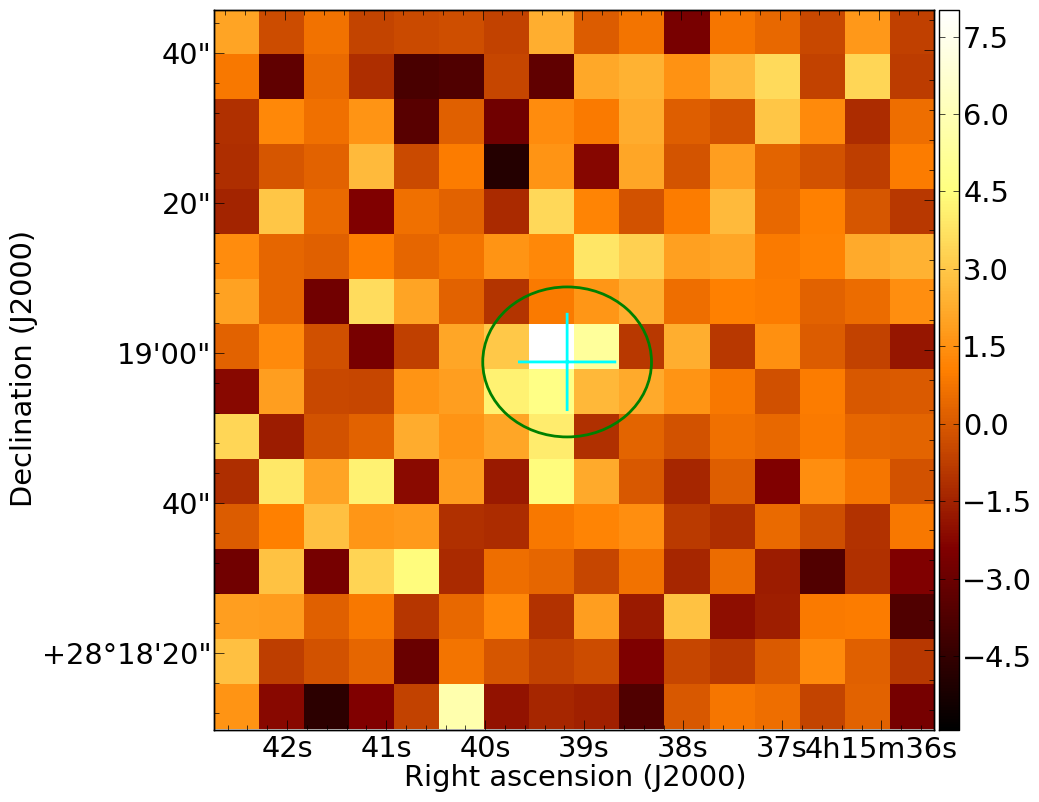}
\includegraphics[width=4cm]{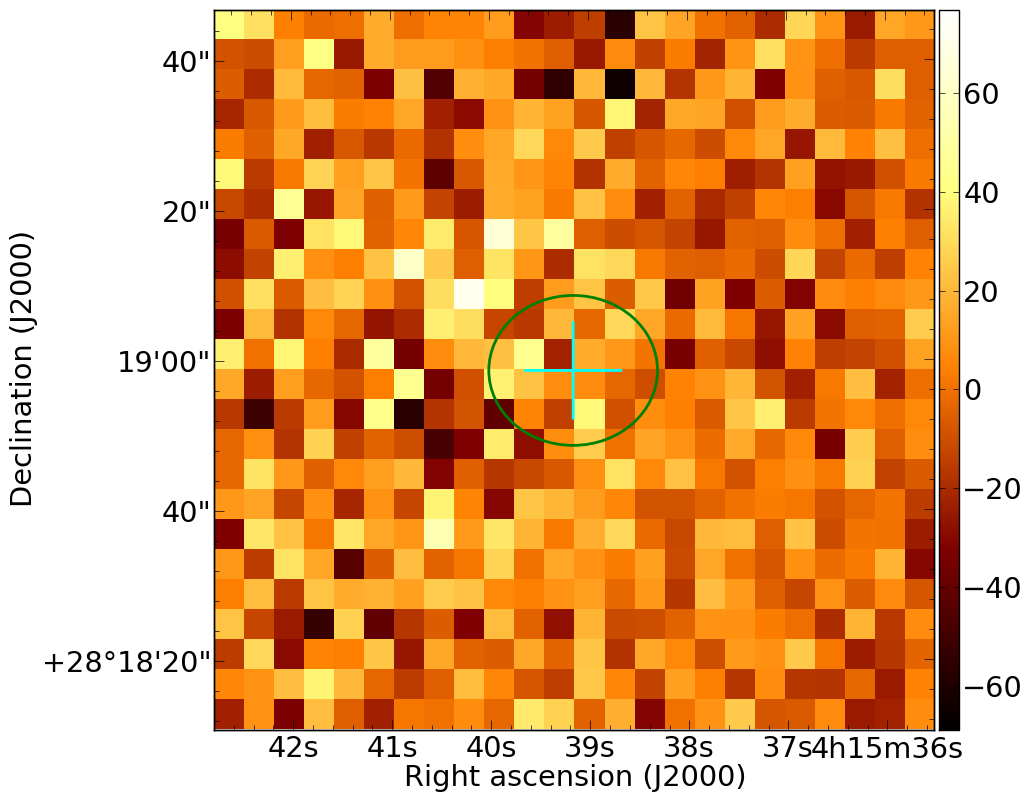}
\caption{as Fig.~\ref{diskfig0} J04153916+2818586 Class II \label{diskfig18}}
 \end{figure}

\begin{figure}
\includegraphics[width=4cm]{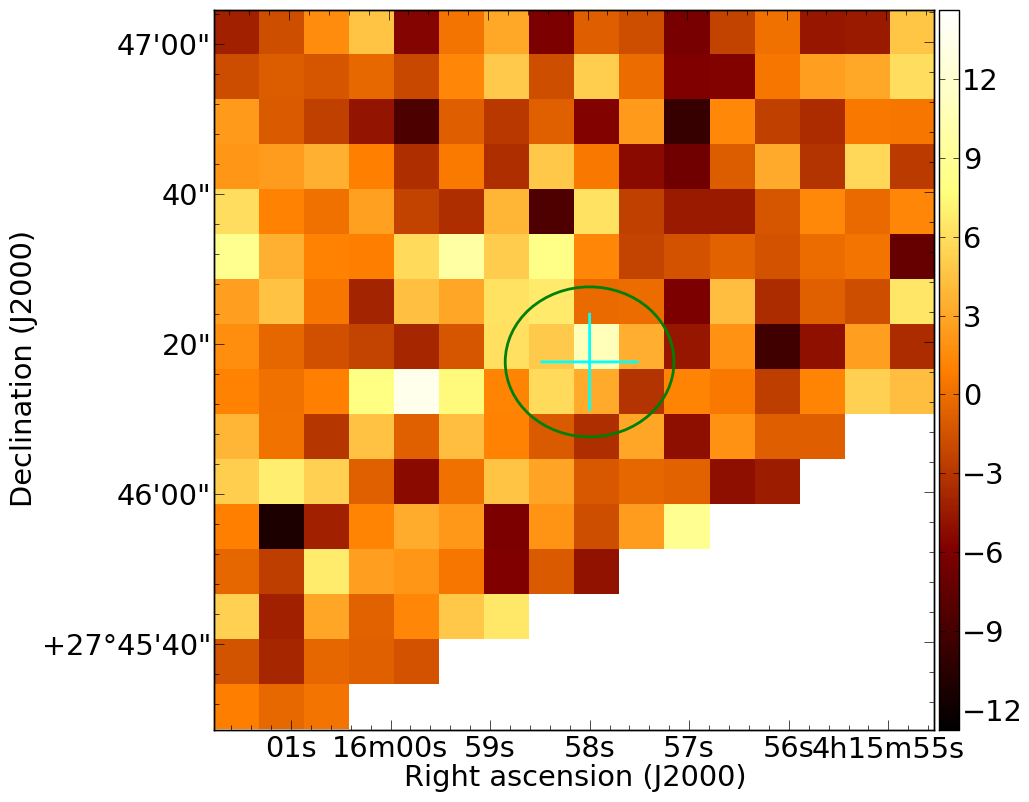}
\includegraphics[width=4cm]{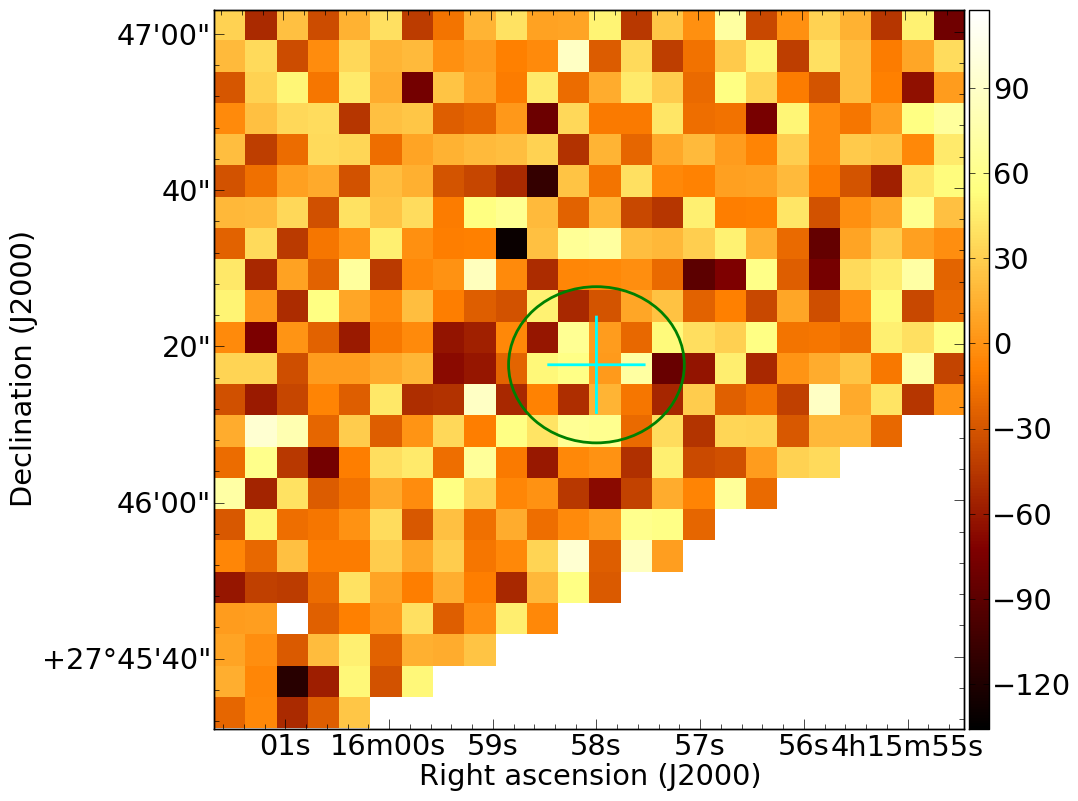}
\caption{as Fig.~\ref{diskfig0} J04155799+2746175 Class II \label{diskfig19}}
 \end{figure}

\begin{figure}
\includegraphics[width=4cm]{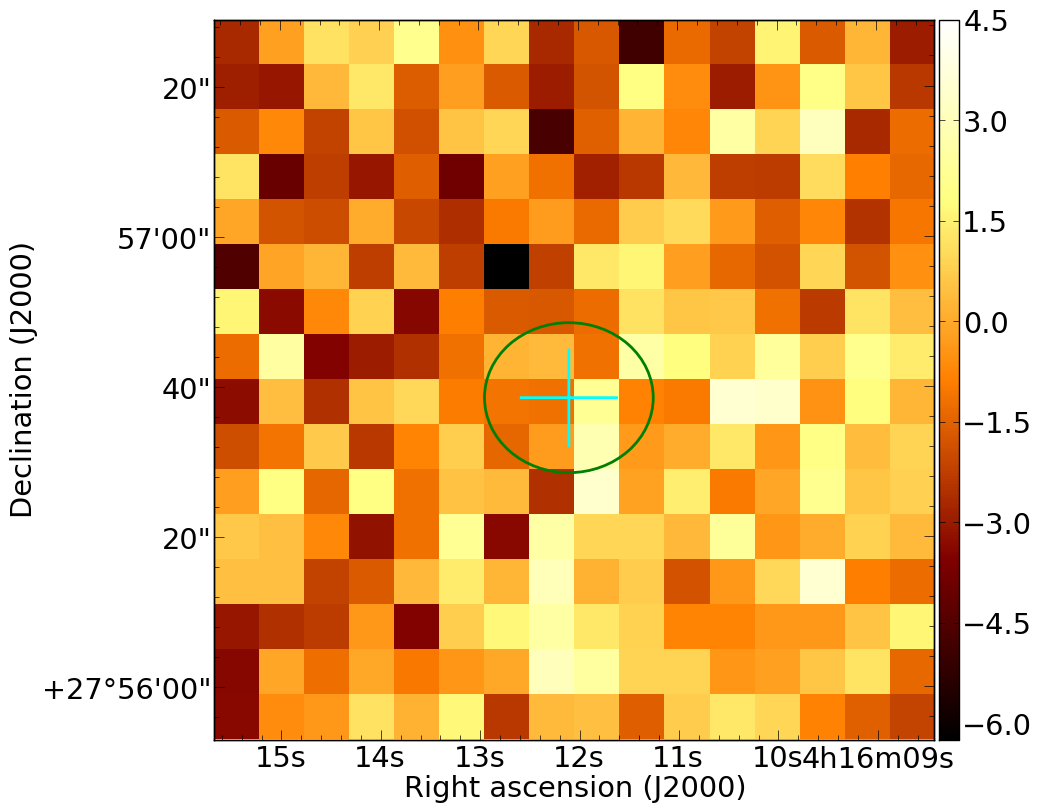}
\includegraphics[width=4cm]{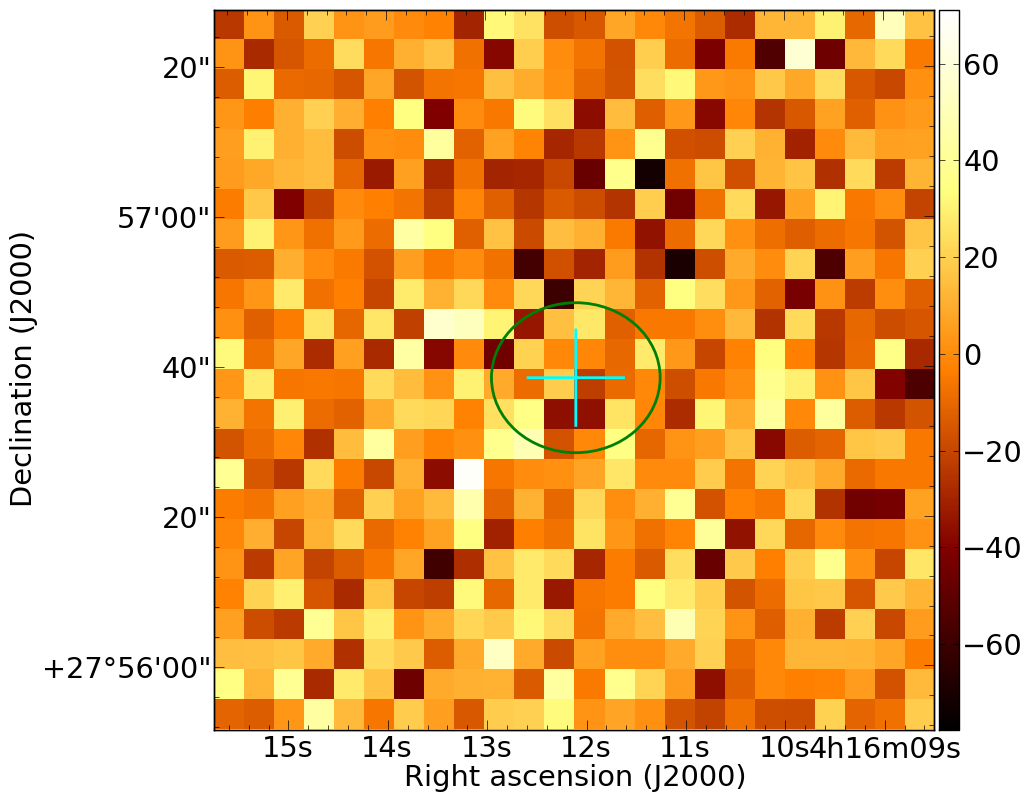}
\caption{as Fig.~\ref{diskfig0} J04161210+2756385 Class II \label{diskfig20}}
 \end{figure}

\begin{figure}
\includegraphics[width=4cm]{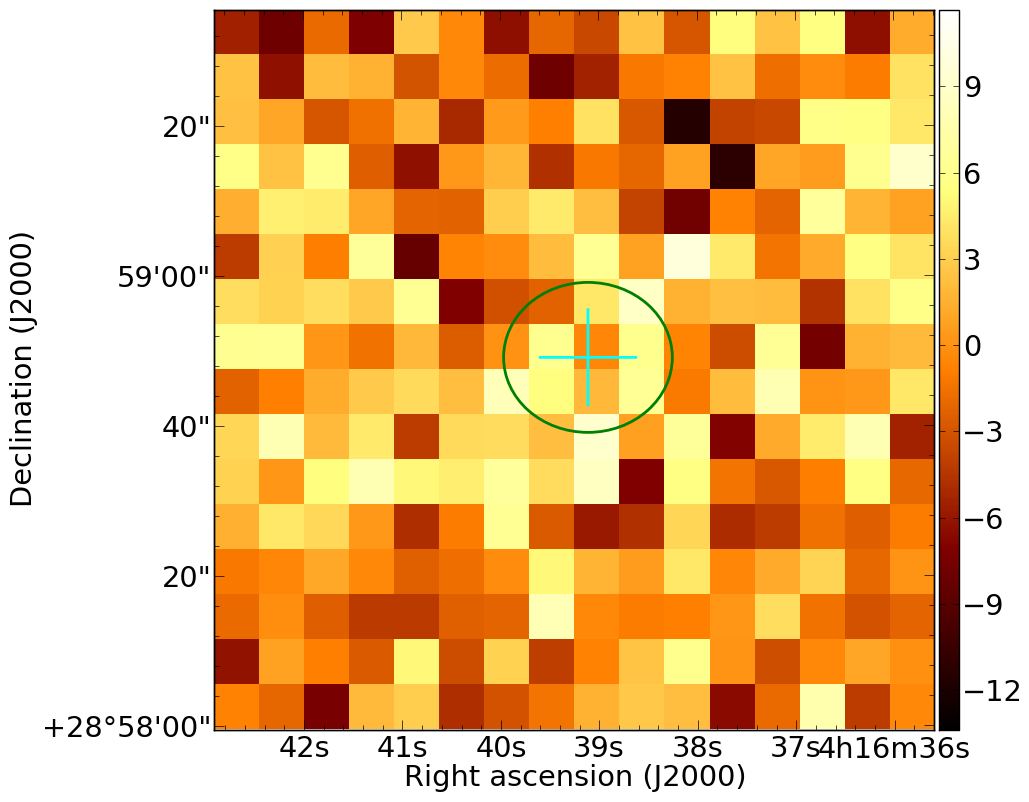}
\includegraphics[width=4cm]{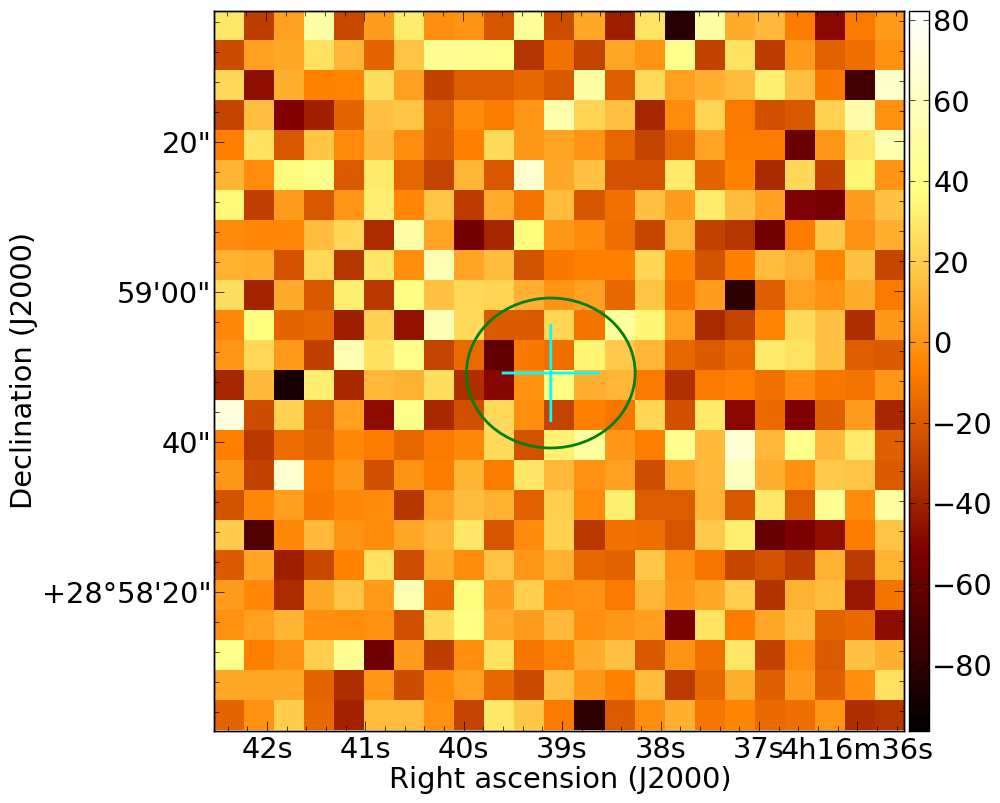}
\caption{as Fig.~\ref{diskfig0} J04163911+2858491 Class II \label{diskfig21}}
 \end{figure}

\begin{figure}
\includegraphics[width=4cm]{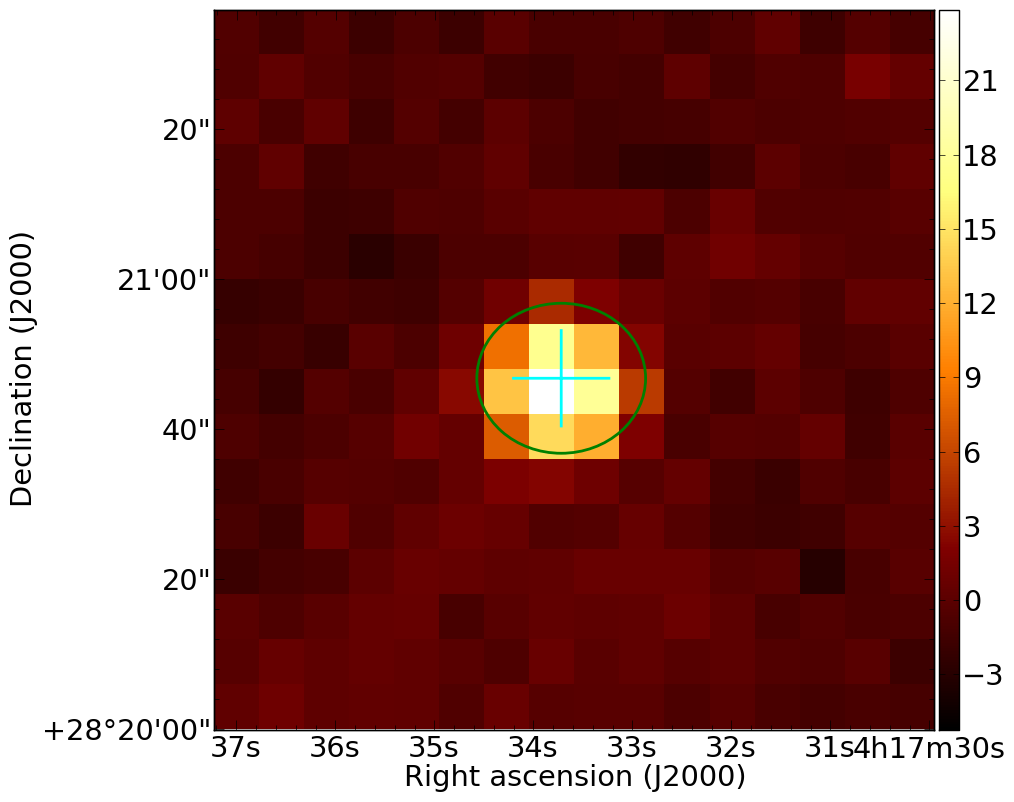}
\includegraphics[width=4cm]{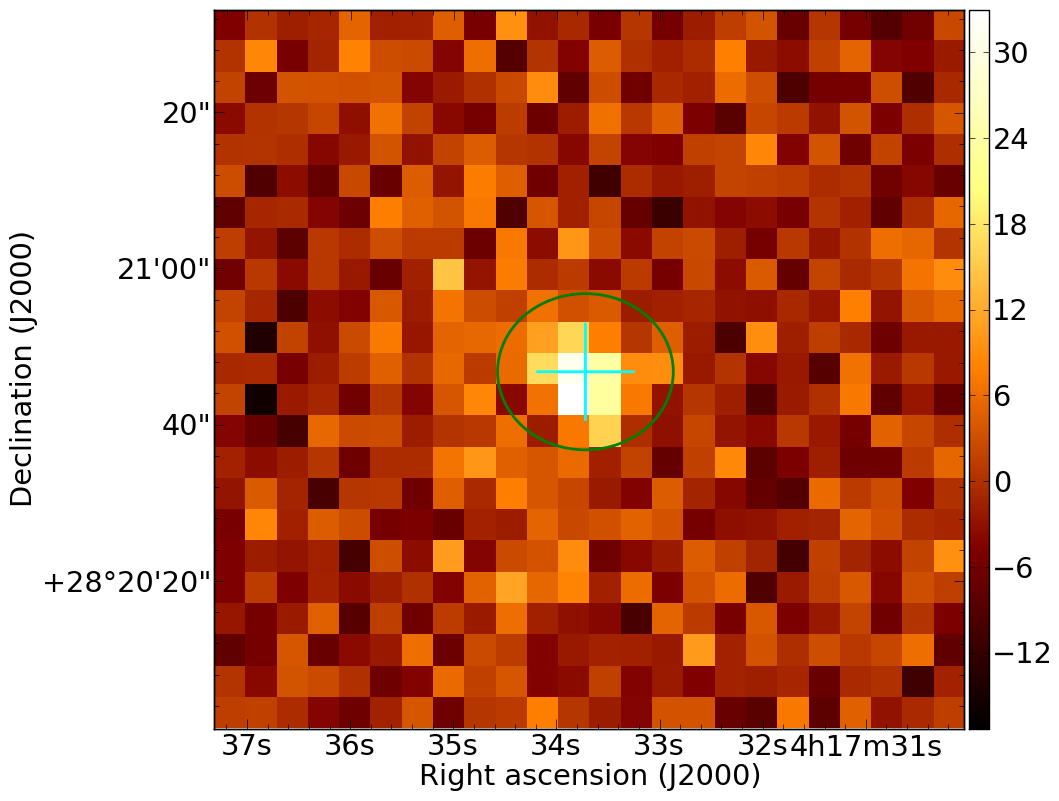}
\caption{as Fig.~\ref{diskfig0} CYTau Class II \label{diskfig22}}
 \end{figure}

\begin{figure}
\includegraphics[width=4cm]{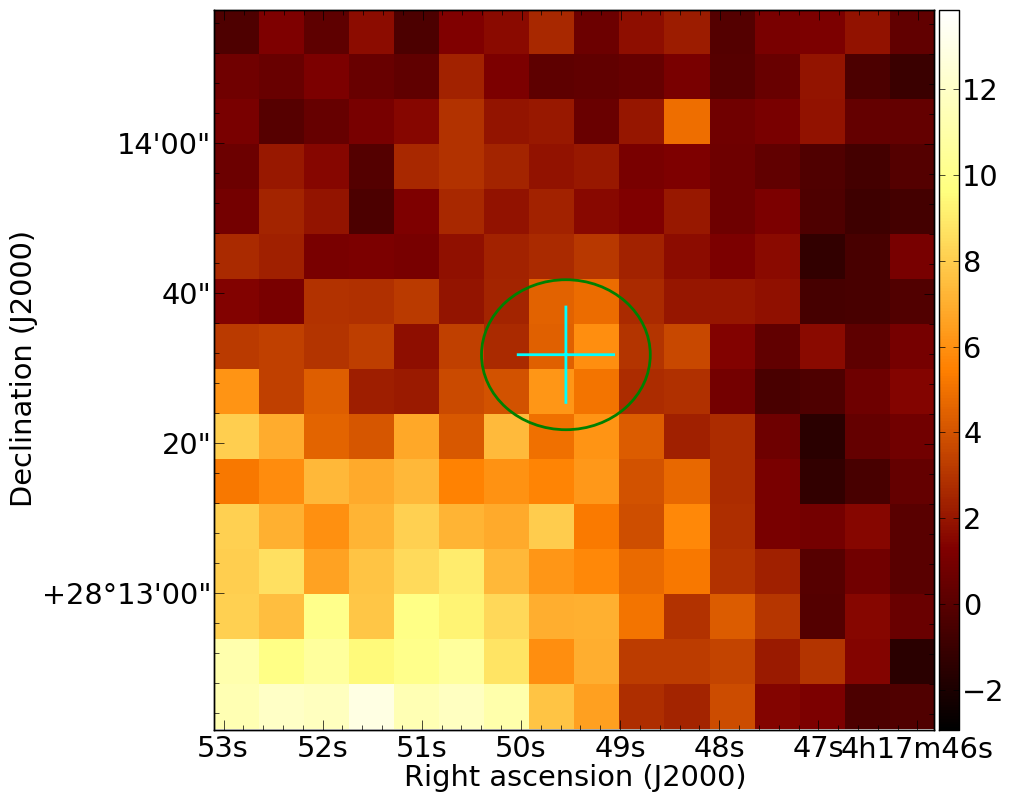}
\includegraphics[width=4cm]{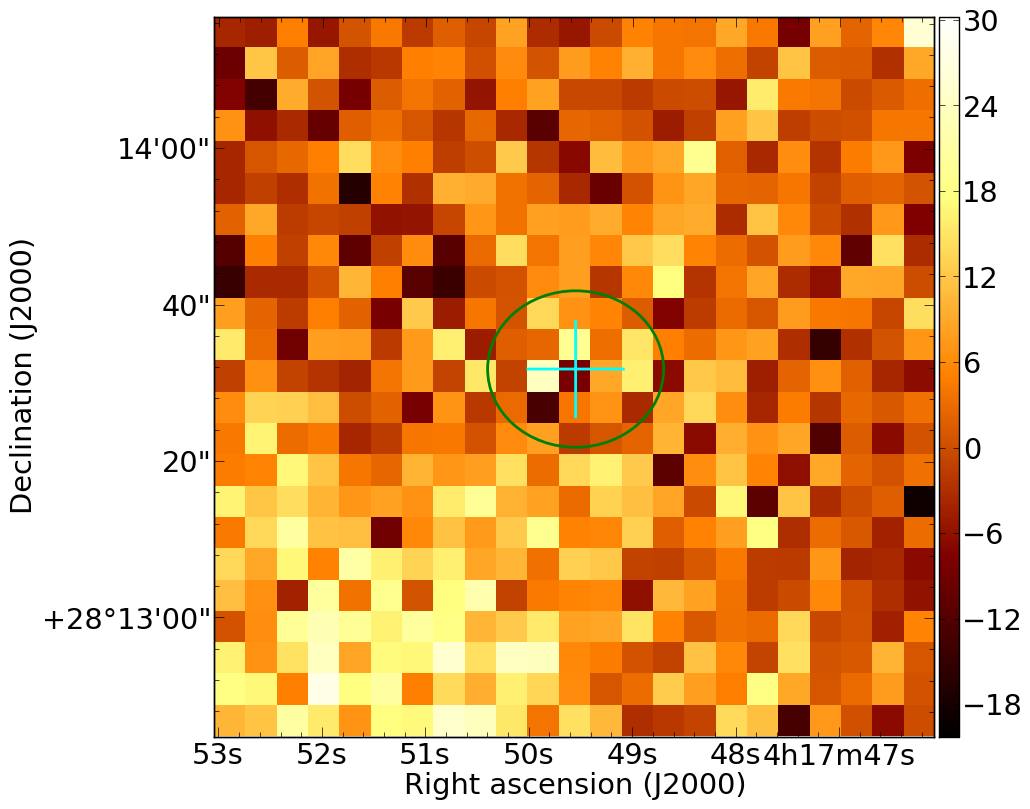}
\caption{as Fig.~\ref{diskfig0} KPNO10 Class II \label{diskfig23}}
 \end{figure}

\begin{figure}
\includegraphics[width=4cm]{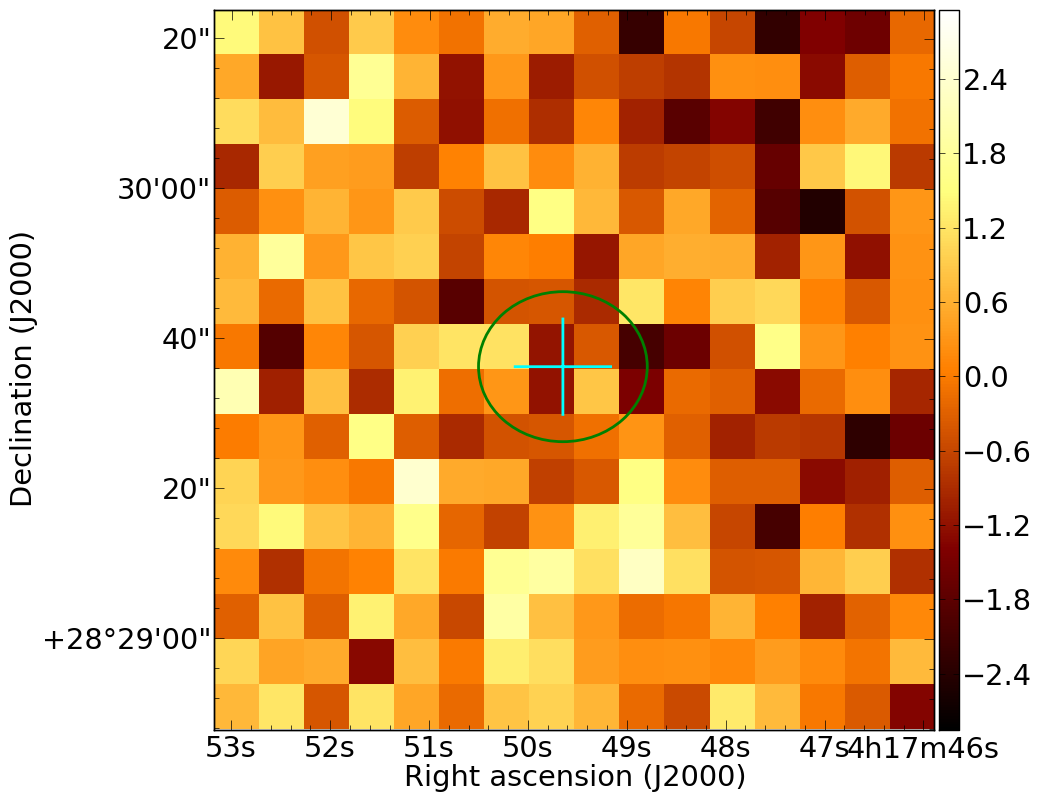}
\includegraphics[width=4cm]{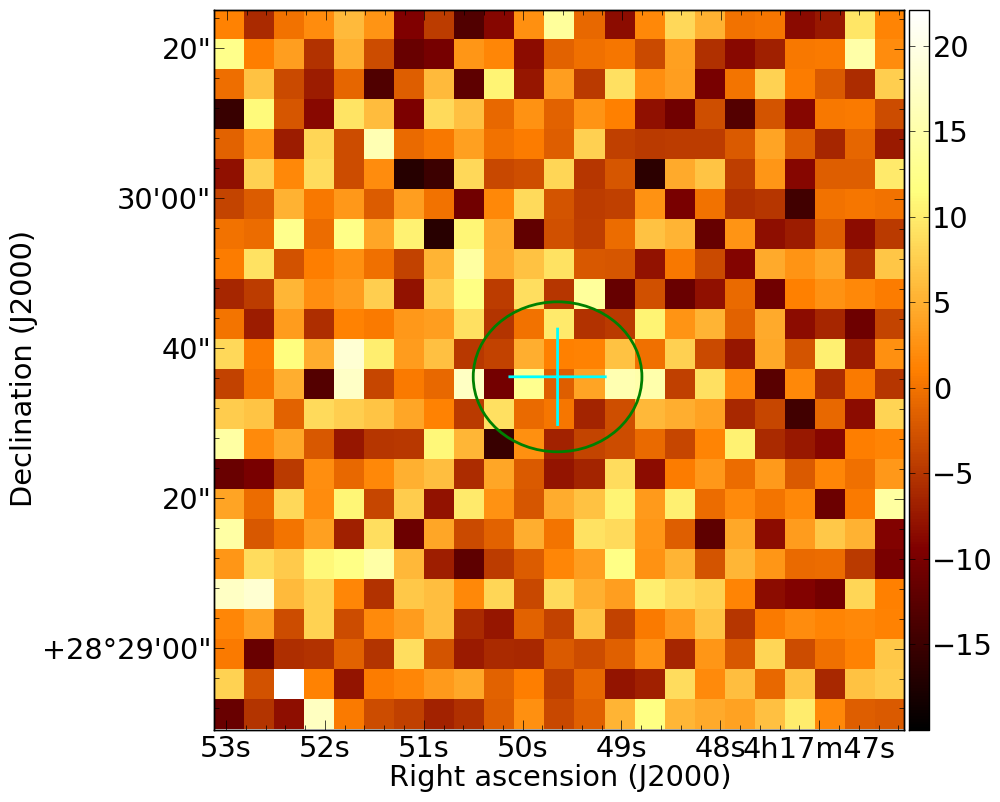}
\caption{as Fig.~\ref{diskfig0} V410X-ray1 Class II \label{diskfig24}}
 \end{figure}

\begin{figure}
\includegraphics[width=4cm]{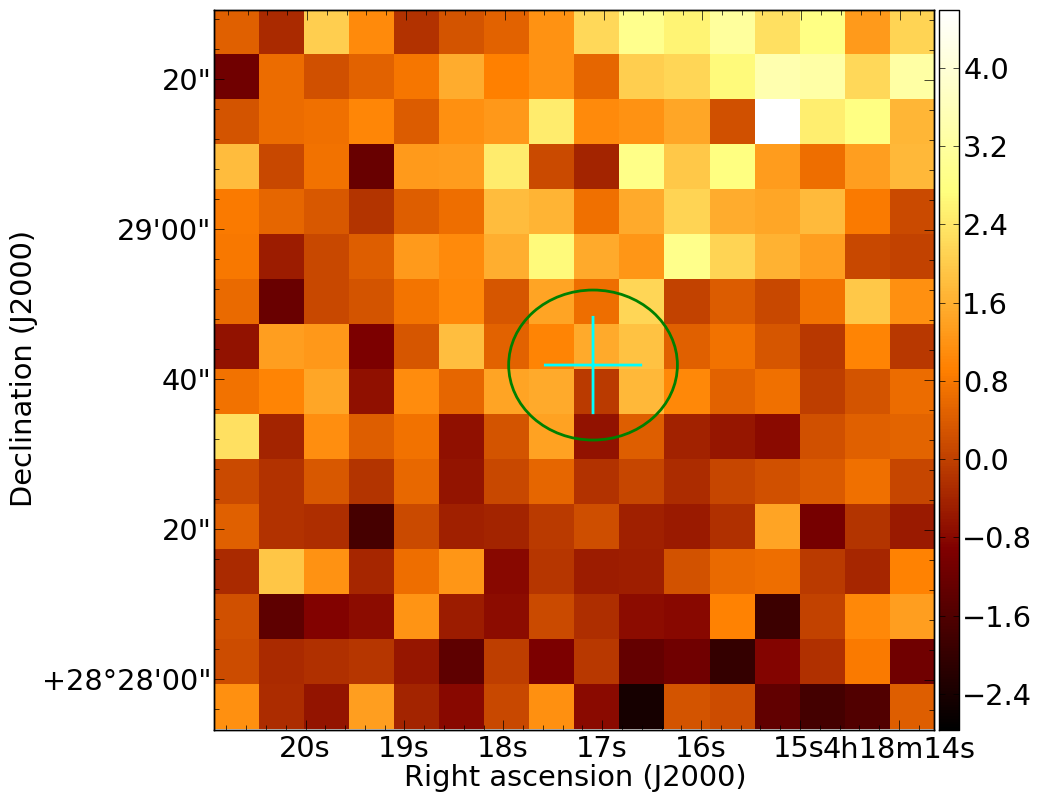}
\includegraphics[width=4cm]{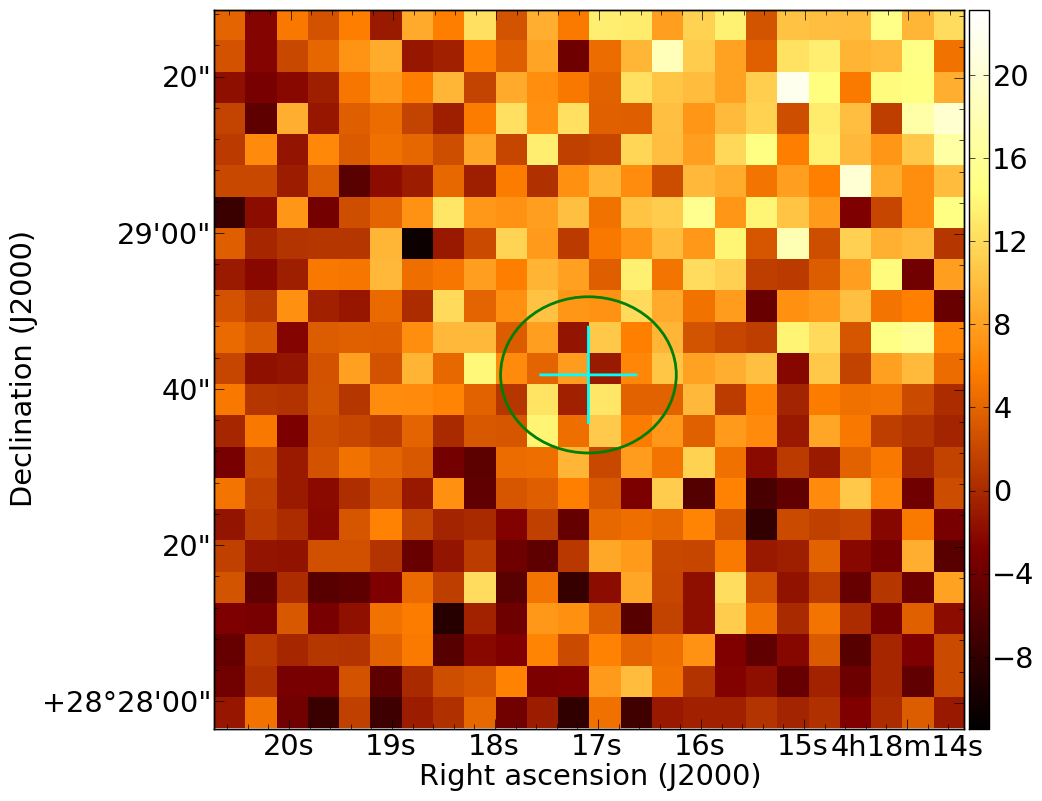}
\caption{as Fig.~\ref{diskfig0} V410Anon13 Class II \label{diskfig25}}
 \end{figure}

\begin{figure}
\includegraphics[width=4cm]{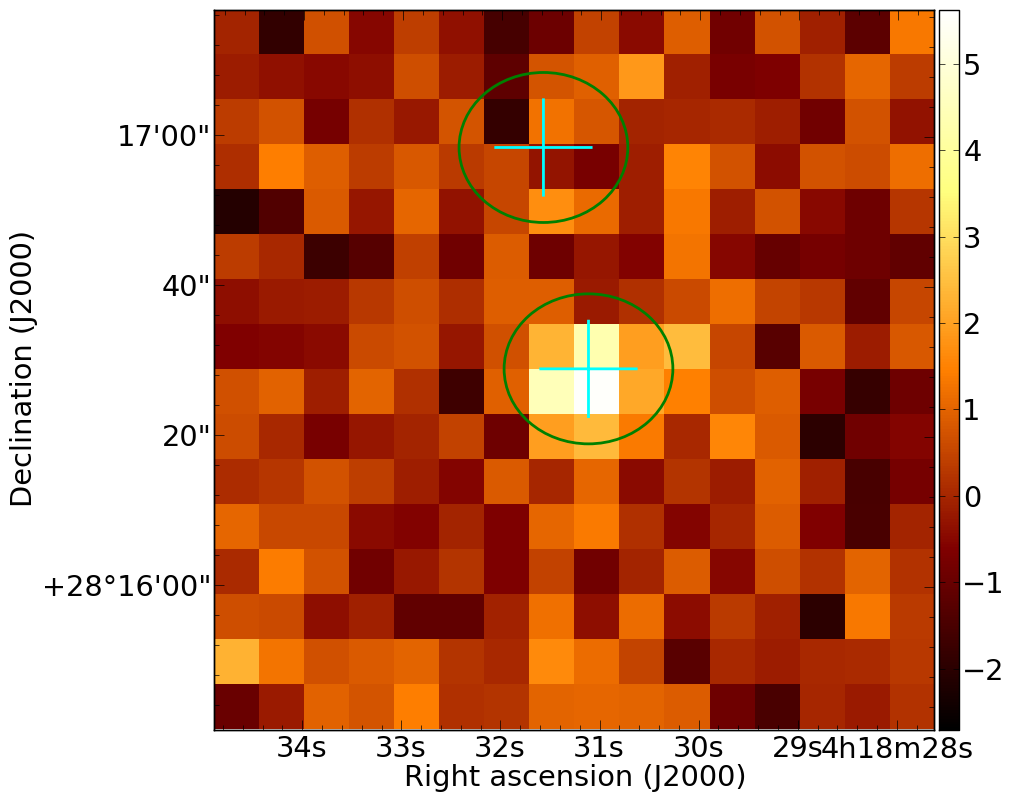}
\includegraphics[width=4cm]{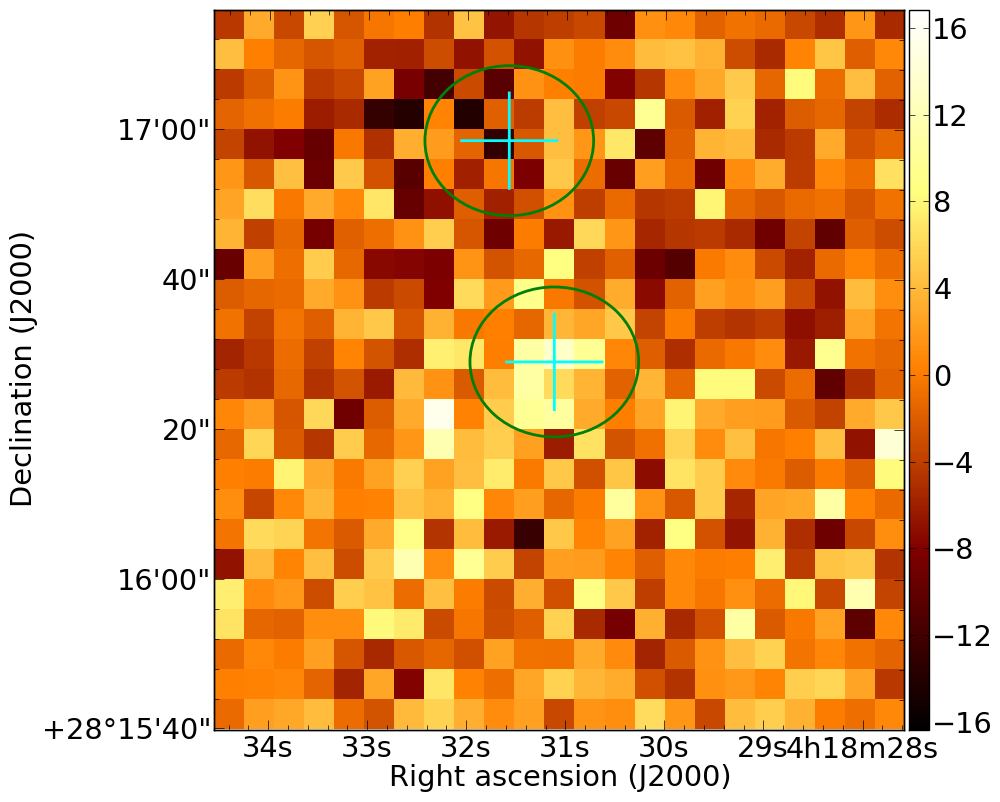}
\caption{as Fig.~\ref{diskfig0} DDTauA+B Class II \label{diskfig26}}
 \end{figure}

\begin{figure}
\includegraphics[width=4cm]{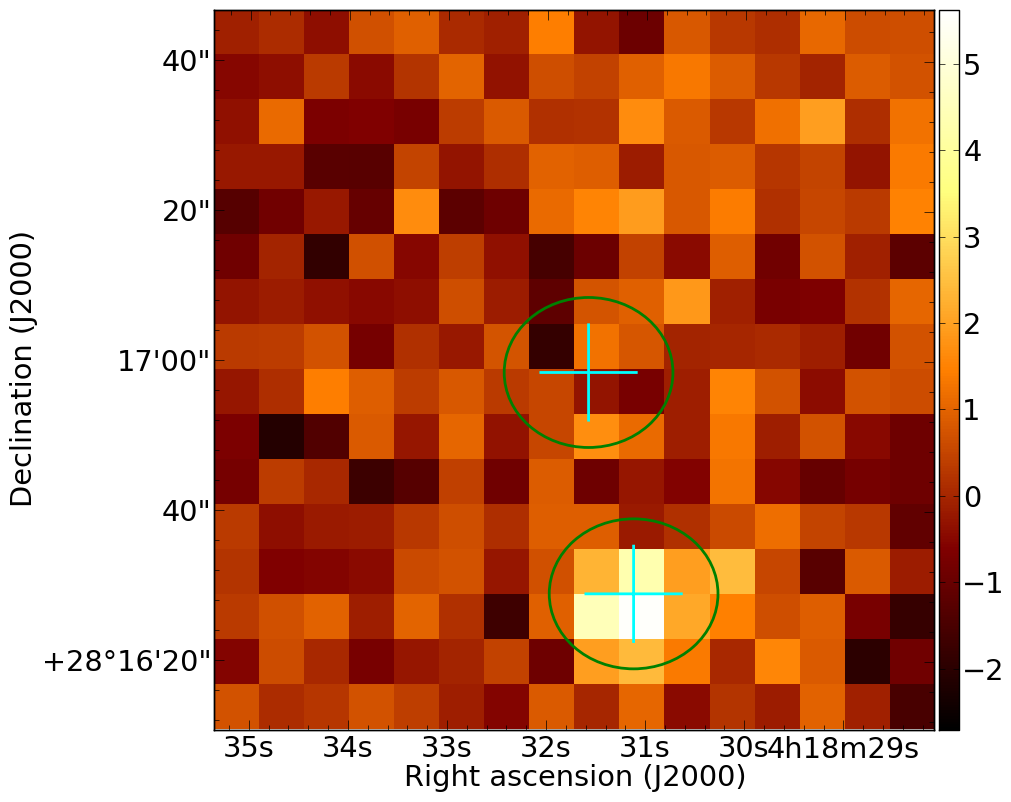}
\includegraphics[width=4cm]{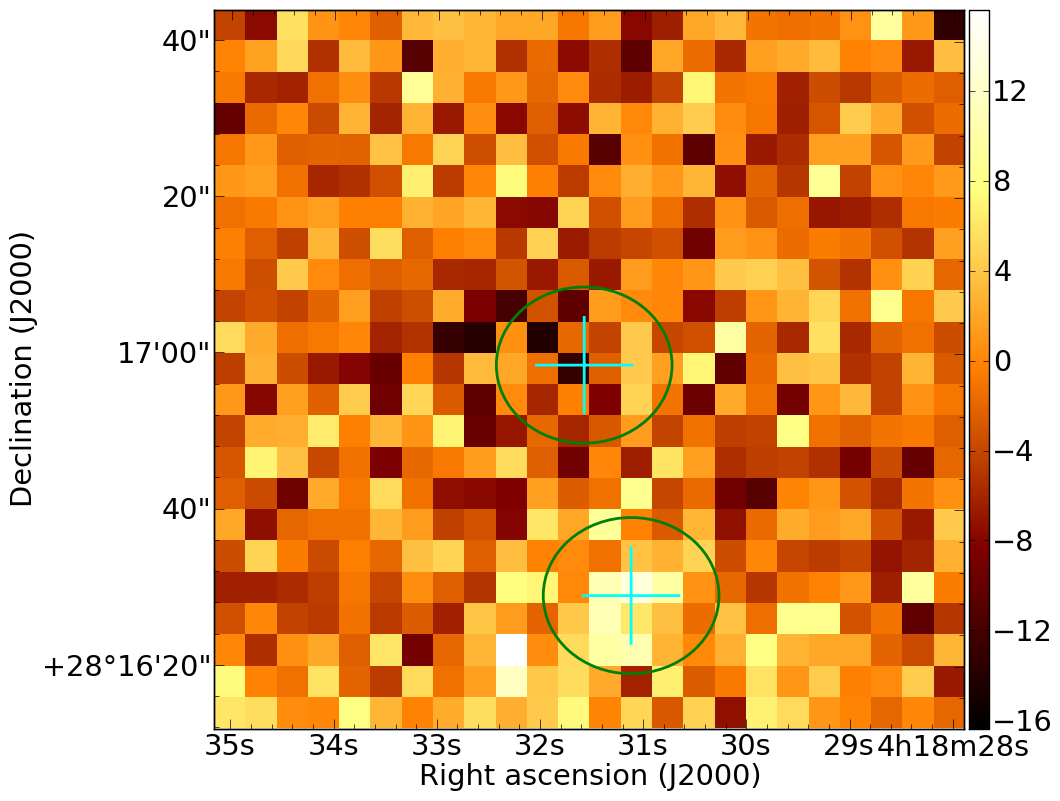}
\caption{as Fig.~\ref{diskfig0} CZTauA+B Class II \label{diskfig27}}
 \end{figure}

\begin{figure}
\includegraphics[width=4cm]{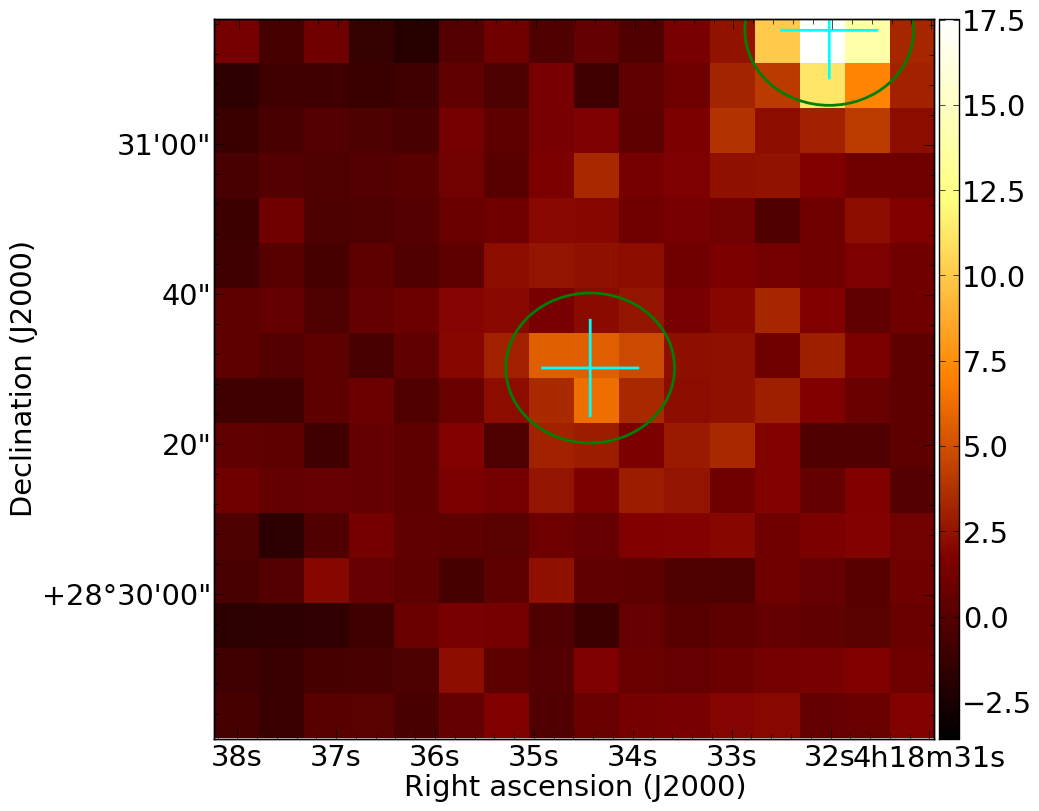}
\includegraphics[width=4cm]{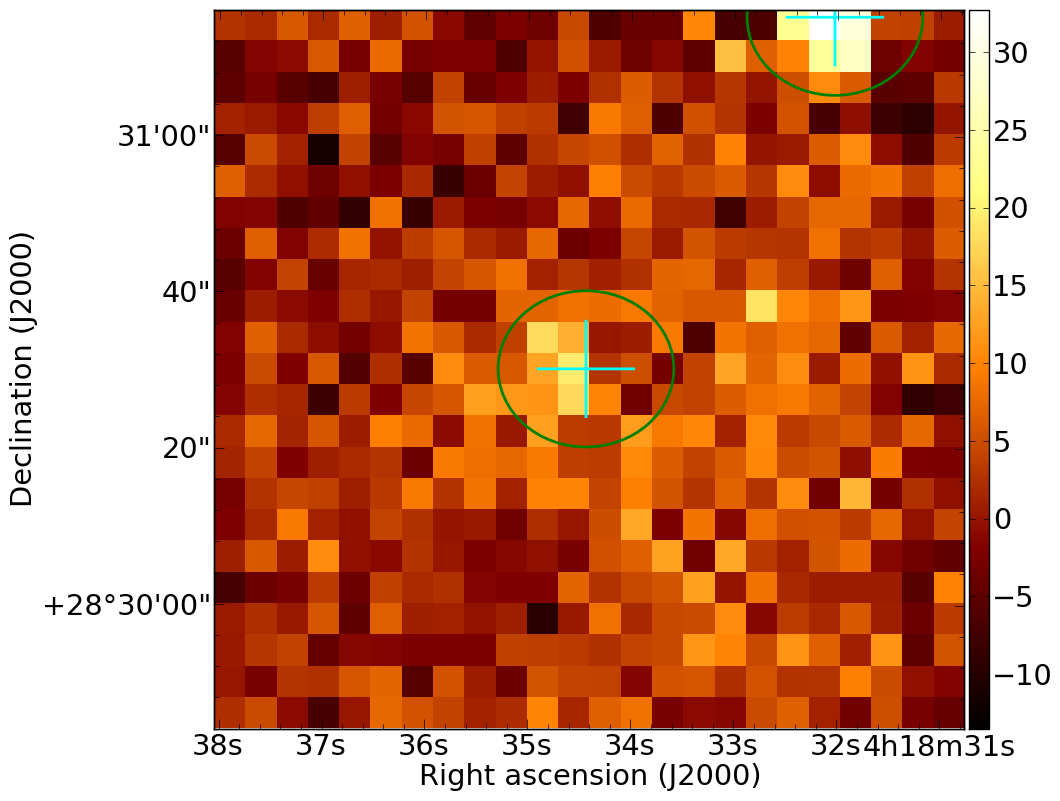}
\caption{as Fig.~\ref{diskfig0} V410X-ray2 Class II \label{diskfig28}}
 \end{figure}

\begin{figure}
\includegraphics[width=4cm]{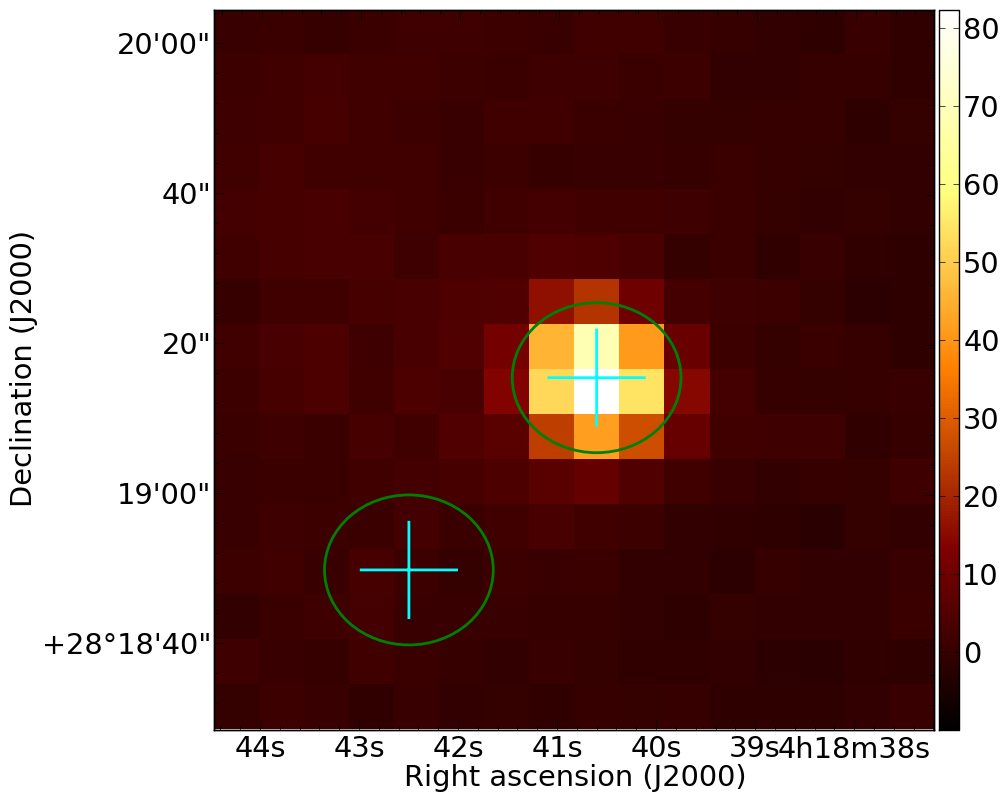}
\includegraphics[width=4cm]{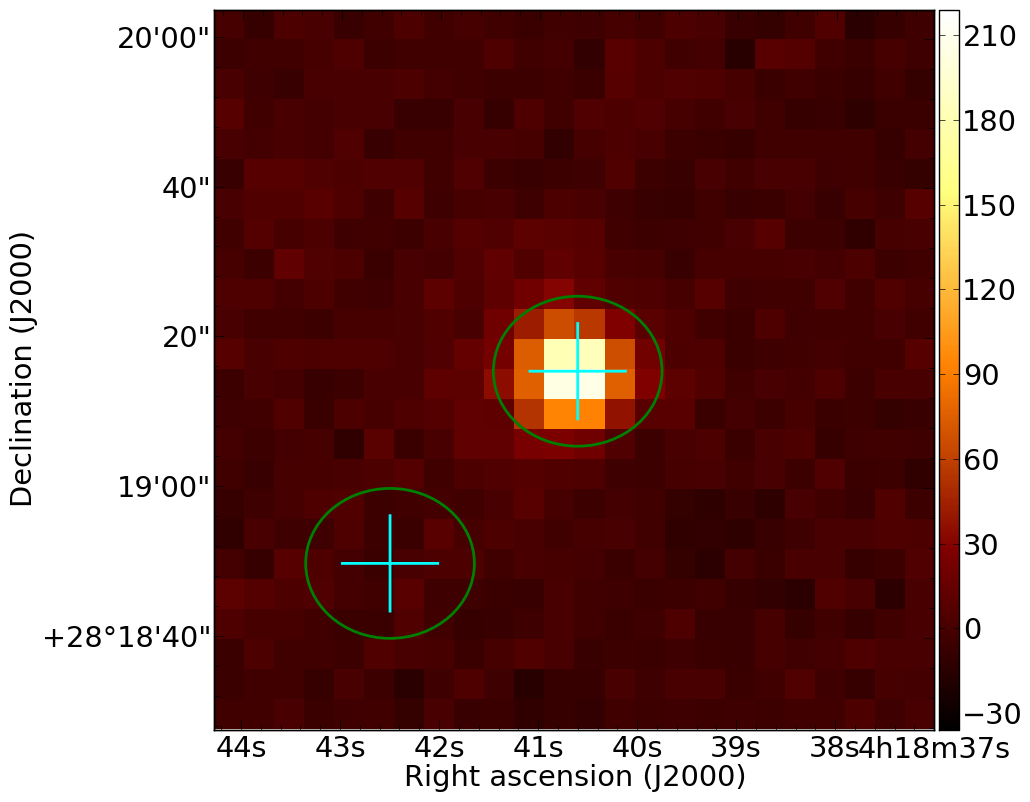}
\caption{as Fig.~\ref{diskfig0} V892Tau Class II \label{diskfig29}}
 \end{figure}

\begin{figure}
\includegraphics[width=4cm]{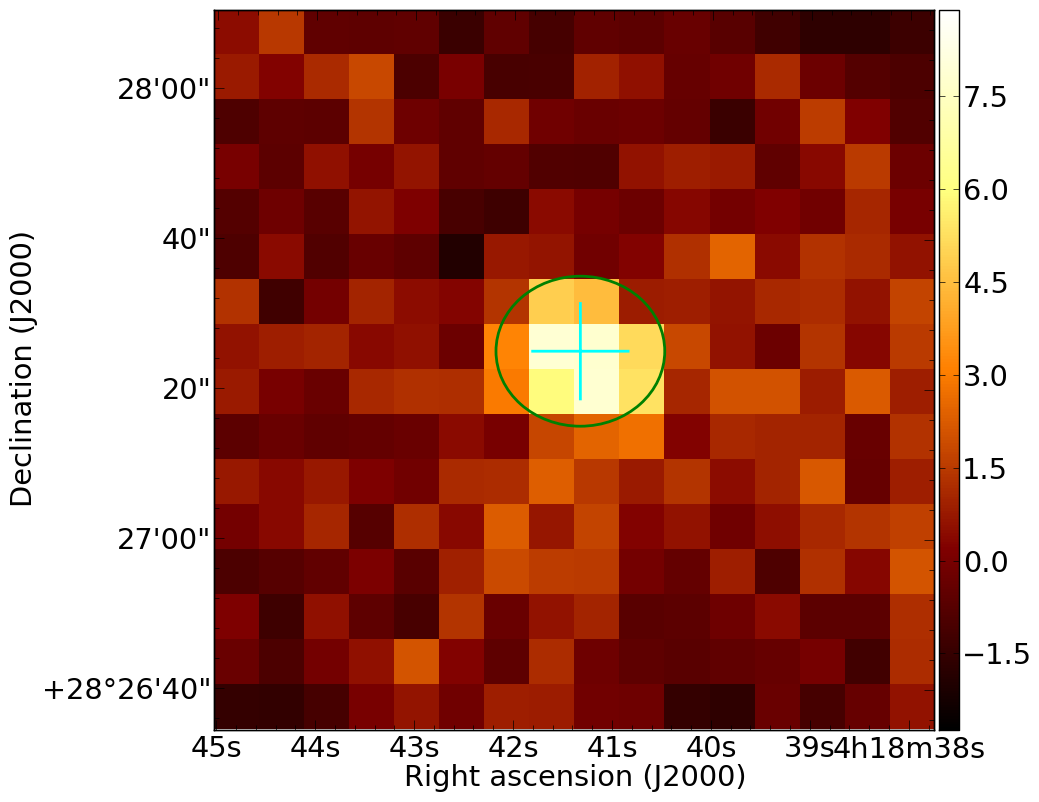}
\includegraphics[width=4cm]{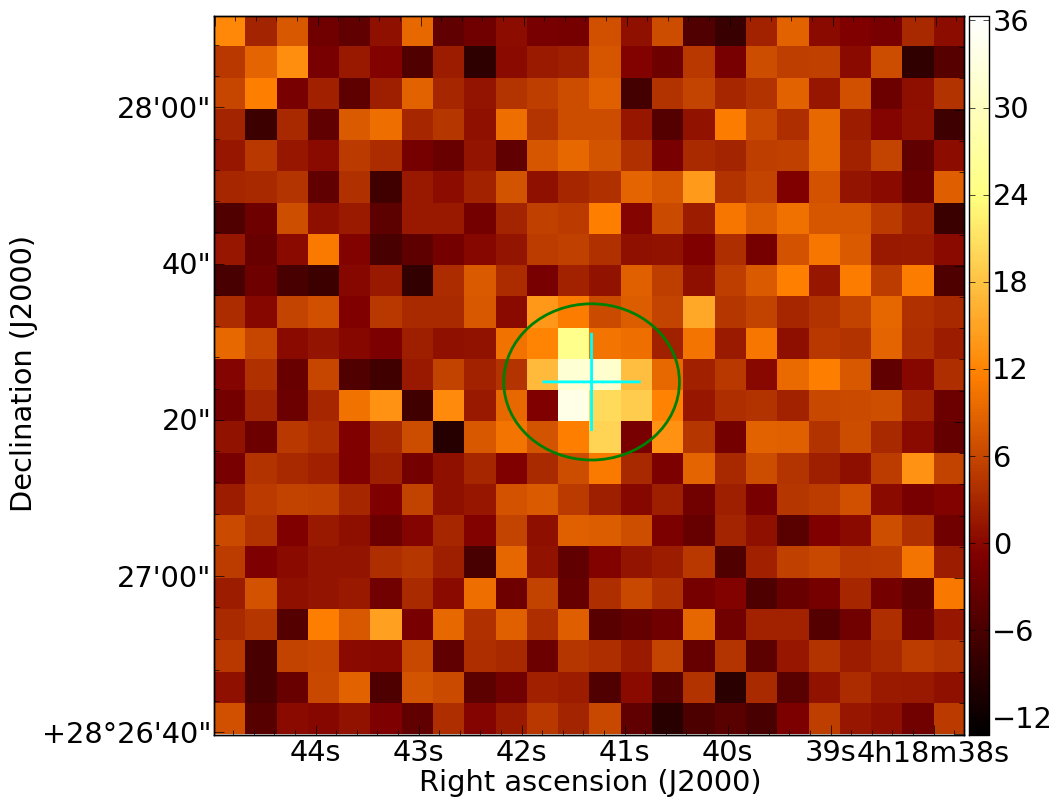}
\caption{as Fig.~\ref{diskfig0} LR1 Class II \label{diskfig30}}
 \end{figure}

\begin{figure}
\includegraphics[width=4cm]{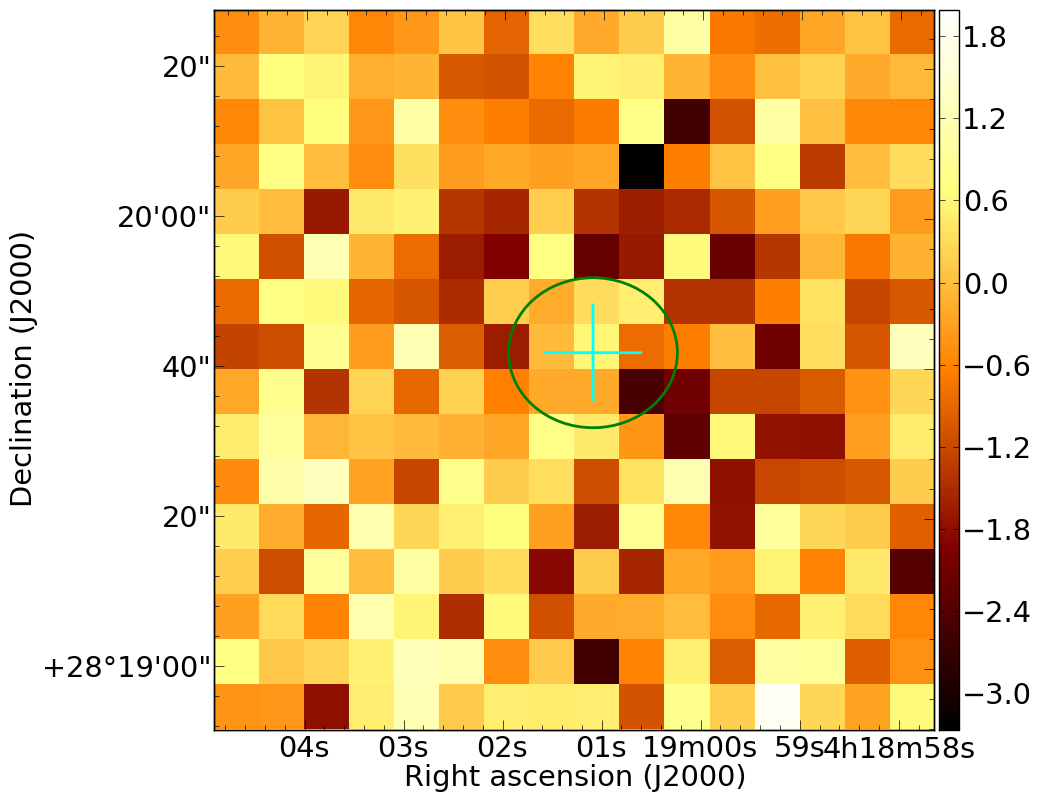}
\includegraphics[width=4cm]{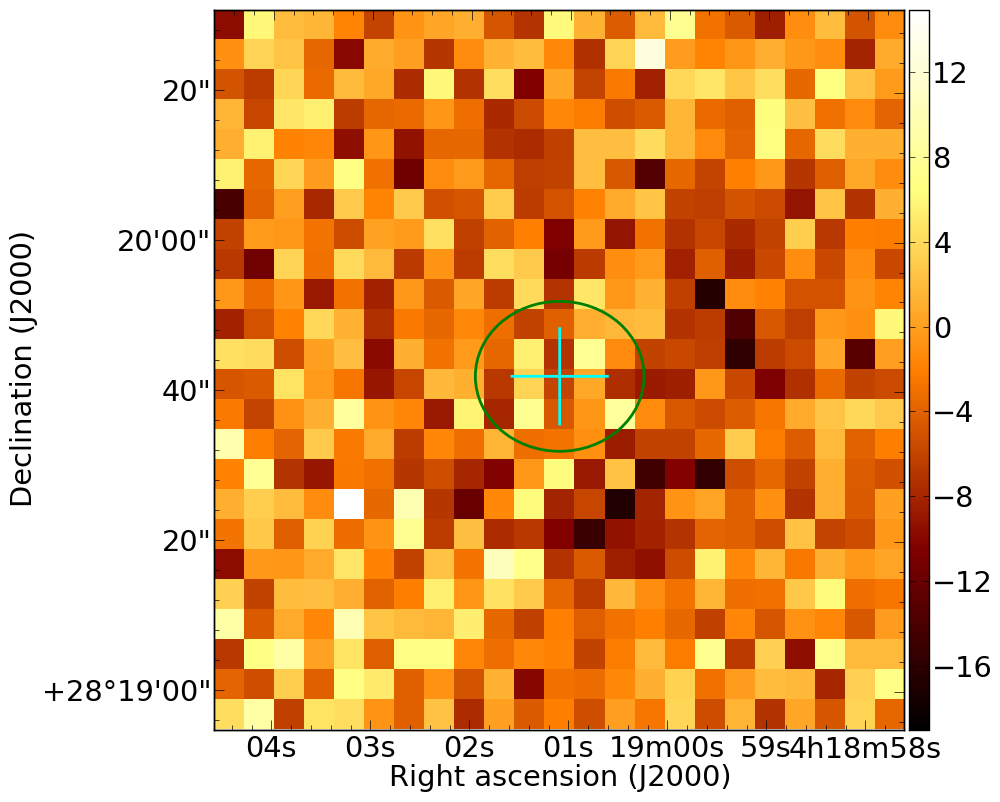}
\caption{as Fig.~\ref{diskfig0} V410X-ray6 Class II \label{diskfig31}}
 \end{figure}

\begin{figure}
\includegraphics[width=4cm]{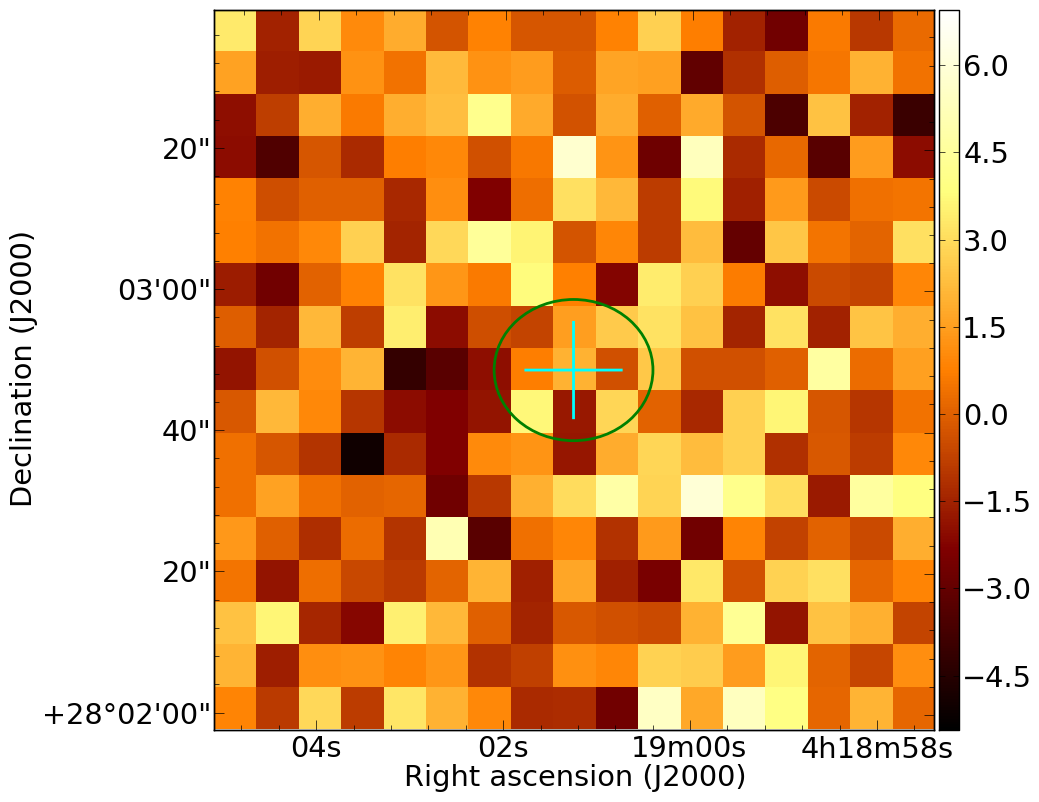}
\includegraphics[width=4cm]{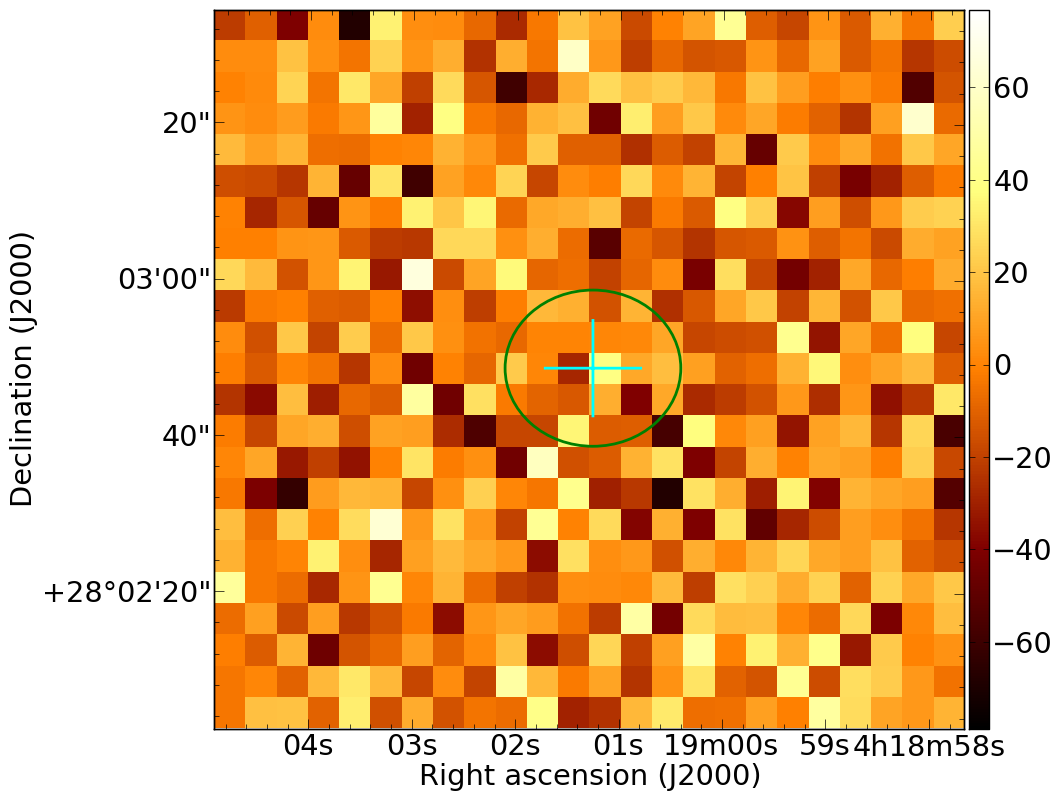}
\caption{as Fig.~\ref{diskfig0} KPNO12 Class II \label{diskfig32}}
 \end{figure}

\begin{figure}
\includegraphics[width=4cm]{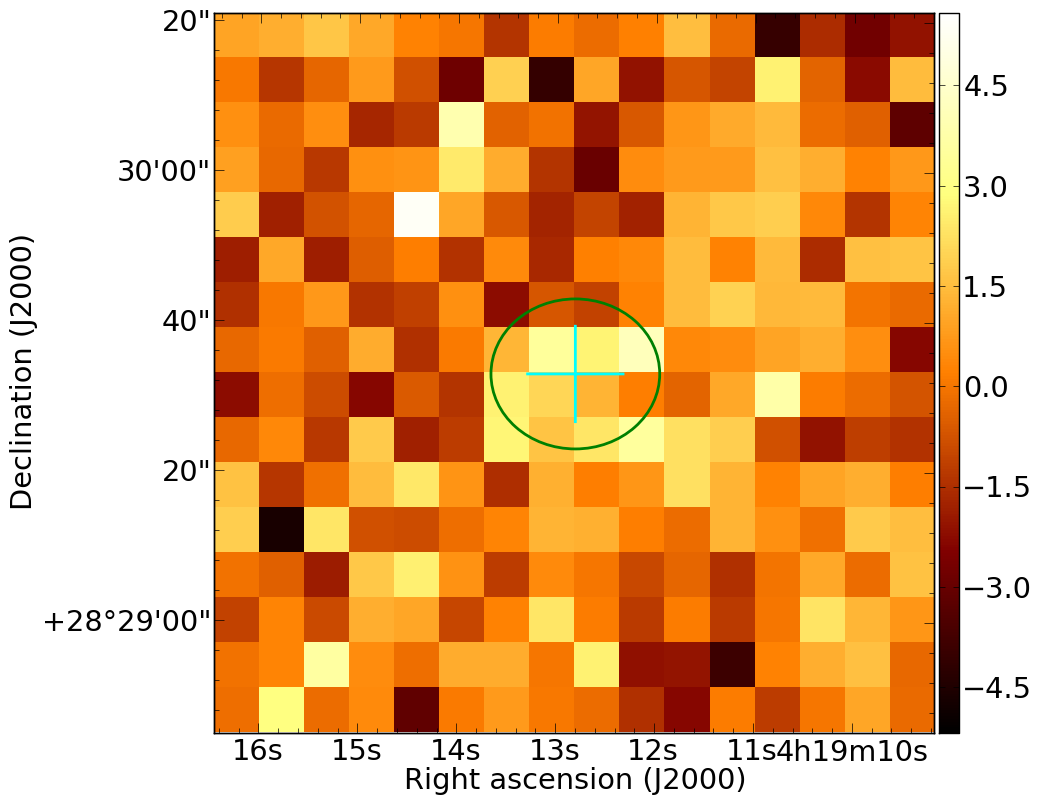}
\includegraphics[width=4cm]{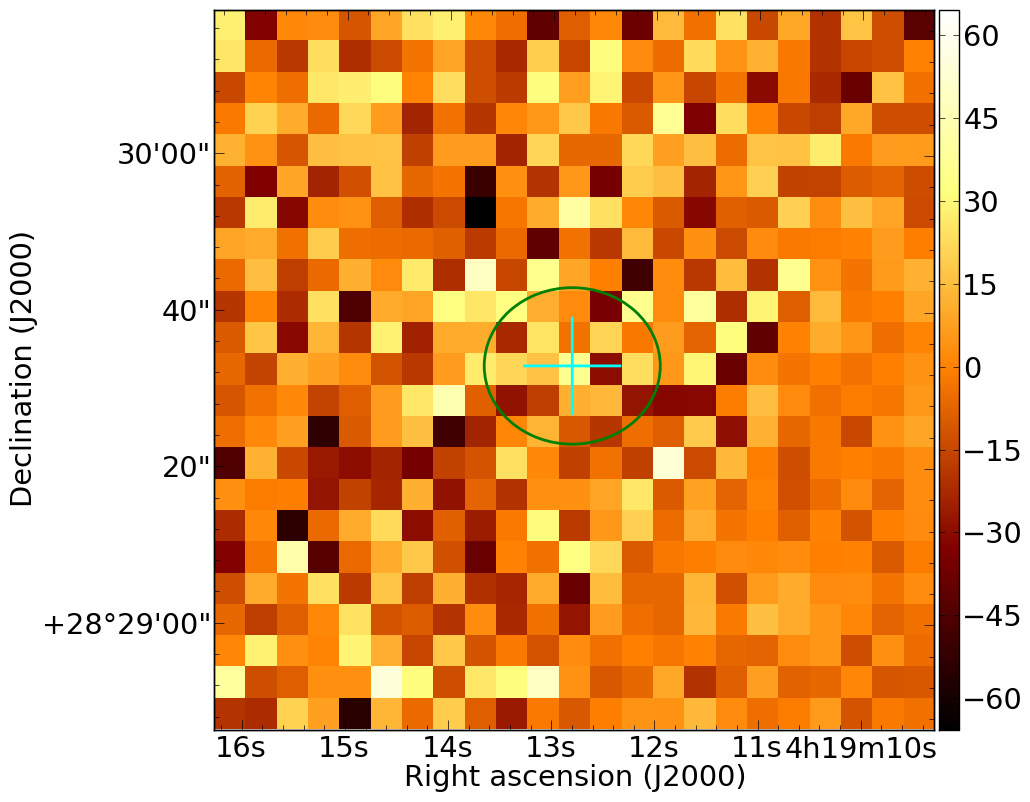}
\caption{as Fig.~\ref{diskfig0} FQTauA+B Class II \label{diskfig33}}
 \end{figure}

\begin{figure}
\includegraphics[width=4cm]{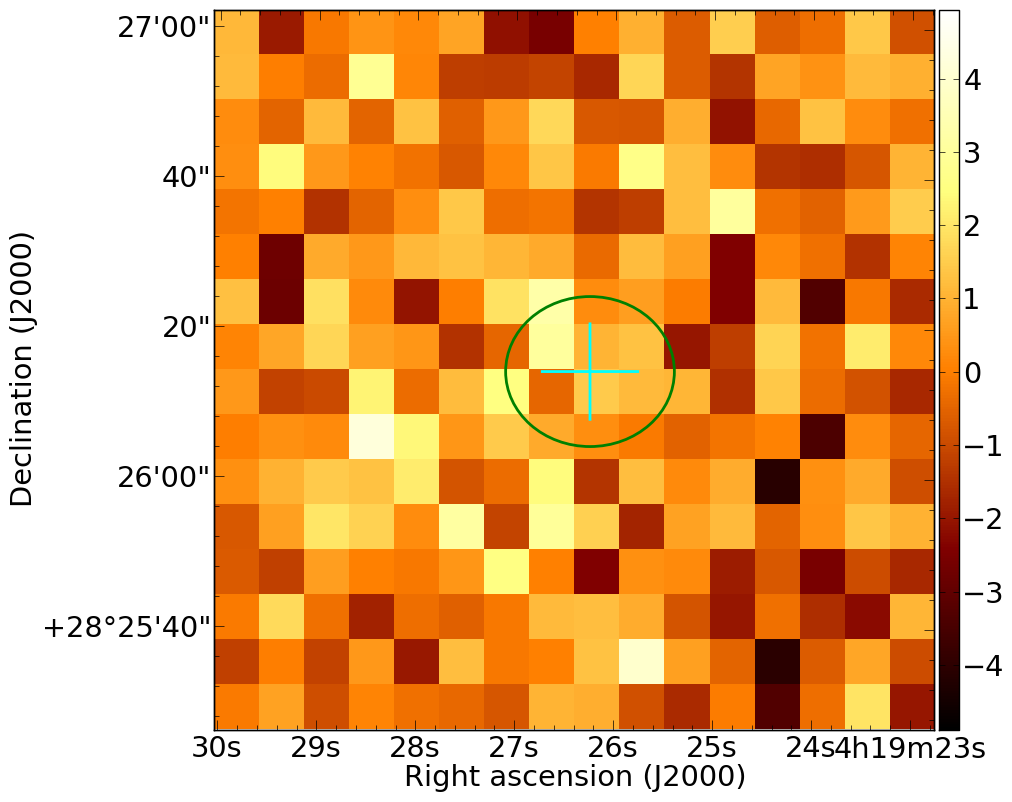}
\includegraphics[width=4cm]{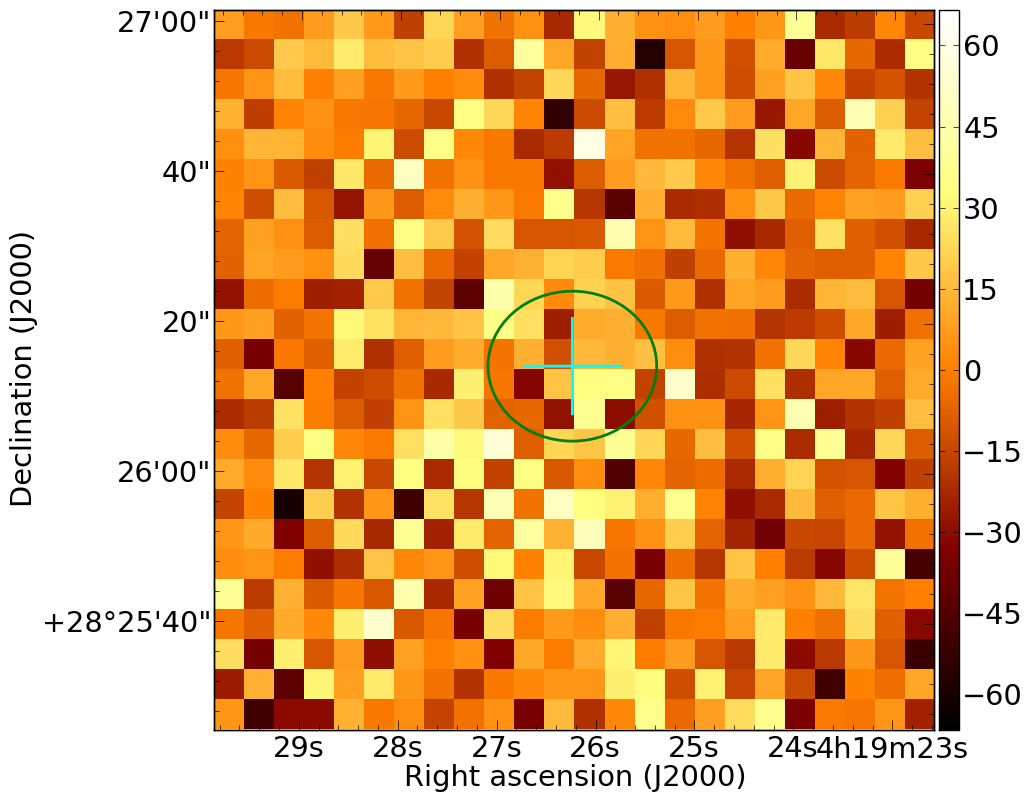}
\caption{as Fig.~\ref{diskfig0} V819Tau Class II \label{diskfig34}}
 \end{figure}

\begin{figure}
\includegraphics[width=4cm]{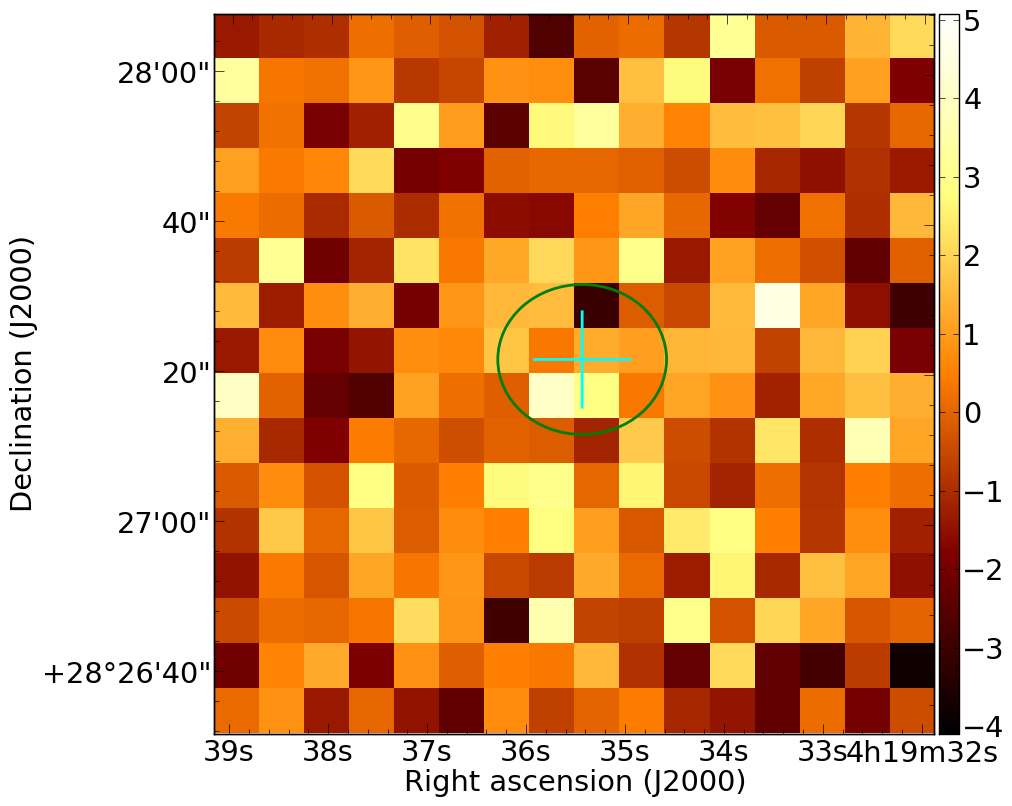}
\includegraphics[width=4cm]{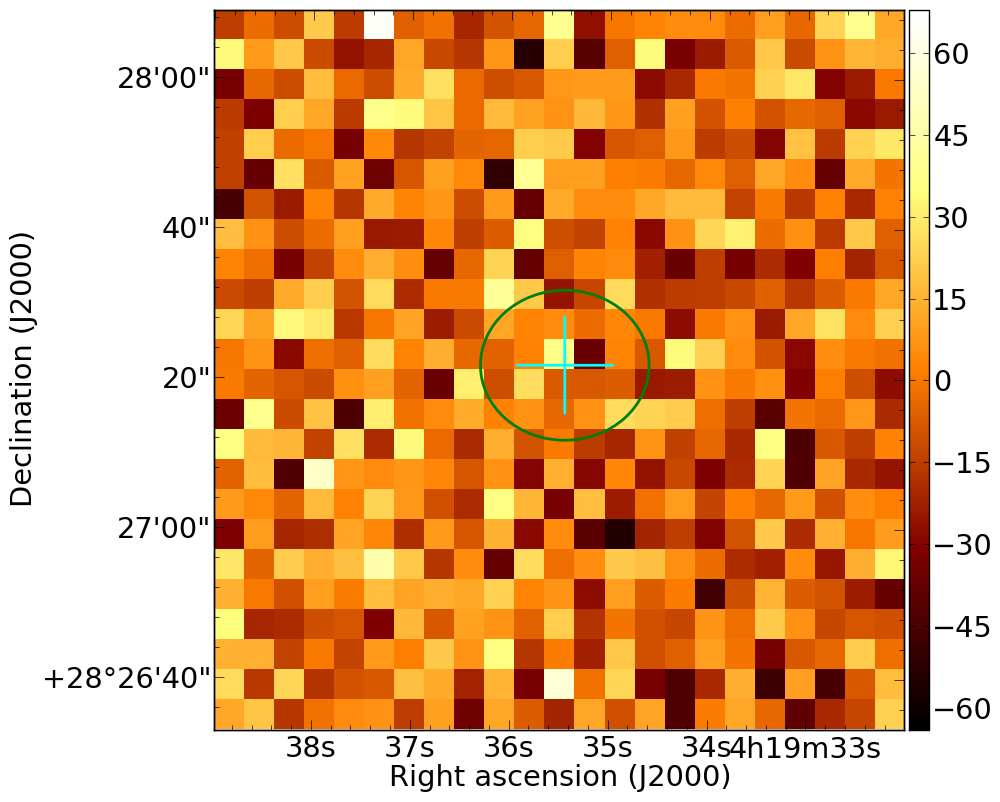}
\caption{as Fig.~\ref{diskfig0} FRTau Class II \label{diskfig35}}
 \end{figure}

\begin{figure}
\includegraphics[width=4cm]{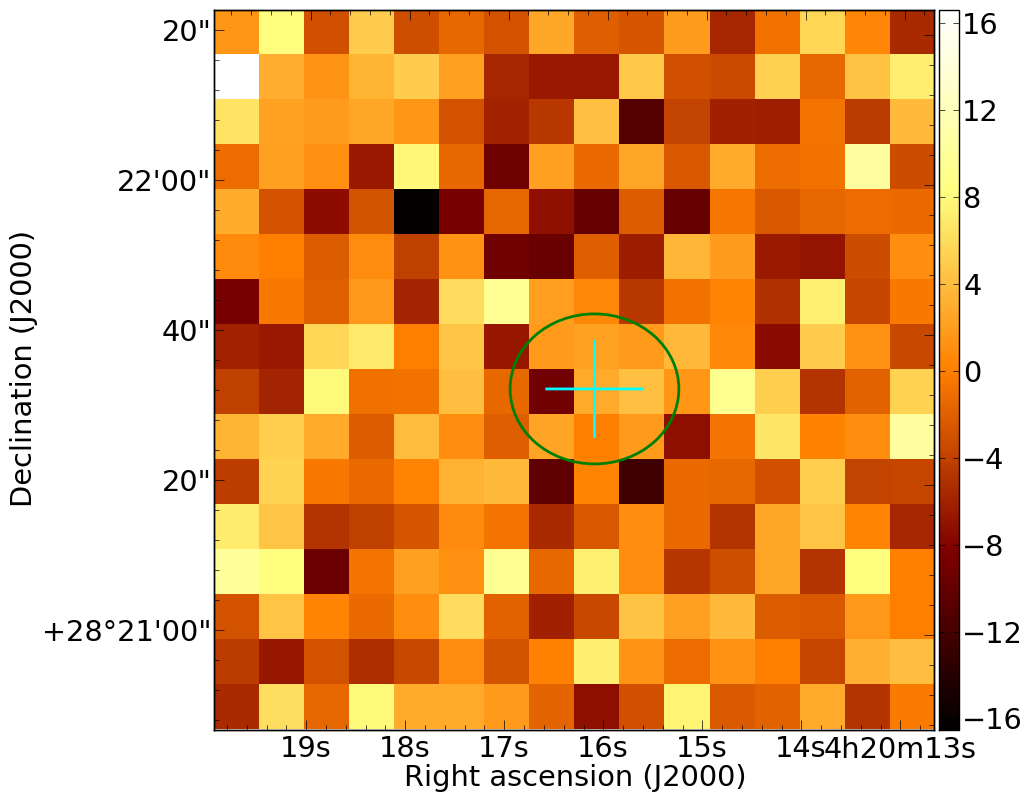}
\includegraphics[width=4cm]{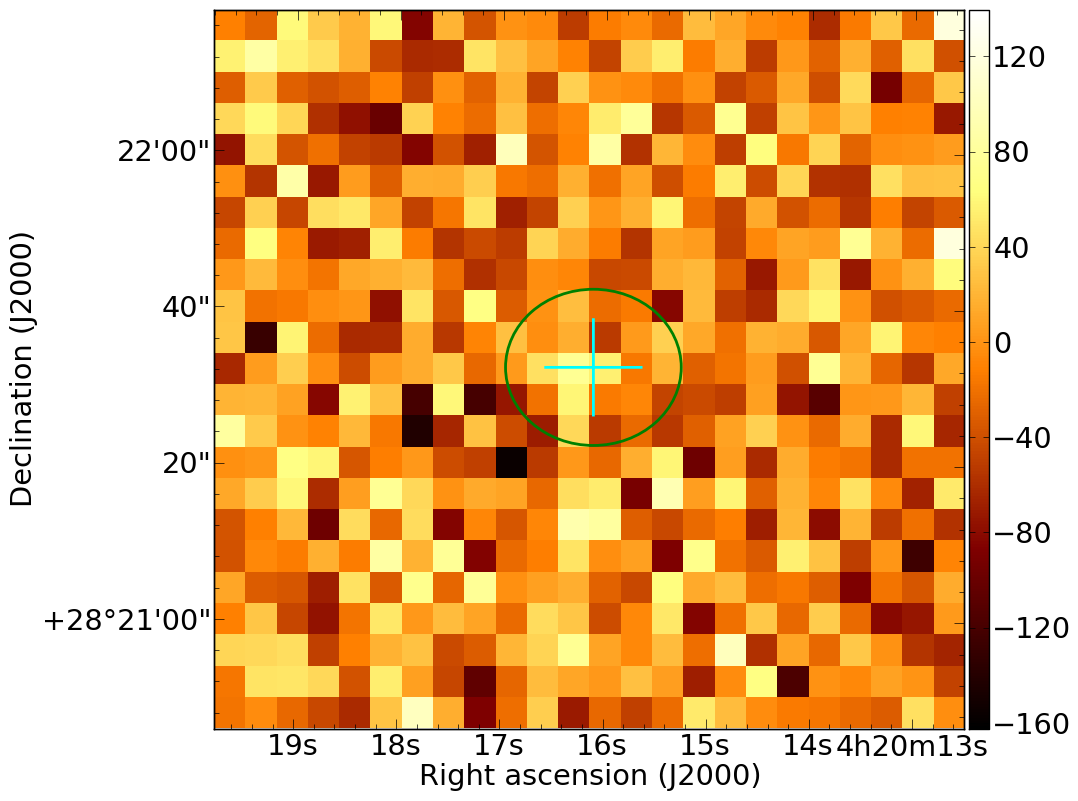}
\caption{as Fig.~\ref{diskfig0} J04201611+2821325 Class II \label{diskfig36}}
 \end{figure}

\begin{figure}
\includegraphics[width=4cm]{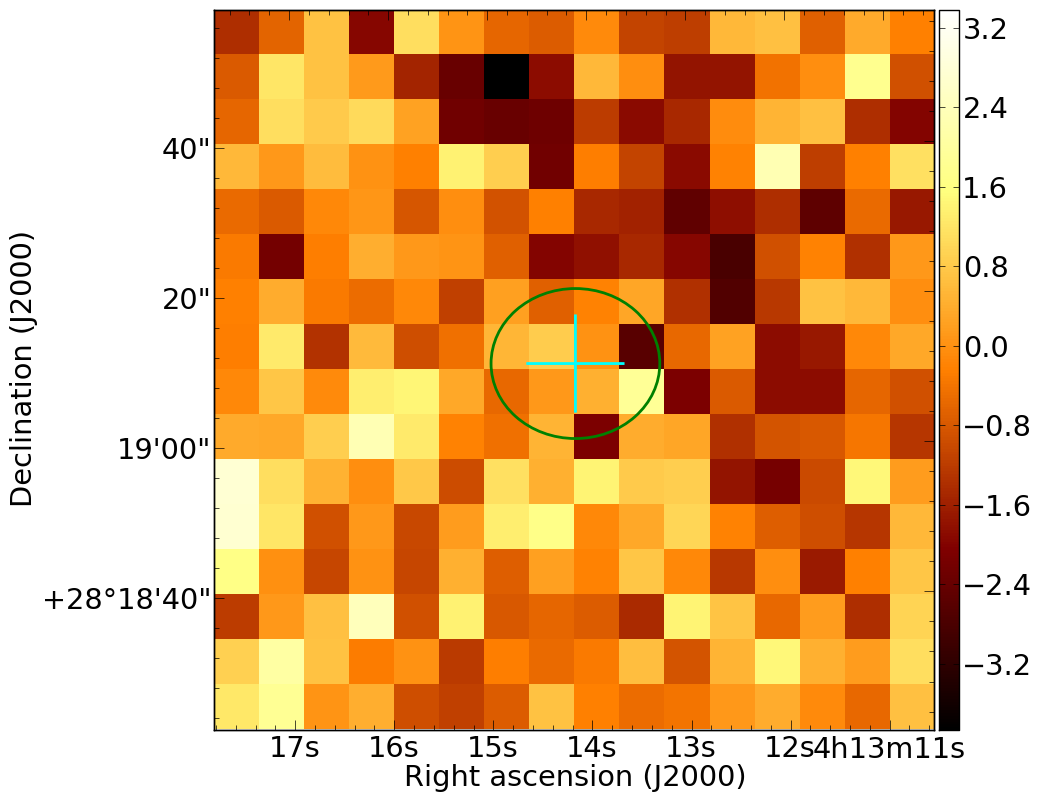}
\includegraphics[width=4cm]{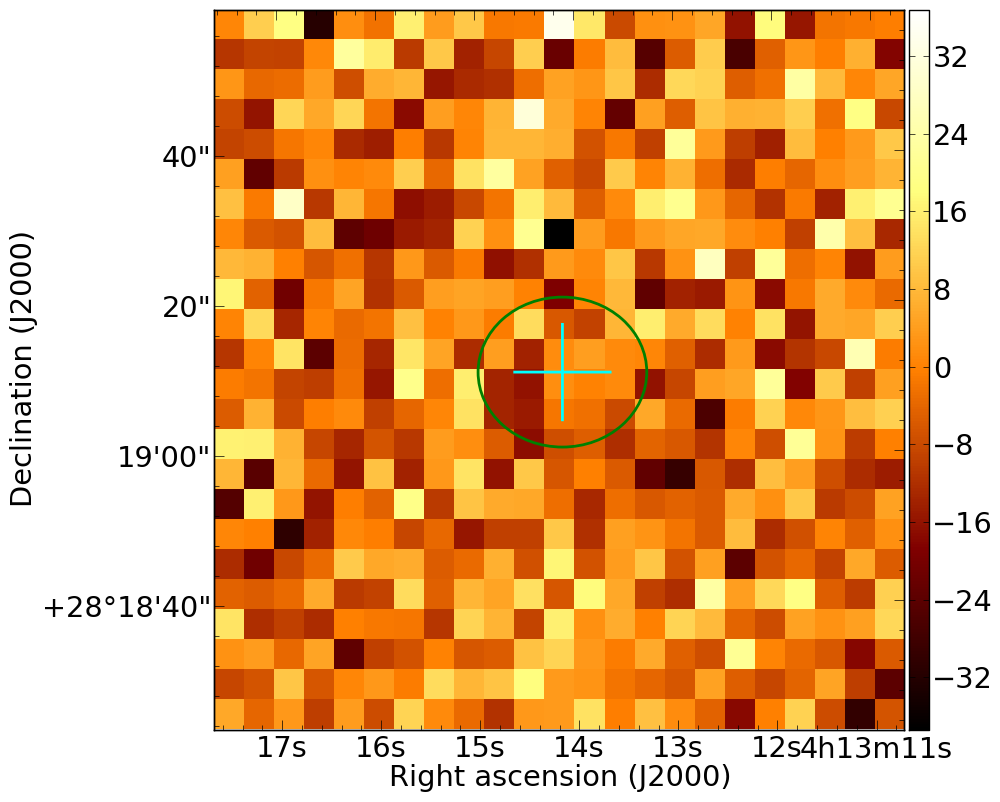}
\caption{as Fig.~\ref{diskfig0} LkCa1 Class III \label{diskfig37}}
 \end{figure}

\begin{figure}
\includegraphics[width=4cm]{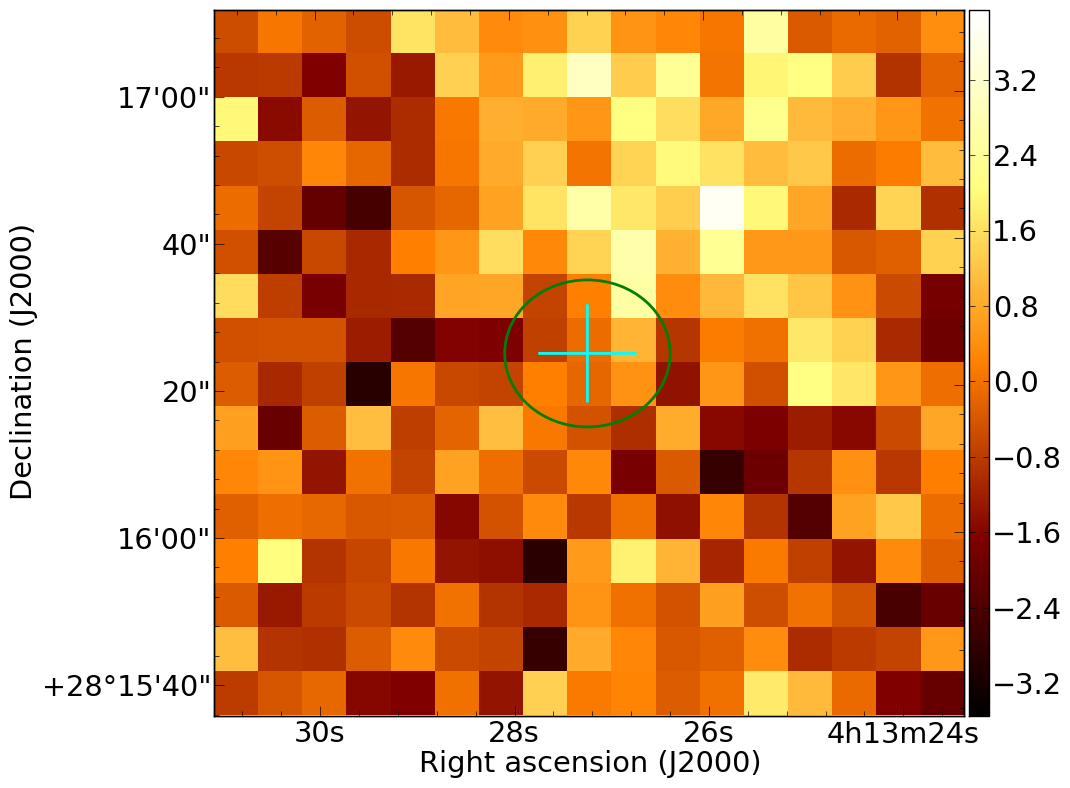}
\includegraphics[width=4cm]{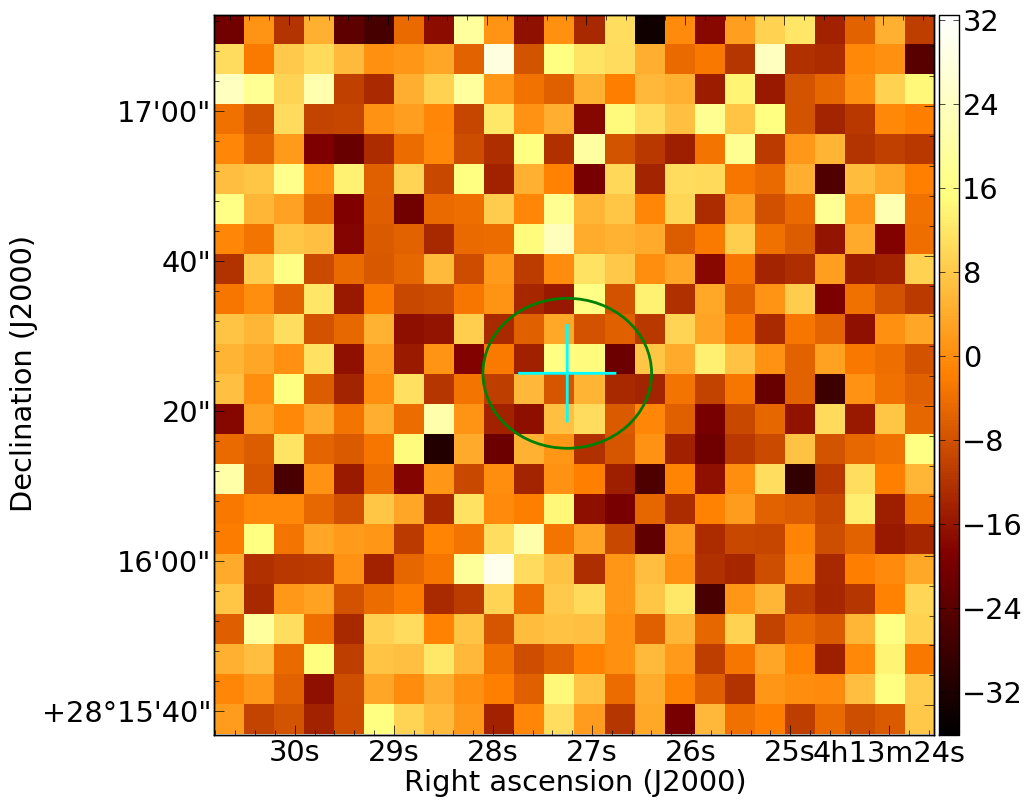}
\caption{as Fig.~\ref{diskfig0} Anon1 Class III \label{diskfig38}}
 \end{figure}

\begin{figure}
\includegraphics[width=4cm]{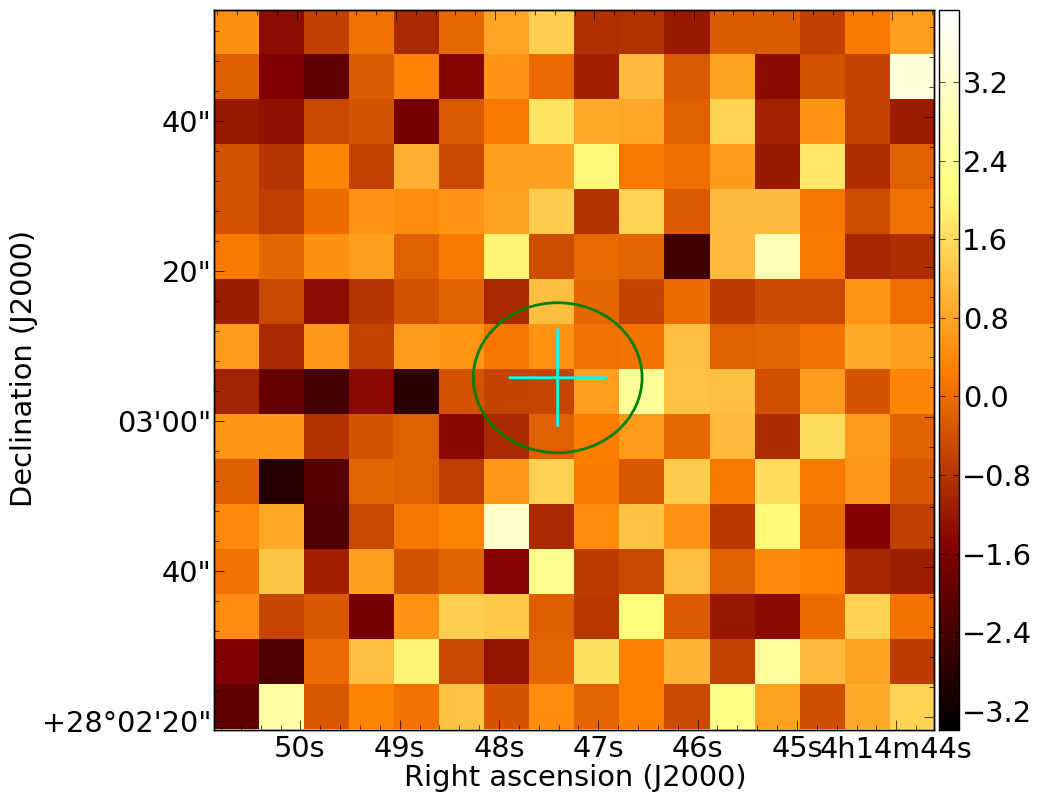}
\includegraphics[width=4cm]{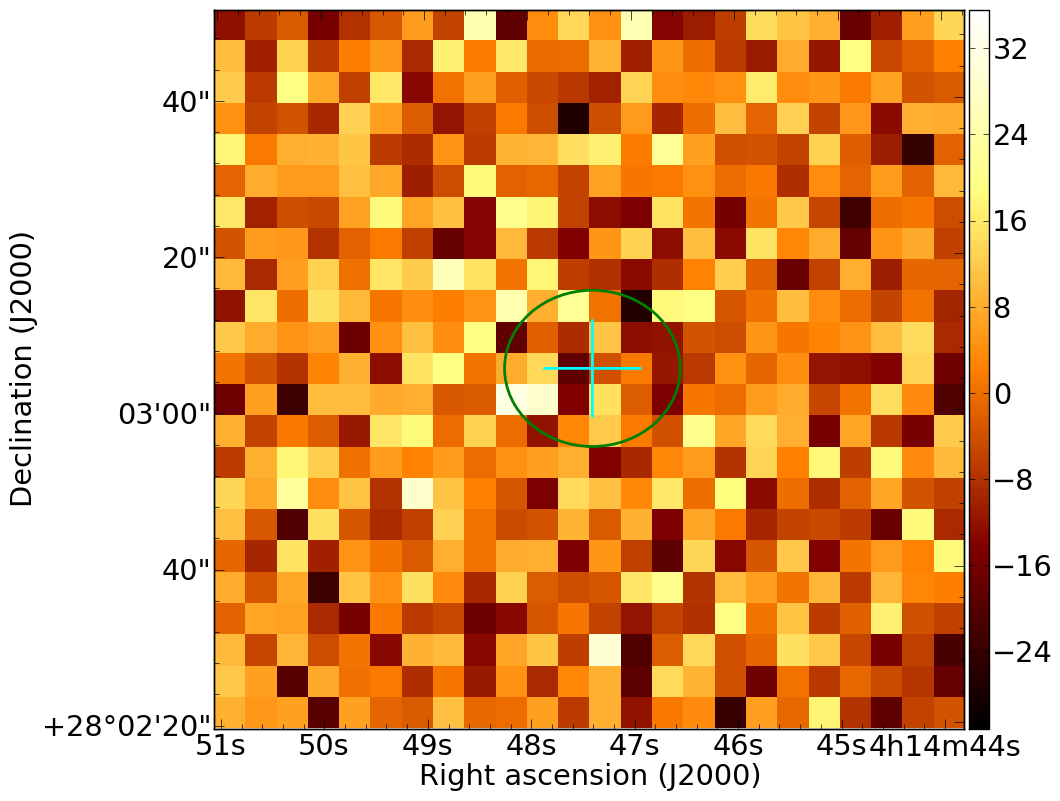}
\caption{as Fig.~\ref{diskfig0} XEST20-066 Class III \label{diskfig39}}
 \end{figure}

\begin{figure}
\includegraphics[width=4cm]{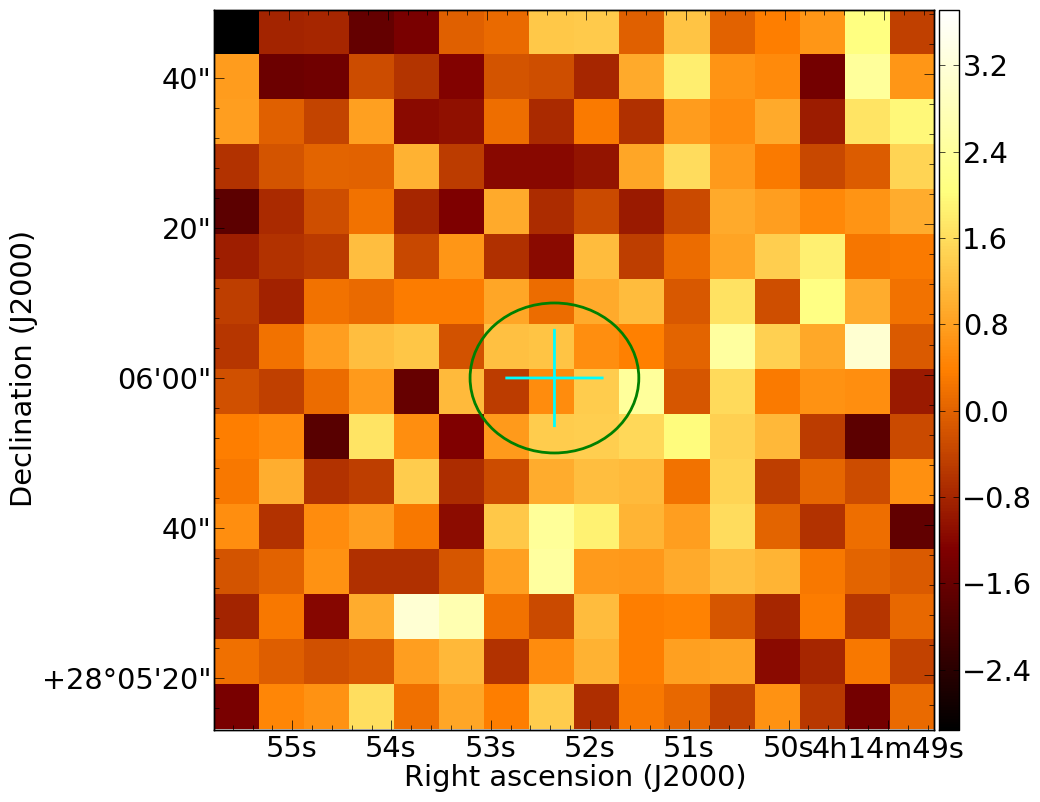}
\includegraphics[width=4cm]{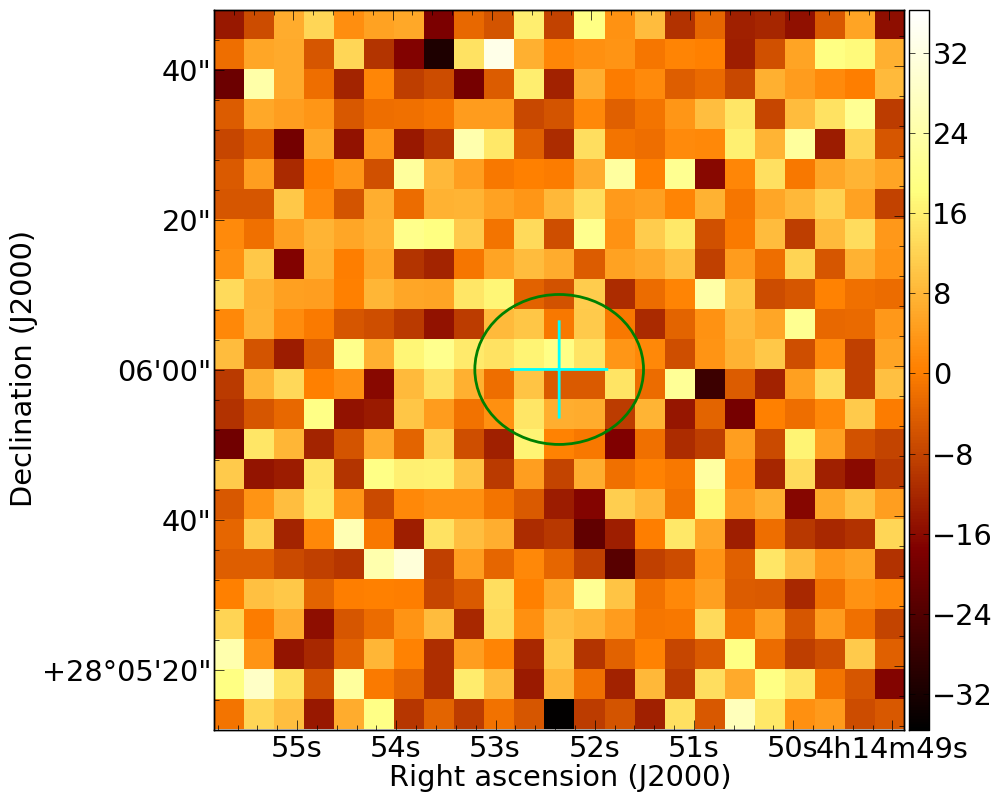}
\caption{as Fig.~\ref{diskfig0} XEST20-071 Class III \label{diskfig40}}
 \end{figure}

\begin{figure}
\includegraphics[width=4cm]{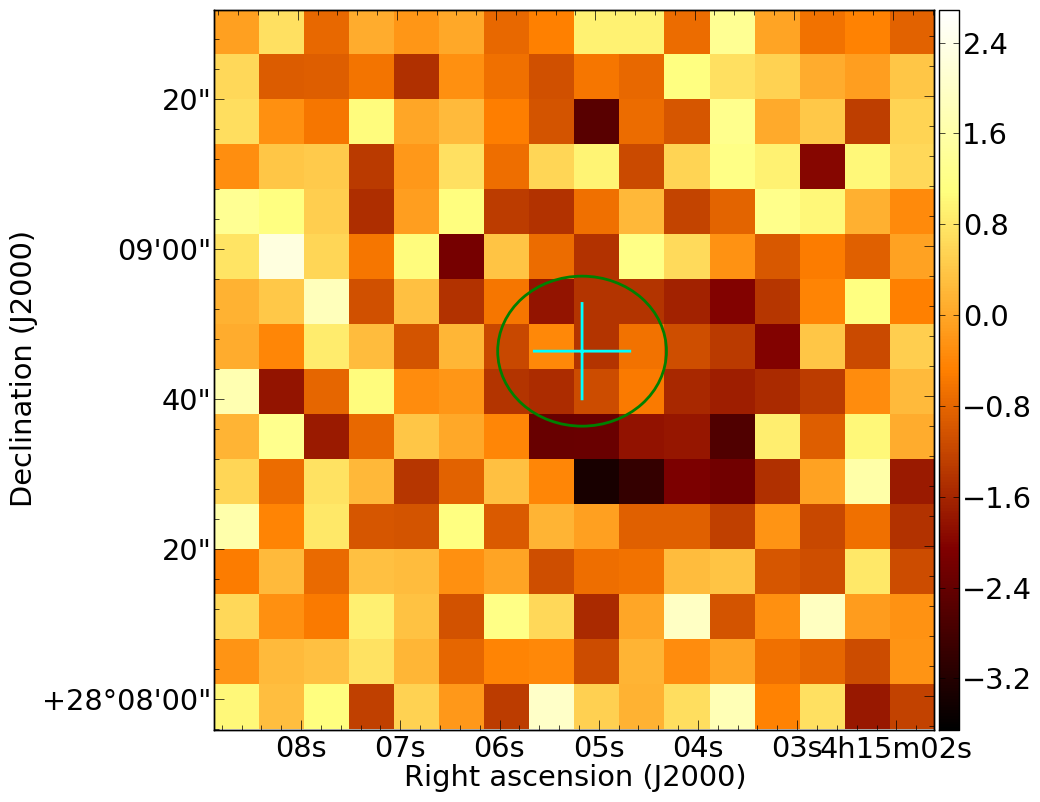}
\includegraphics[width=4cm]{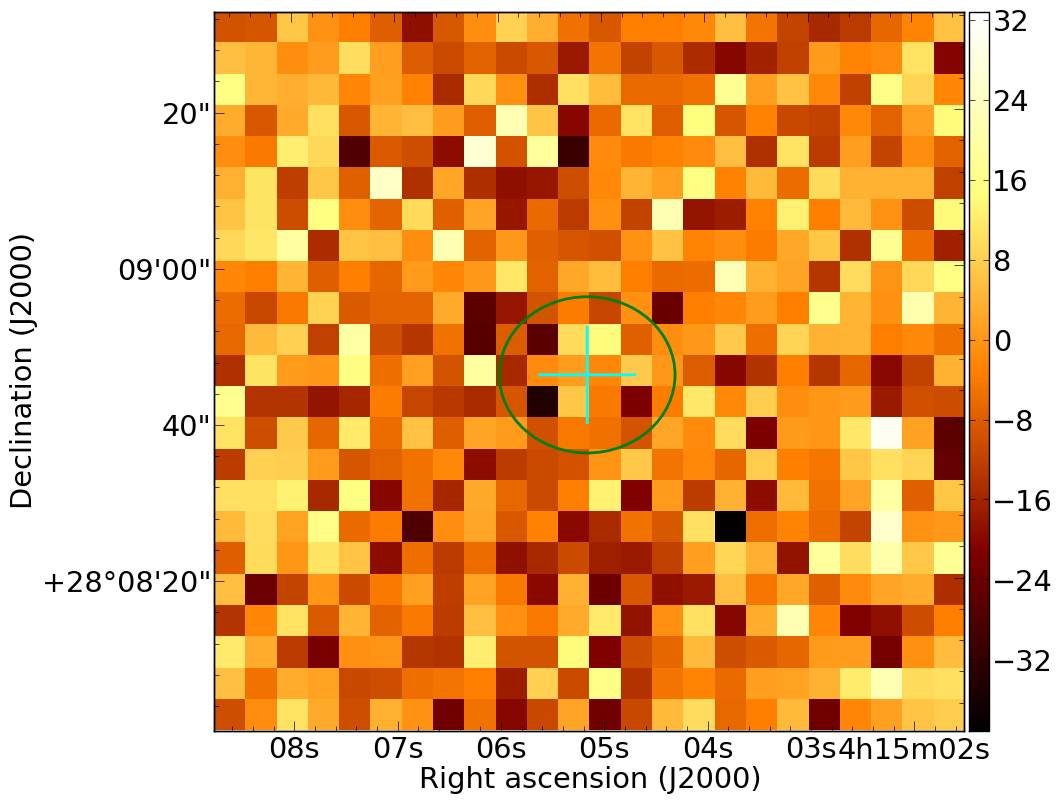}
\caption{as Fig.~\ref{diskfig0} CIDA2 Class III \label{diskfig41}}
 \end{figure}

\begin{figure}
\includegraphics[width=4cm]{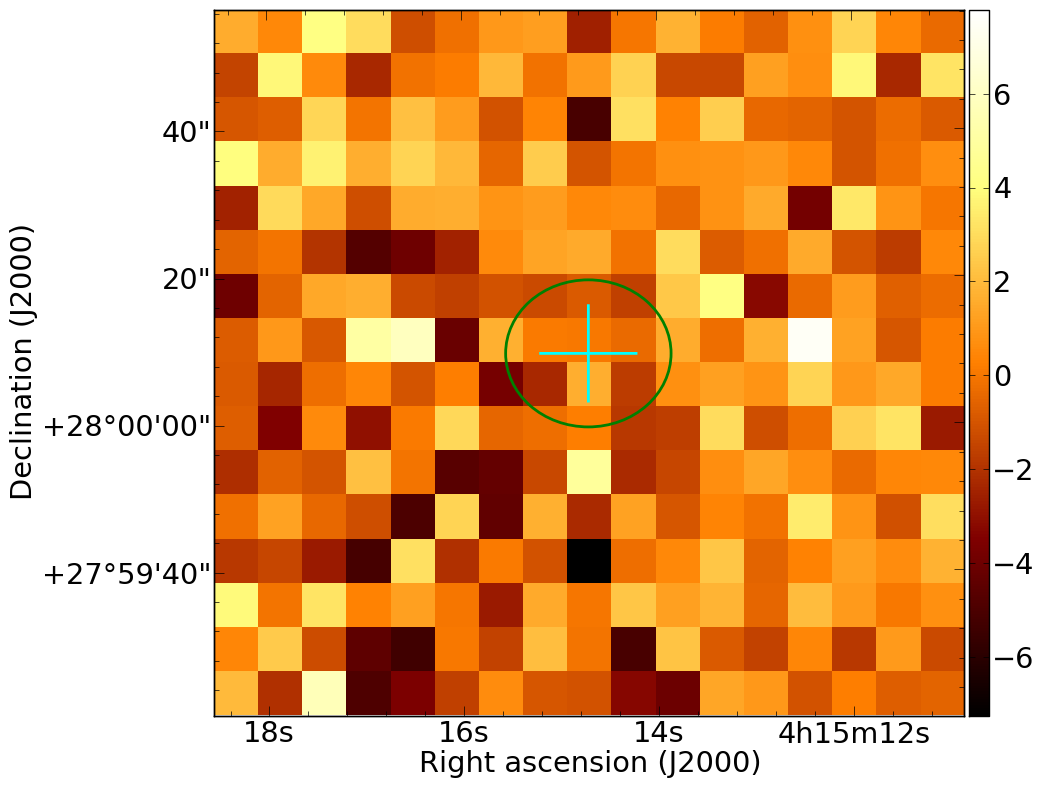}
\includegraphics[width=4cm]{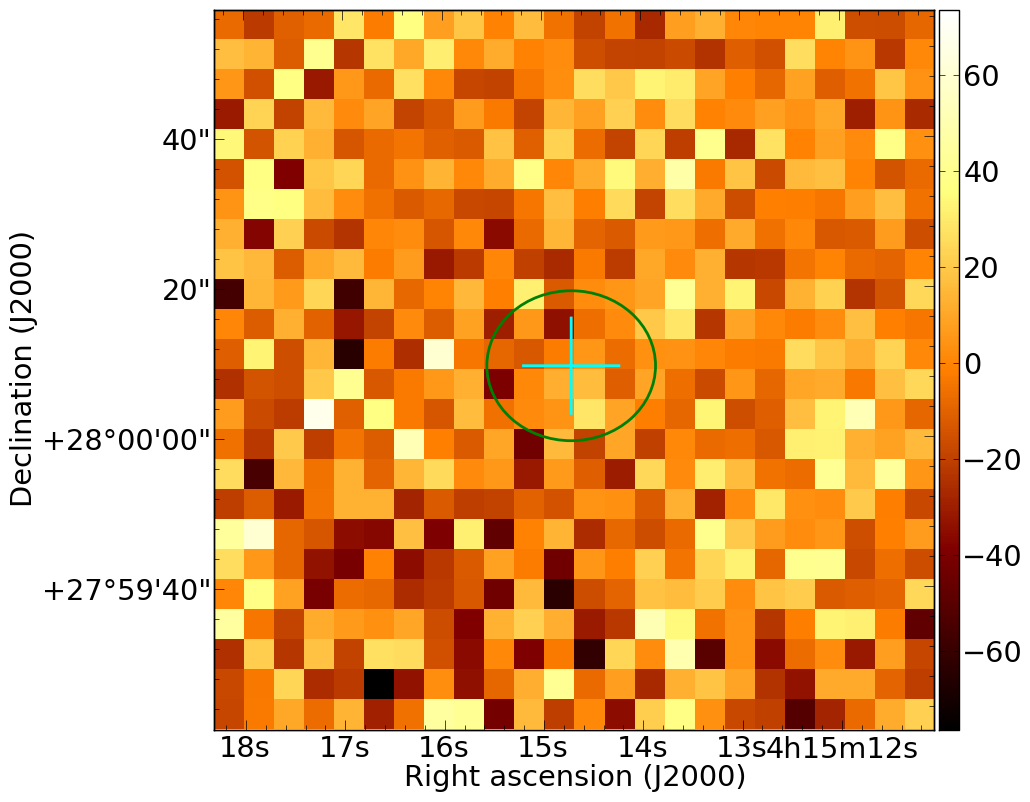}
\caption{as Fig.~\ref{diskfig0} KPNO1 Class III \label{diskfig42}}
 \end{figure}

\begin{figure}
\includegraphics[width=4cm]{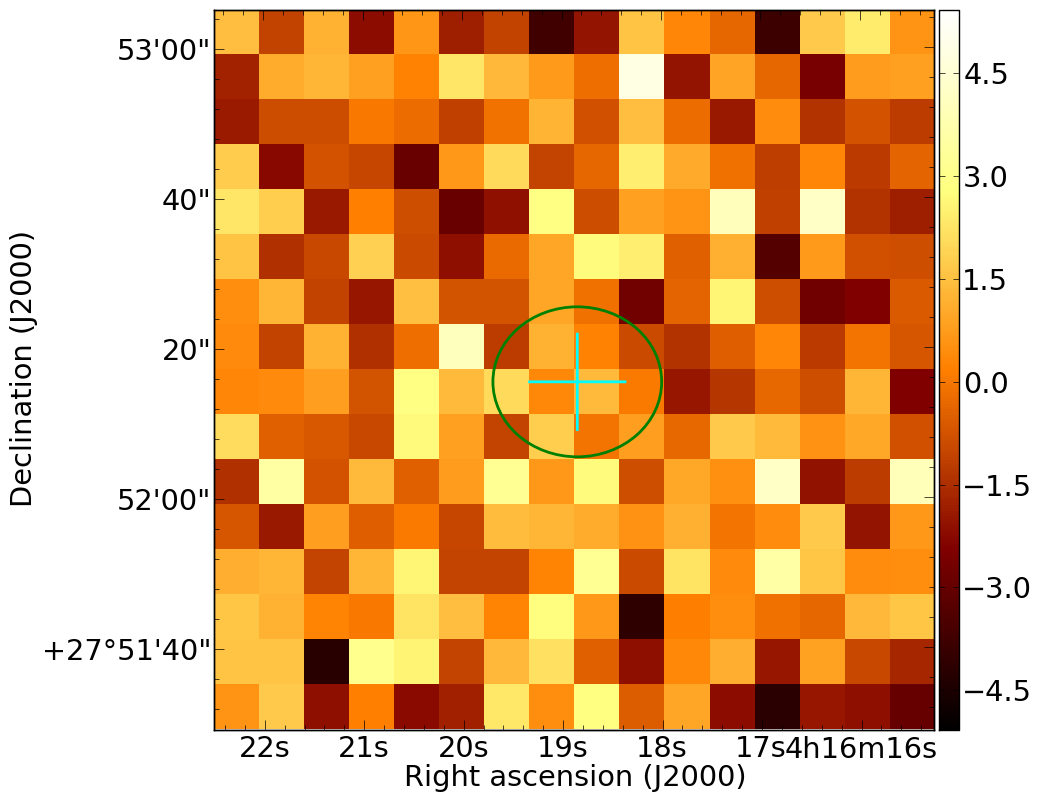}
\includegraphics[width=4cm]{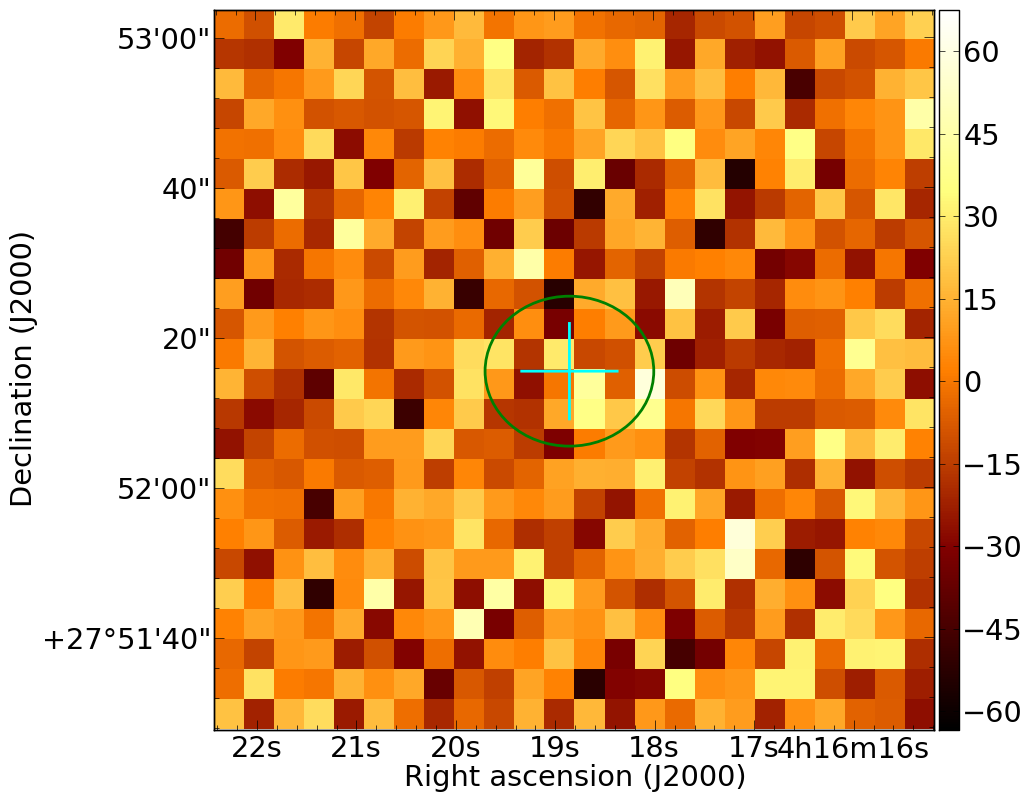}
\caption{as Fig.~\ref{diskfig0} J04161885+2752155 Class III \label{diskfig43}}
 \end{figure}

\begin{figure}
\includegraphics[width=4cm]{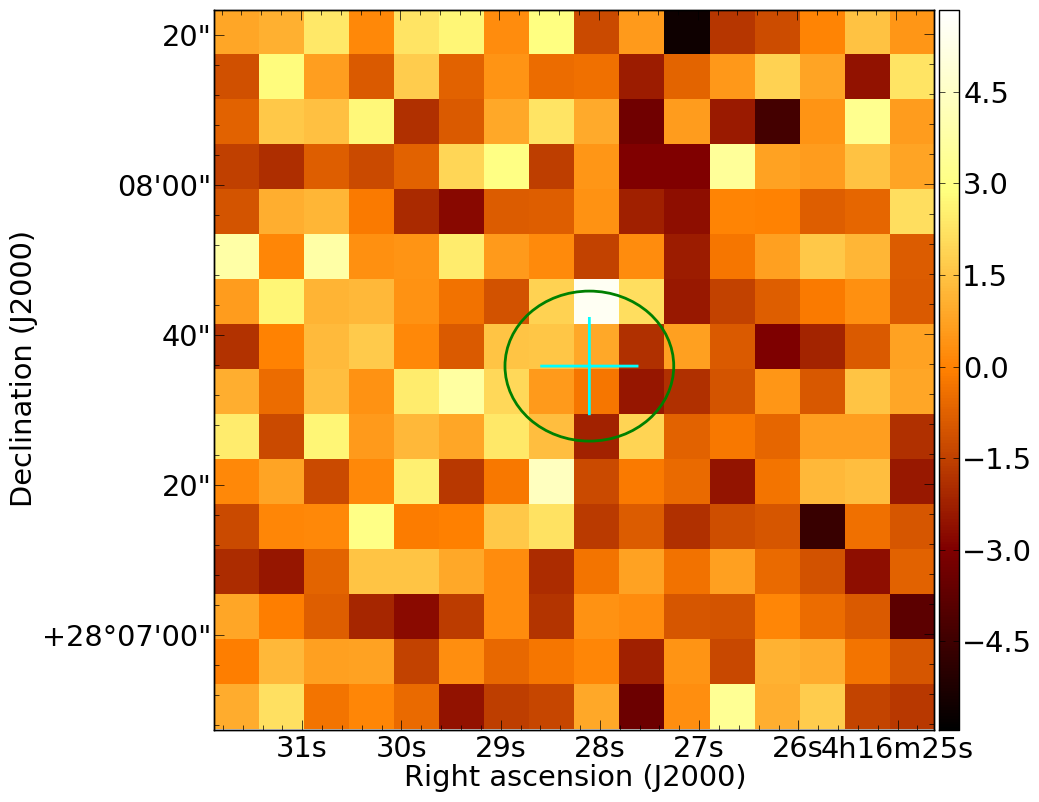}
\includegraphics[width=4cm]{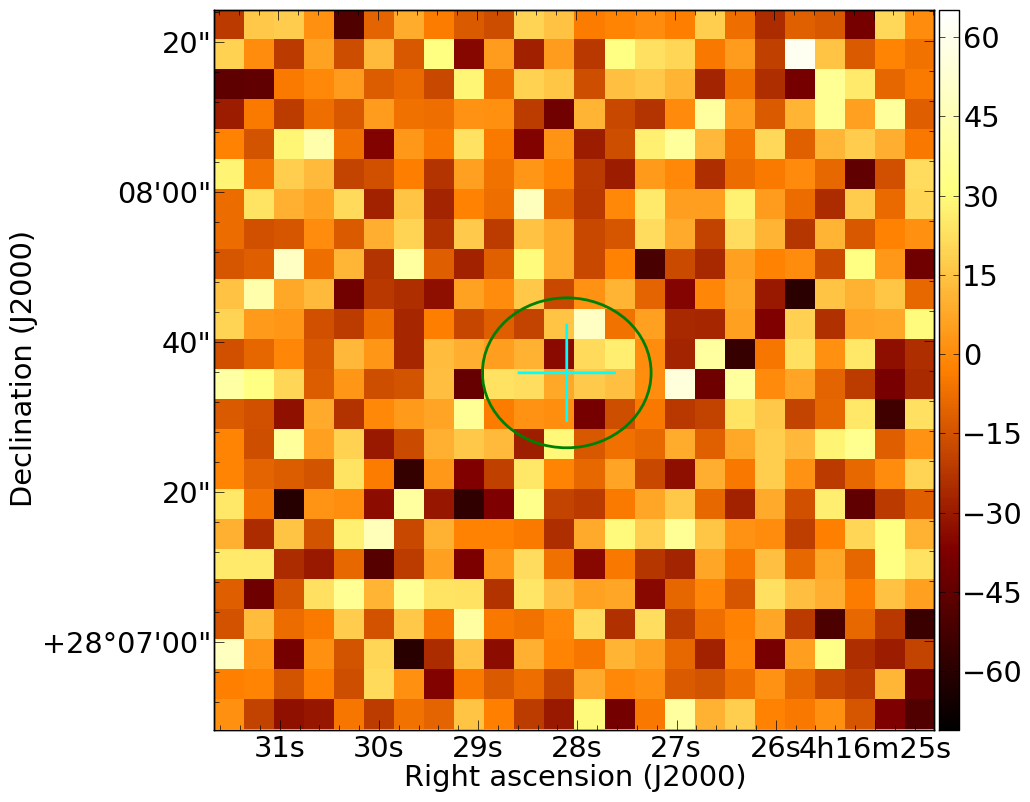}
\caption{as Fig.~\ref{diskfig0} LkCa4 Class III \label{diskfig44}}
 \end{figure}

\begin{figure}
\includegraphics[width=4cm]{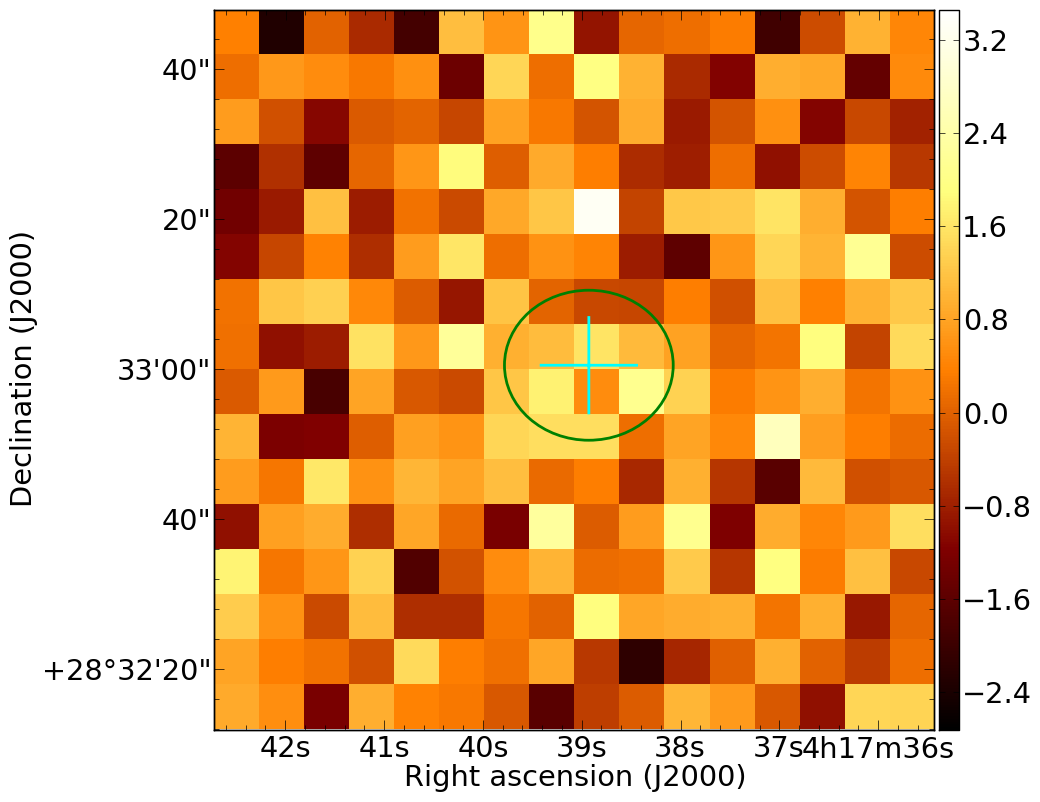}
\includegraphics[width=4cm]{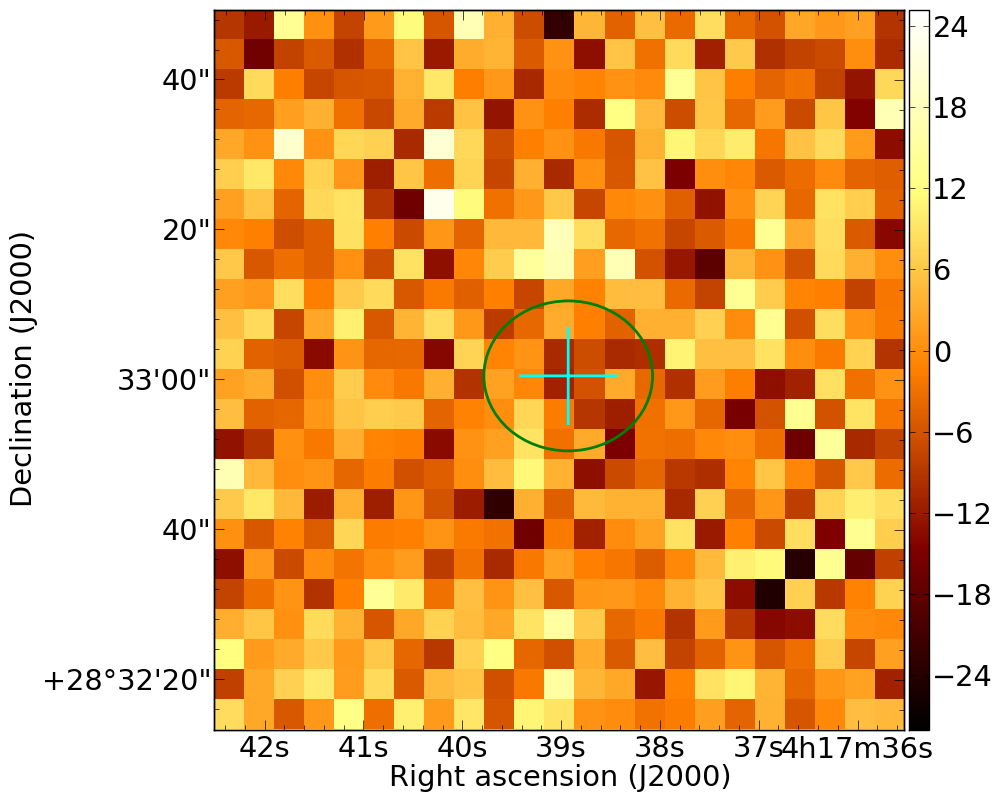}
\caption{as Fig.~\ref{diskfig0} LkCa5 Class III \label{diskfig45}}
 \end{figure}

\begin{figure}
\includegraphics[width=4cm]{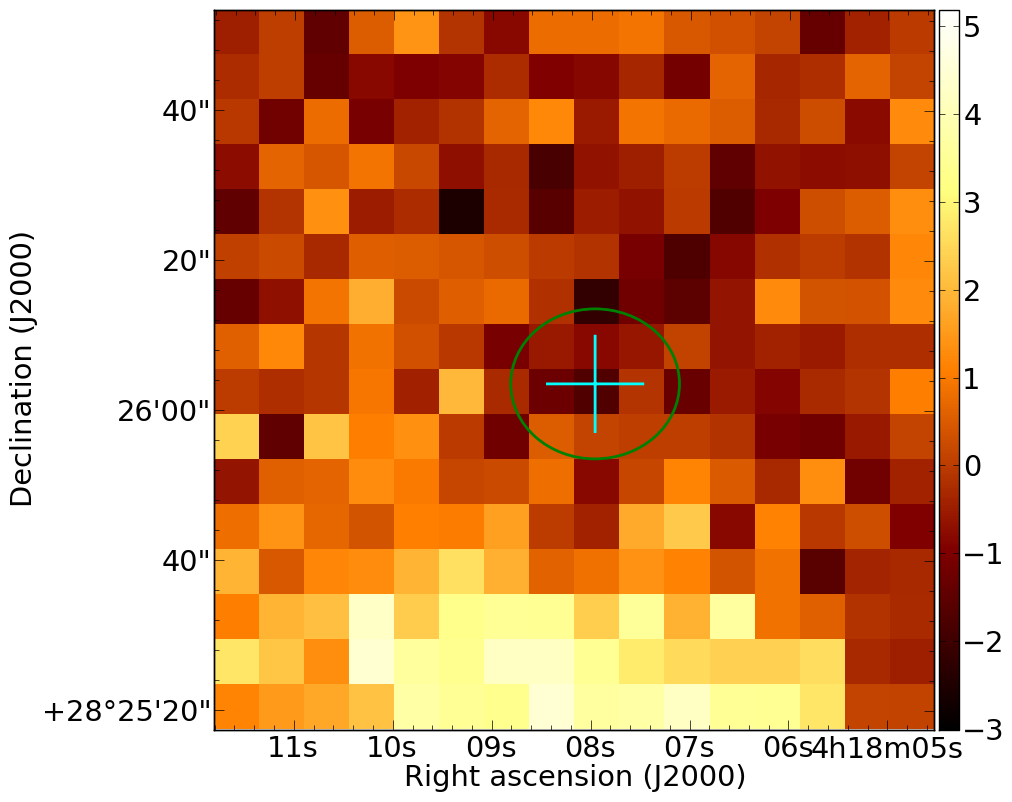}
\includegraphics[width=4cm]{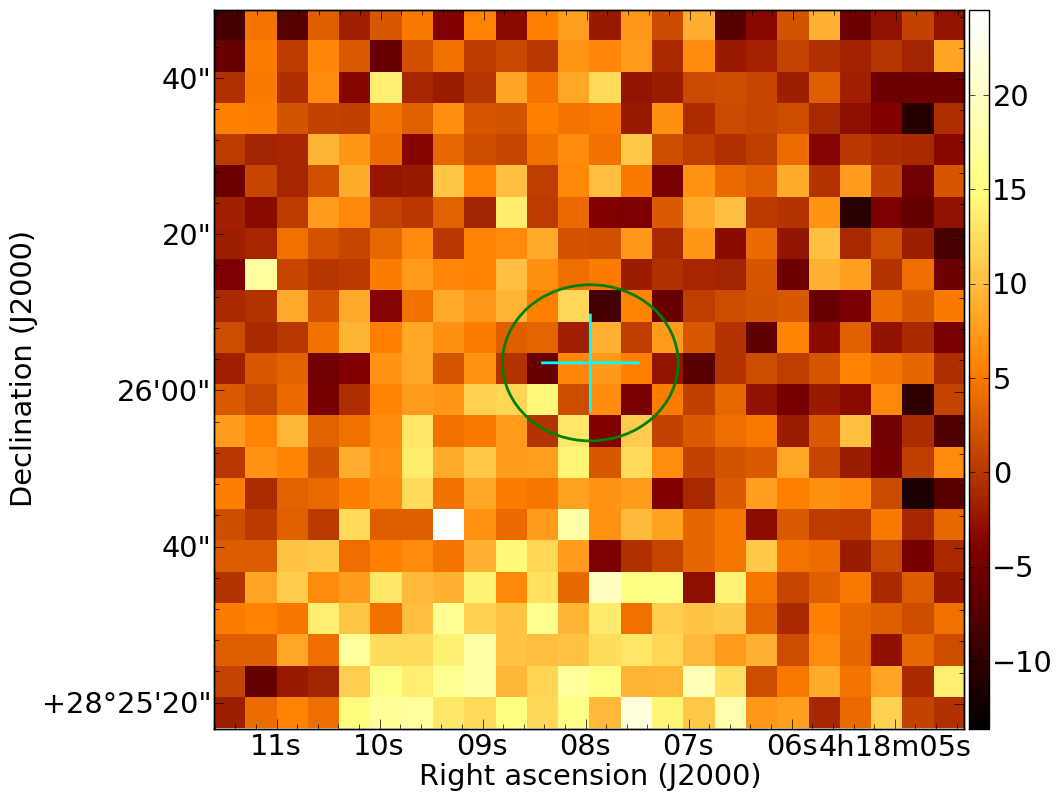}
\caption{as Fig.~\ref{diskfig0} V410Xray3 Class III \label{diskfig46}}
 \end{figure}

\begin{figure}
\includegraphics[width=4cm]{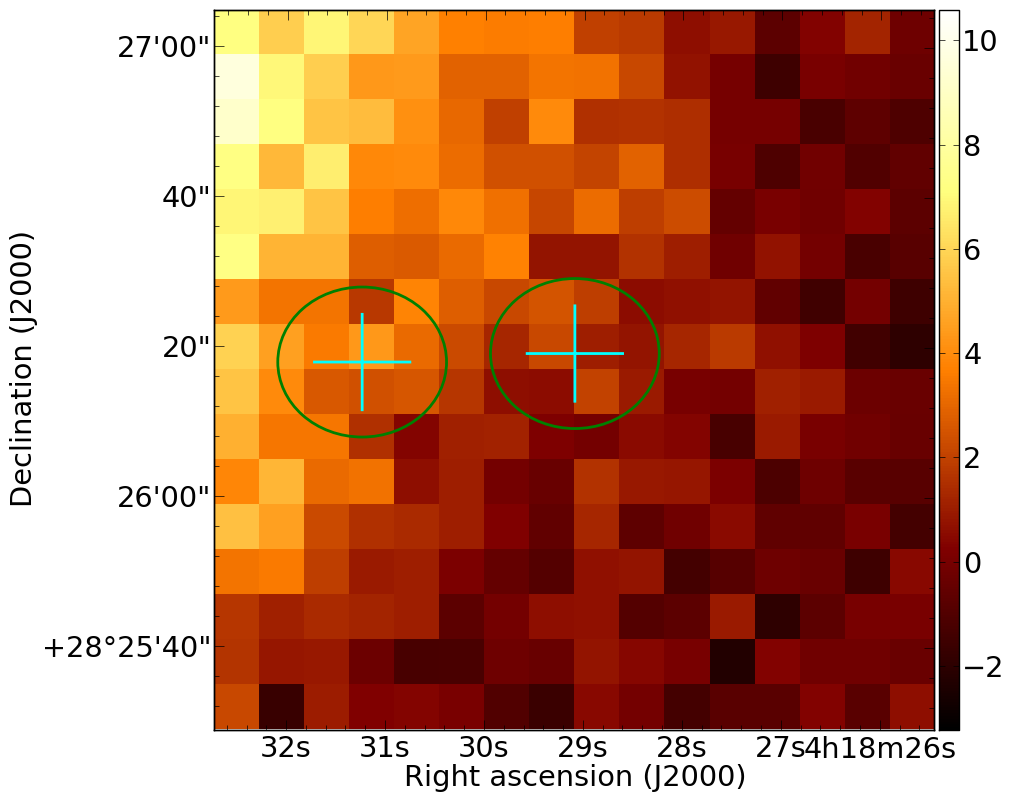}
\includegraphics[width=4cm]{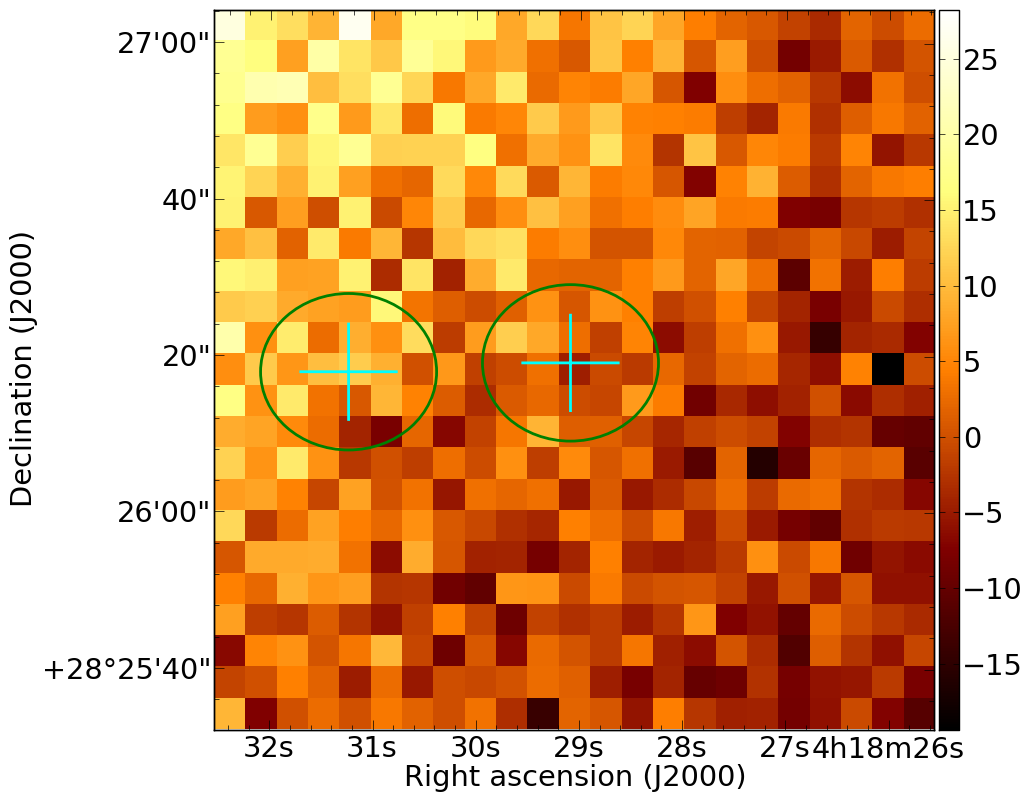}
\caption{as Fig.~\ref{diskfig0} V410Anon25 Class III \label{diskfig47}}
 \end{figure}

\begin{figure}
\includegraphics[width=4cm]{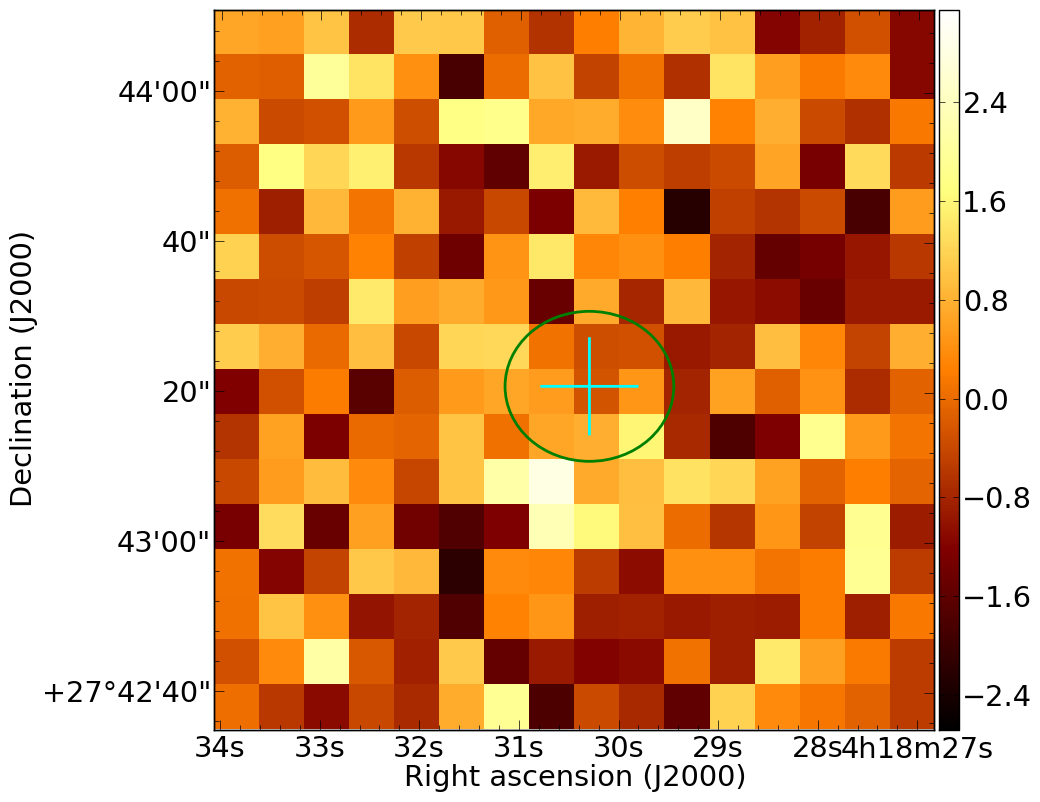}
\includegraphics[width=4cm]{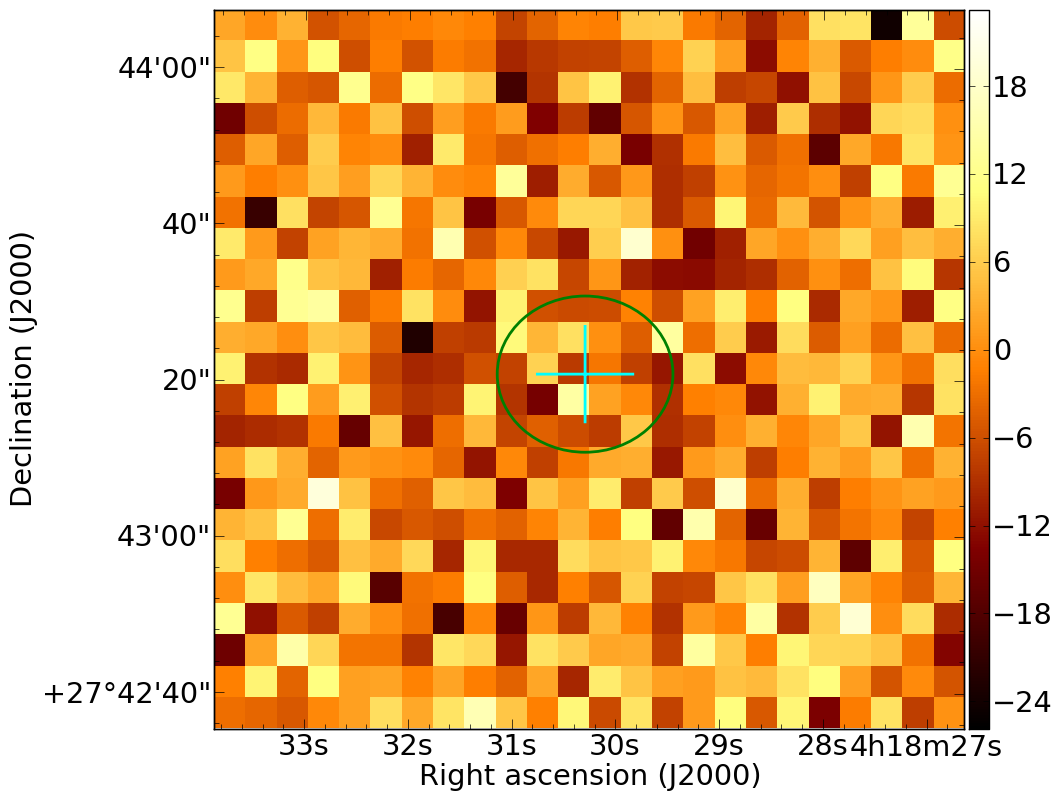}
\caption{as Fig.~\ref{diskfig0} KPNO11 Class III \label{diskfig48}}
 \end{figure}

\begin{figure}
\includegraphics[width=4cm]{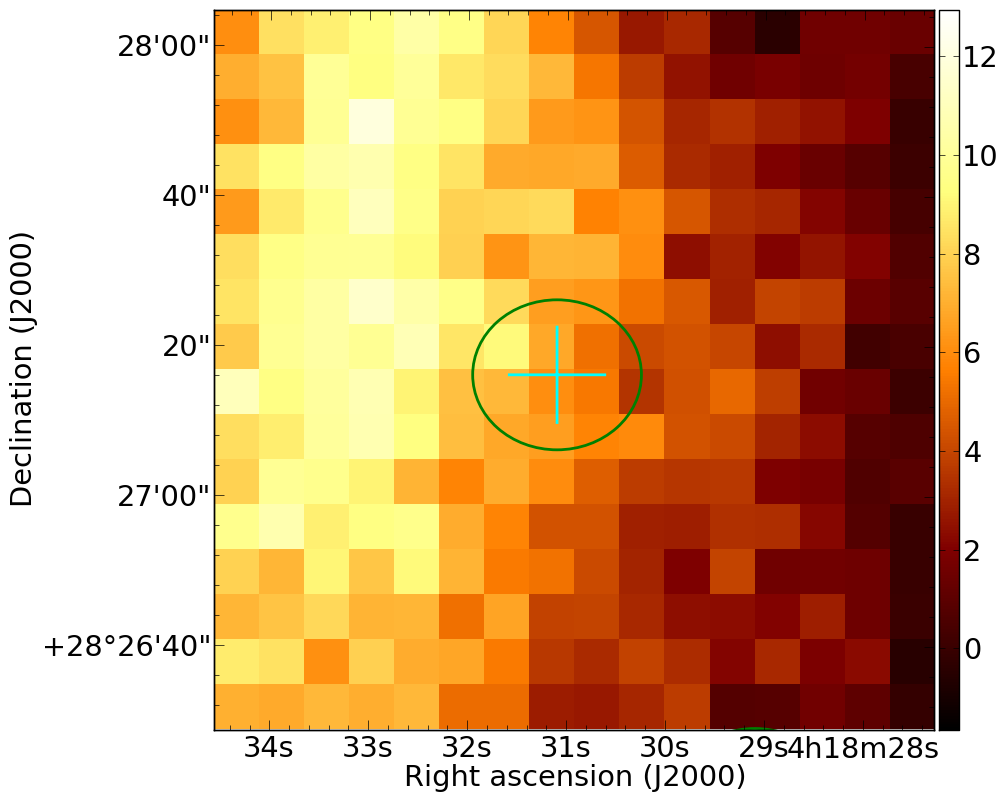}
\includegraphics[width=4cm]{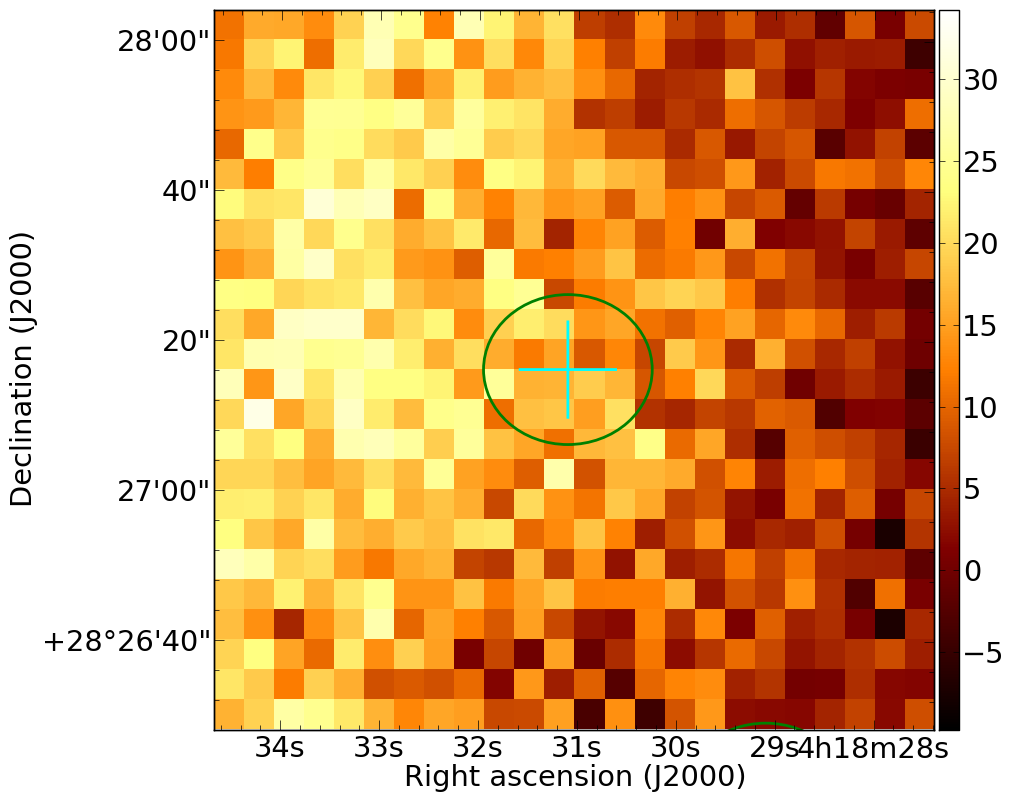}
\caption{as Fig.~\ref{diskfig0} V410TauA+B+C Class III \label{diskfig49}}
 \end{figure}

\begin{figure}
\includegraphics[width=4cm]{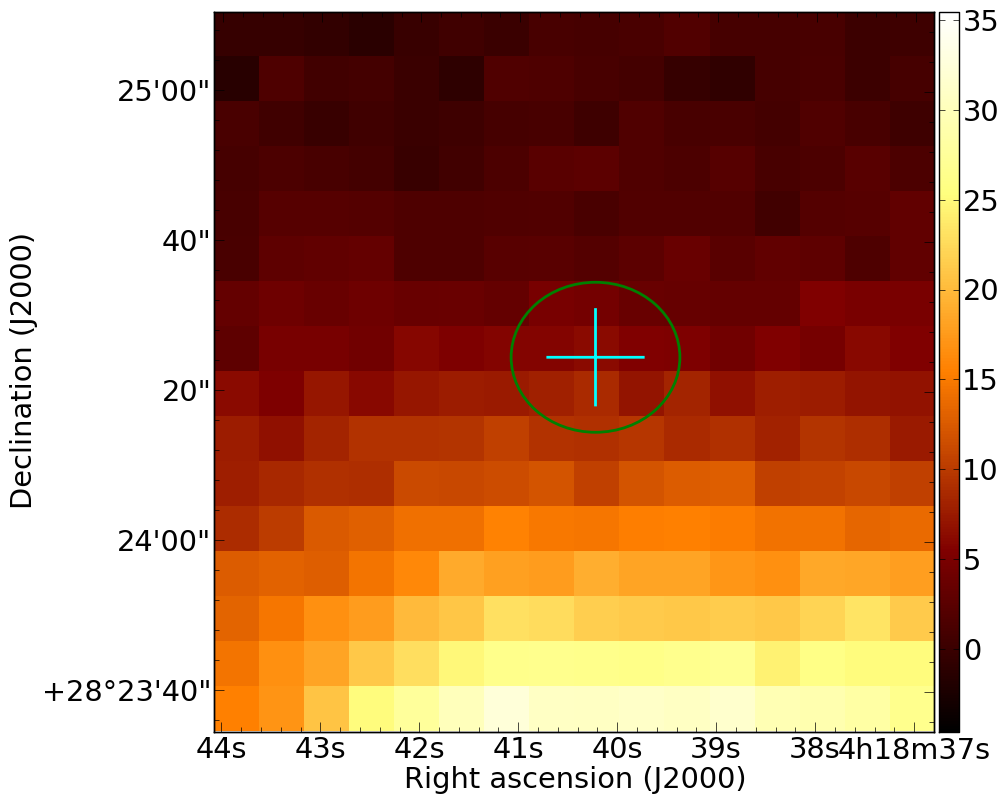}
\includegraphics[width=4cm]{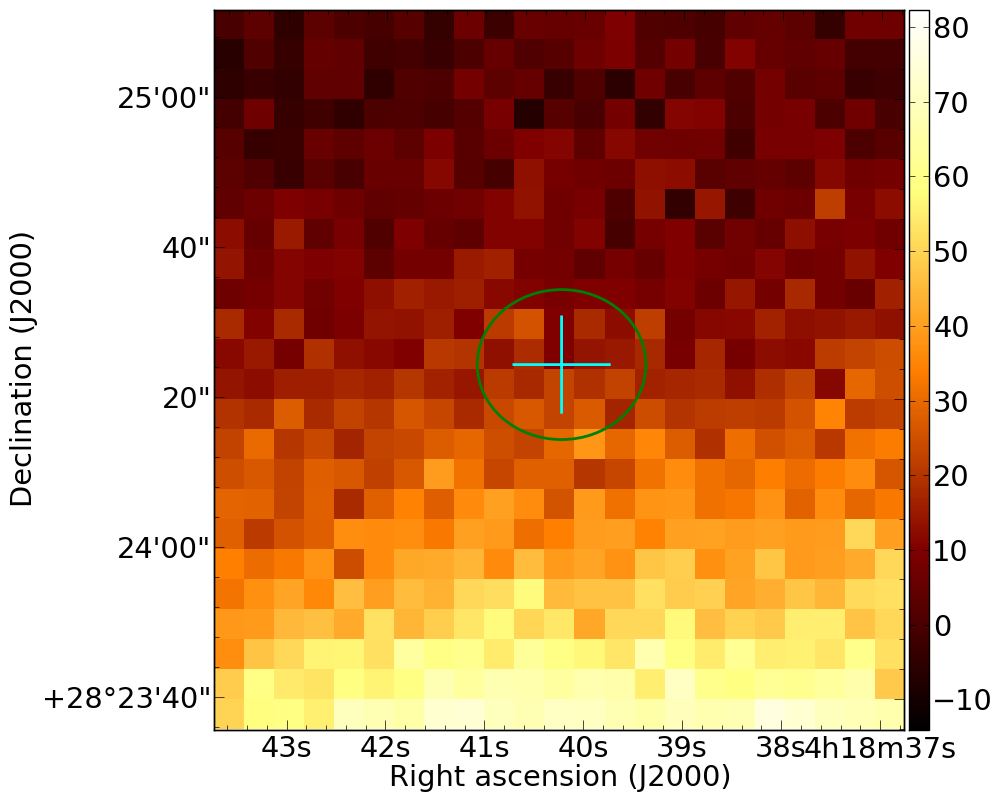}
\caption{as Fig.~\ref{diskfig0} V410X-ray4 Class III \label{diskfig50}}
 \end{figure}

\begin{figure}
\includegraphics[width=4cm]{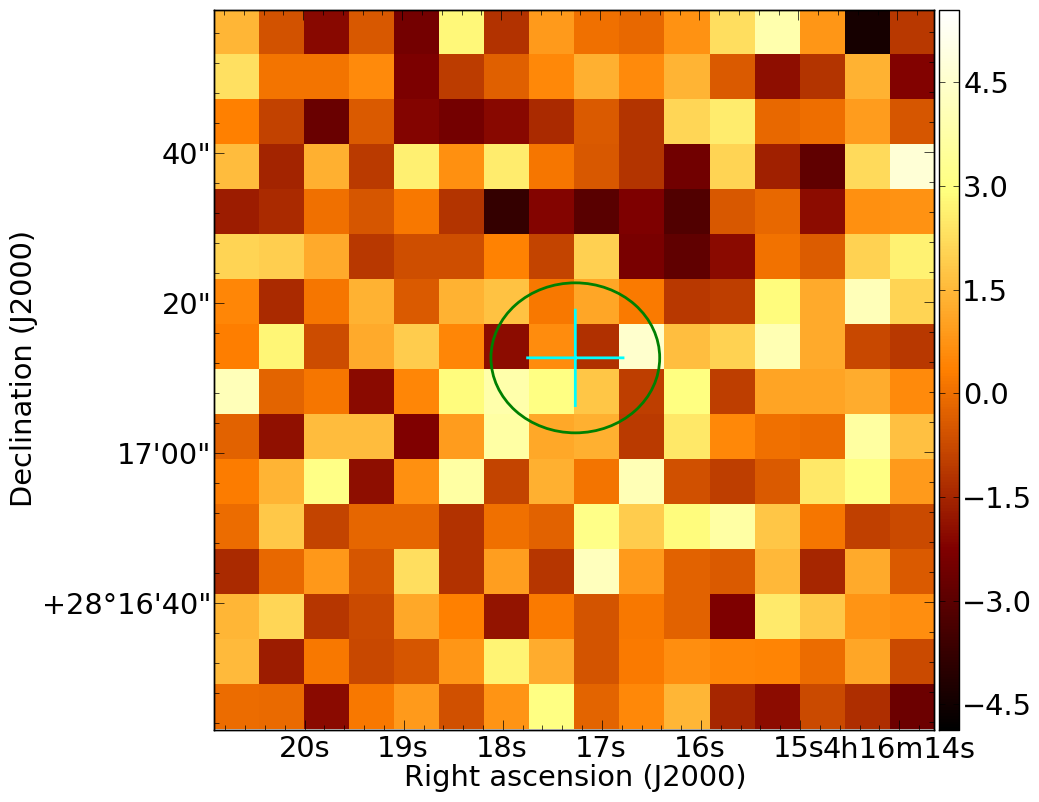}
\includegraphics[width=4cm]{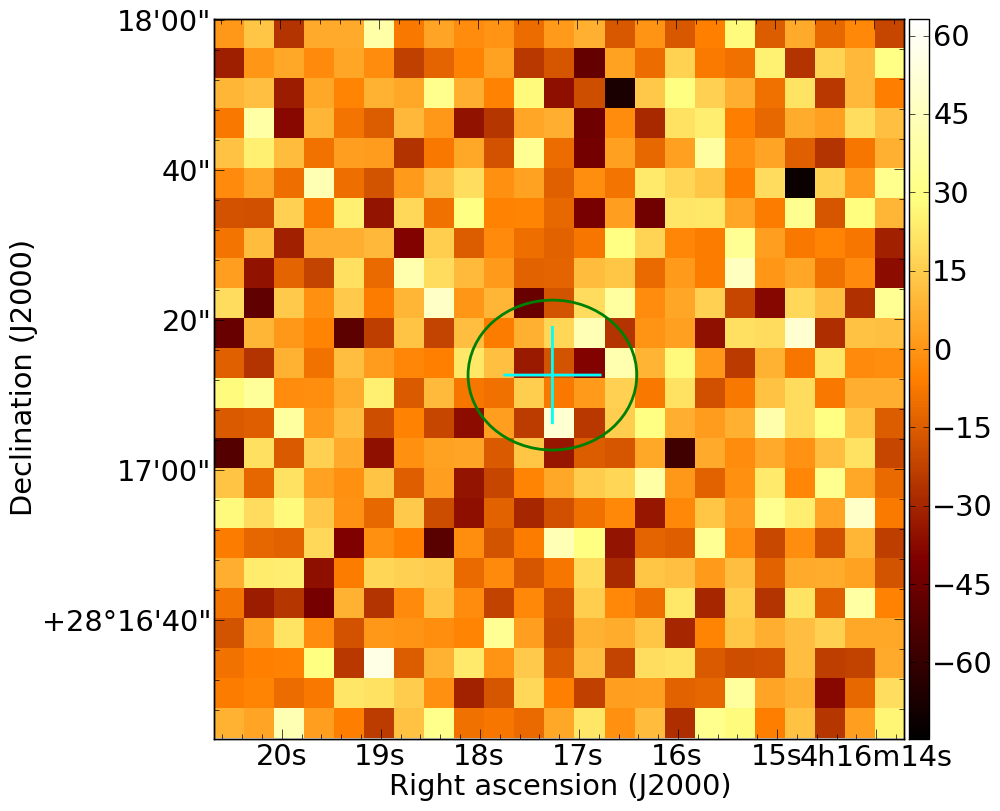}
\caption{as Fig.~\ref{diskfig0} J04161726+2817128 Class III \label{diskfig55}}
 \end{figure}

Figures \ref{diskfig0}--\ref{diskfig55} show thumbnail images of the dust continuum at the nominal position for each disk at 850\,$\mu$m and 450\,$\mu$m.  The flux is shown in mJy\,pixel$^{-1}$, and circles show the 20\arcsec\ aperture used to measure fluxes. The centrally-aligned source is the one named in the caption. 

 
 
 \clearpage 

 
 \clearpage 
 
 
 \clearpage 
 
 
 \clearpage
 

\clearpage

\begin{figure}
\includegraphics[width=4cm]{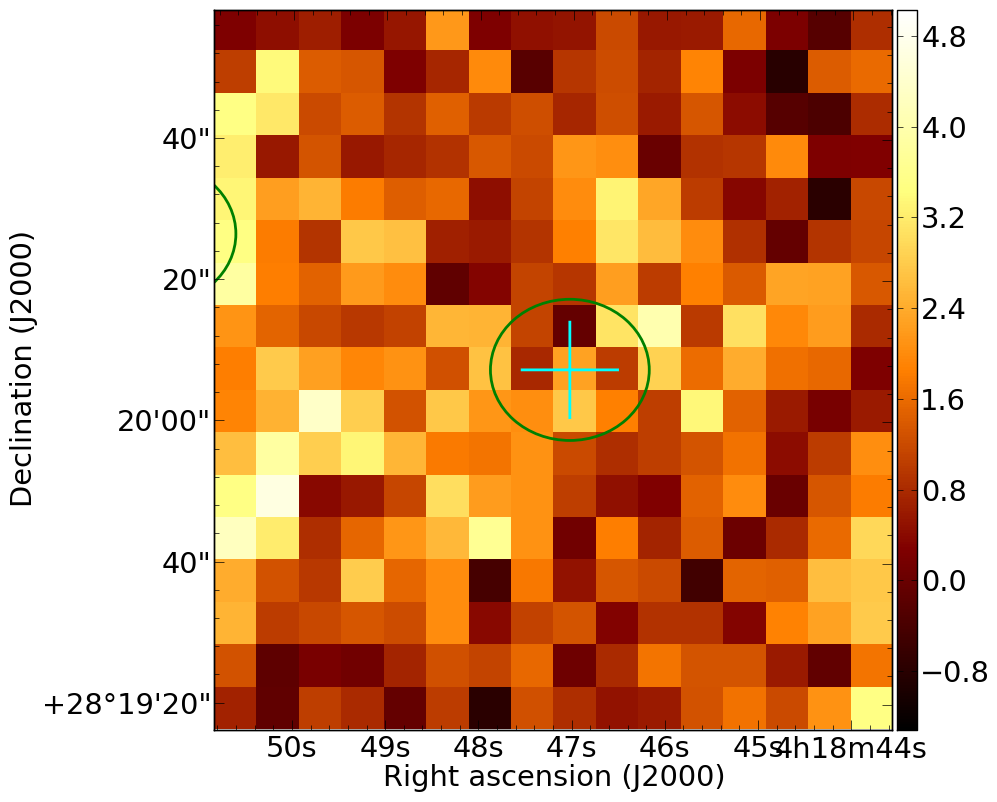}
\includegraphics[width=4cm]{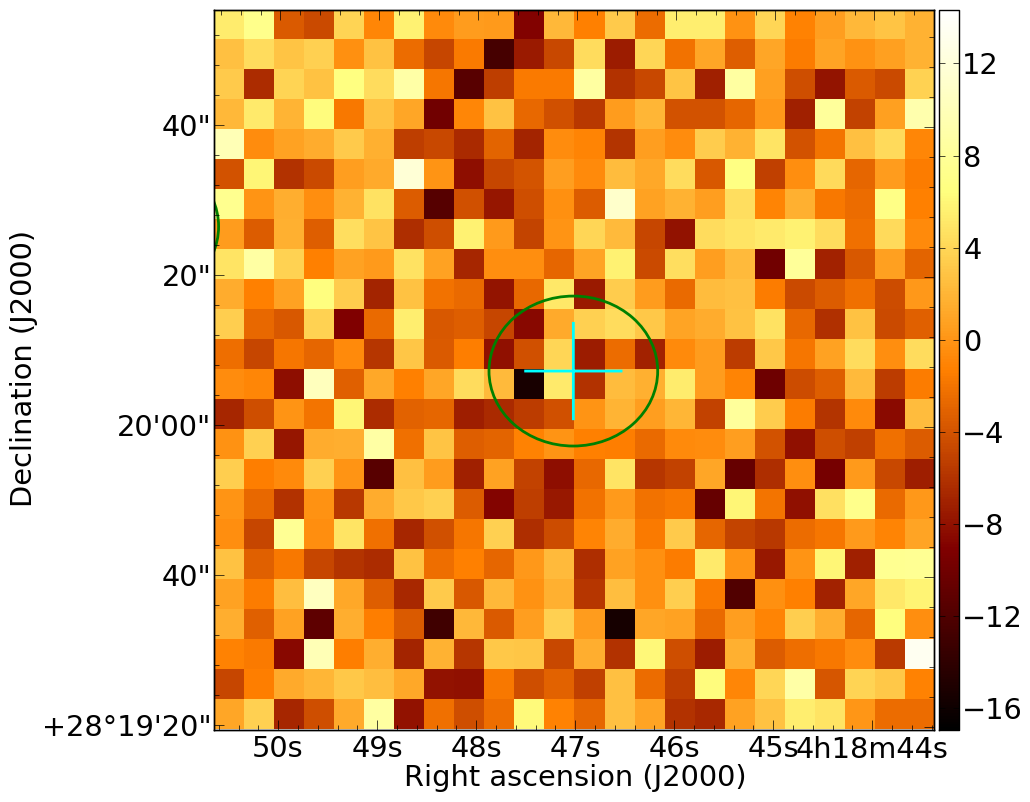}
\caption{as Fig.~\ref{diskfig0} Hubble4 Class III \label{diskfig51}}
 \end{figure} 
\begin{figure}
\includegraphics[width=4cm]{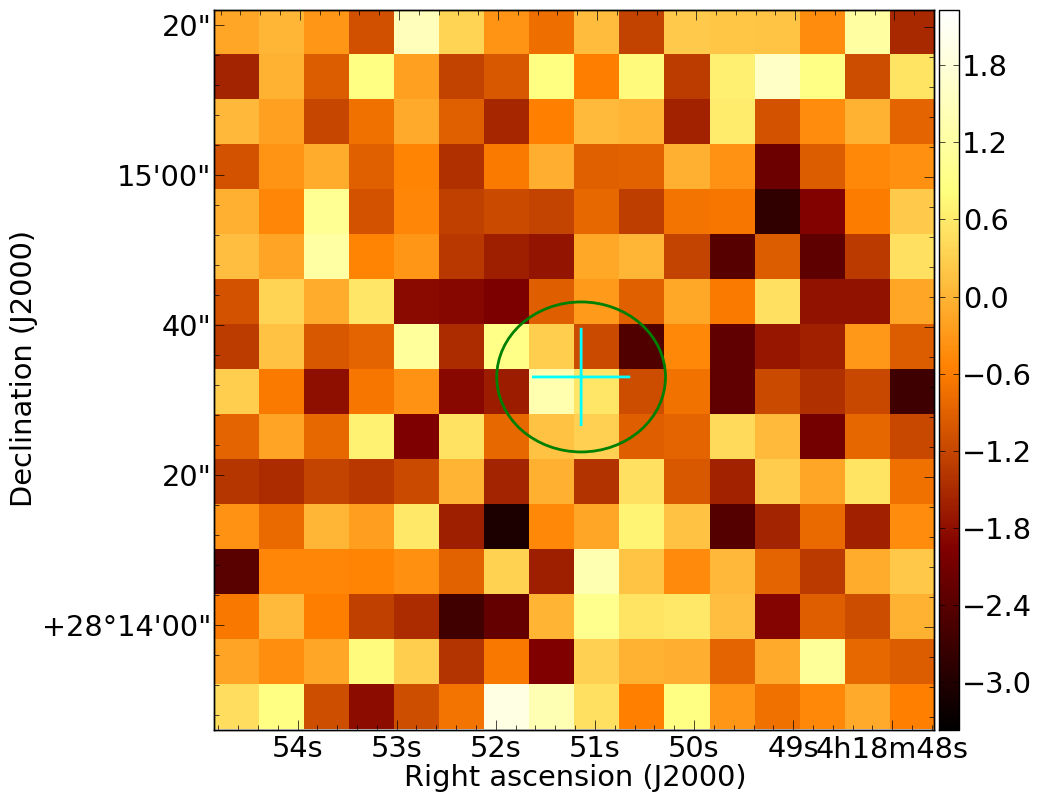}
\includegraphics[width=4cm]{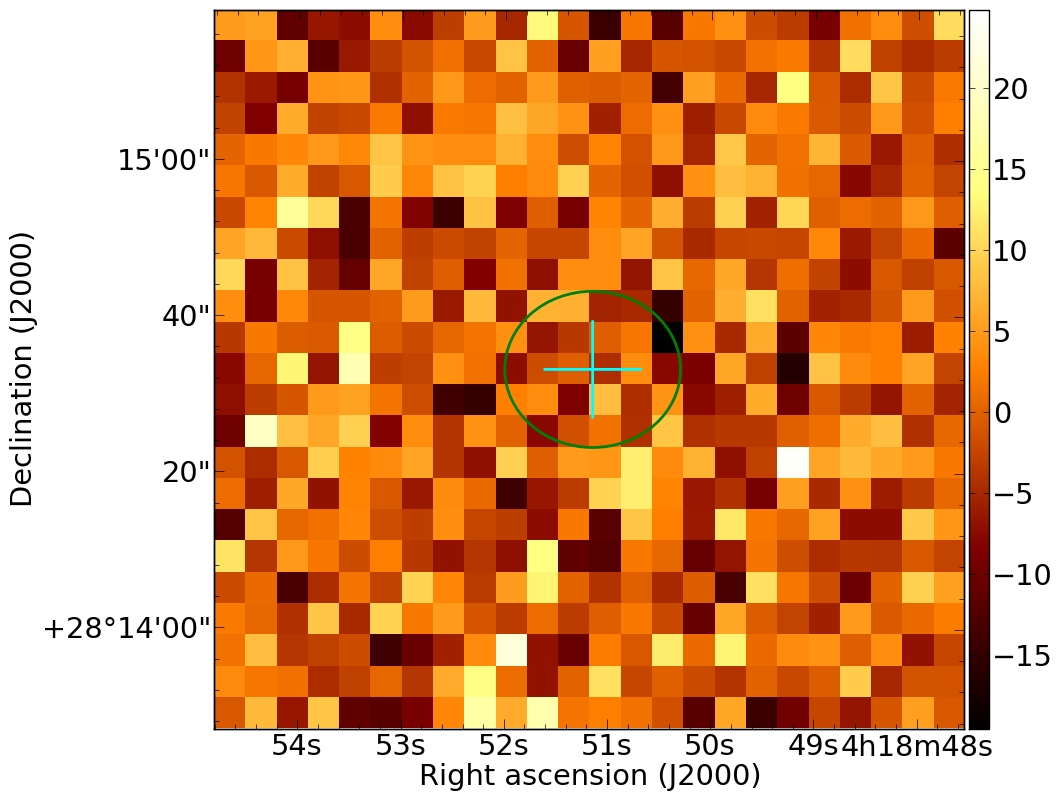}
\caption{as Fig.~\ref{diskfig0} KPNO2 Class III \label{diskfig52}}
 \end{figure} 
\begin{figure}
\includegraphics[width=4cm]{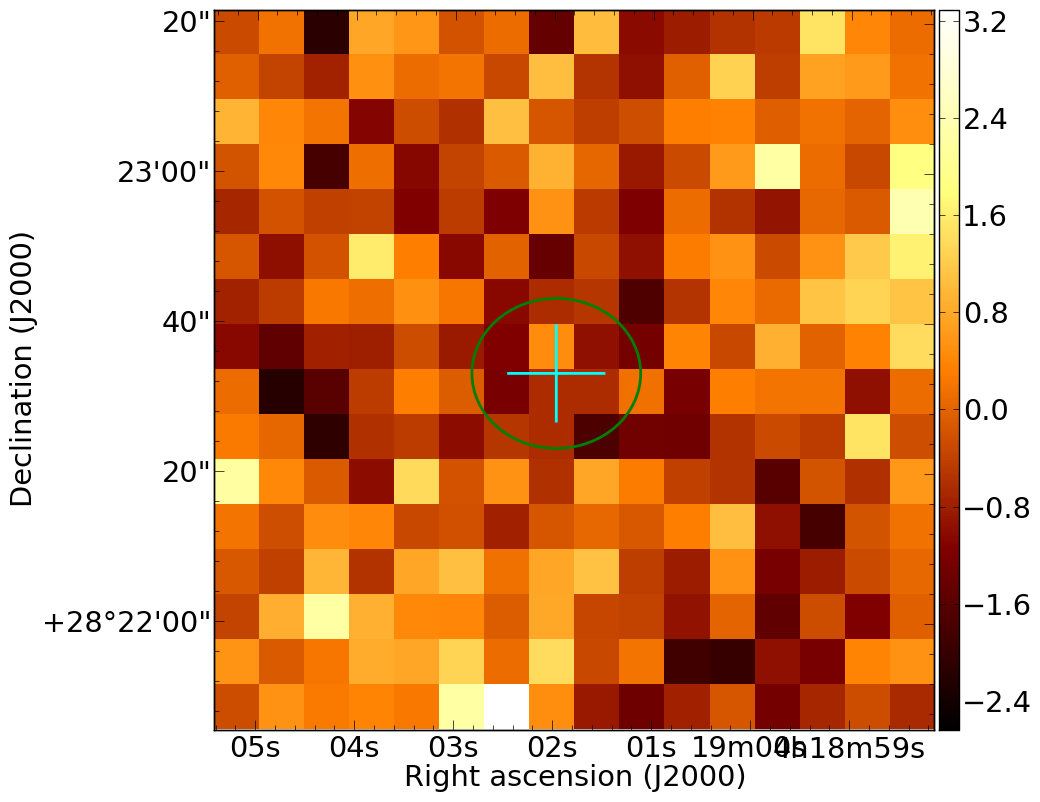}
\includegraphics[width=4cm]{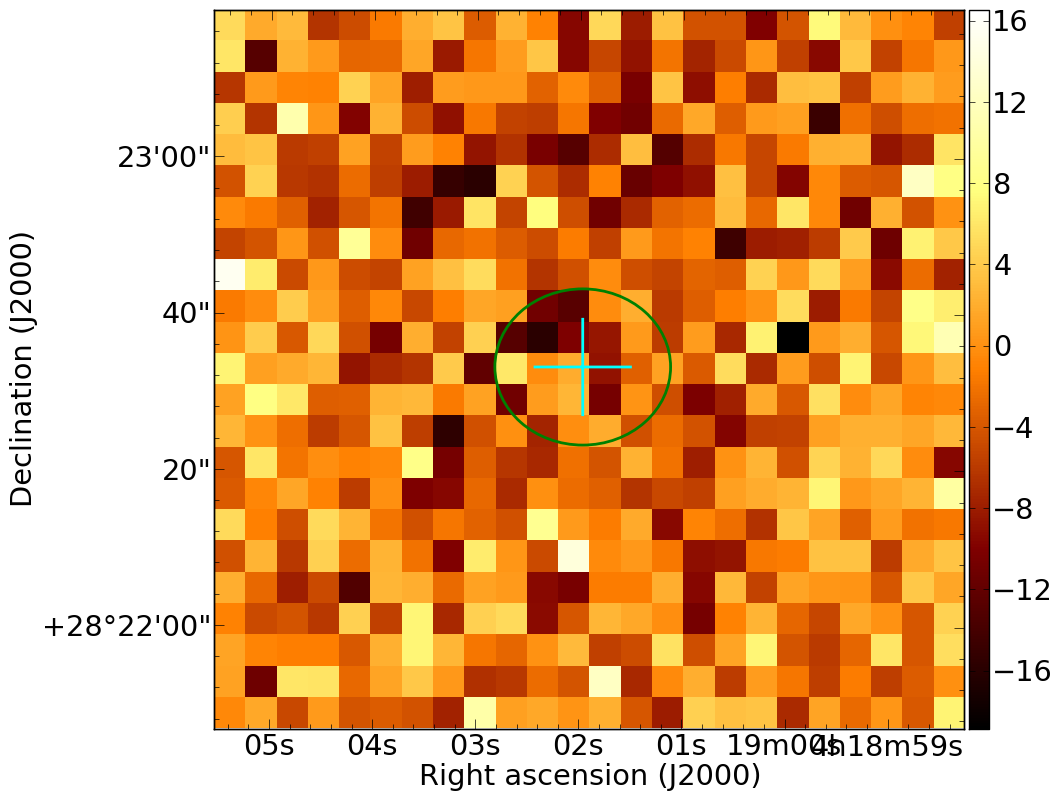}
\caption{as Fig.~\ref{diskfig0} V410X-ray5a Class III \label{diskfig53}}
 \end{figure} 
\begin{figure}
\includegraphics[width=4cm]{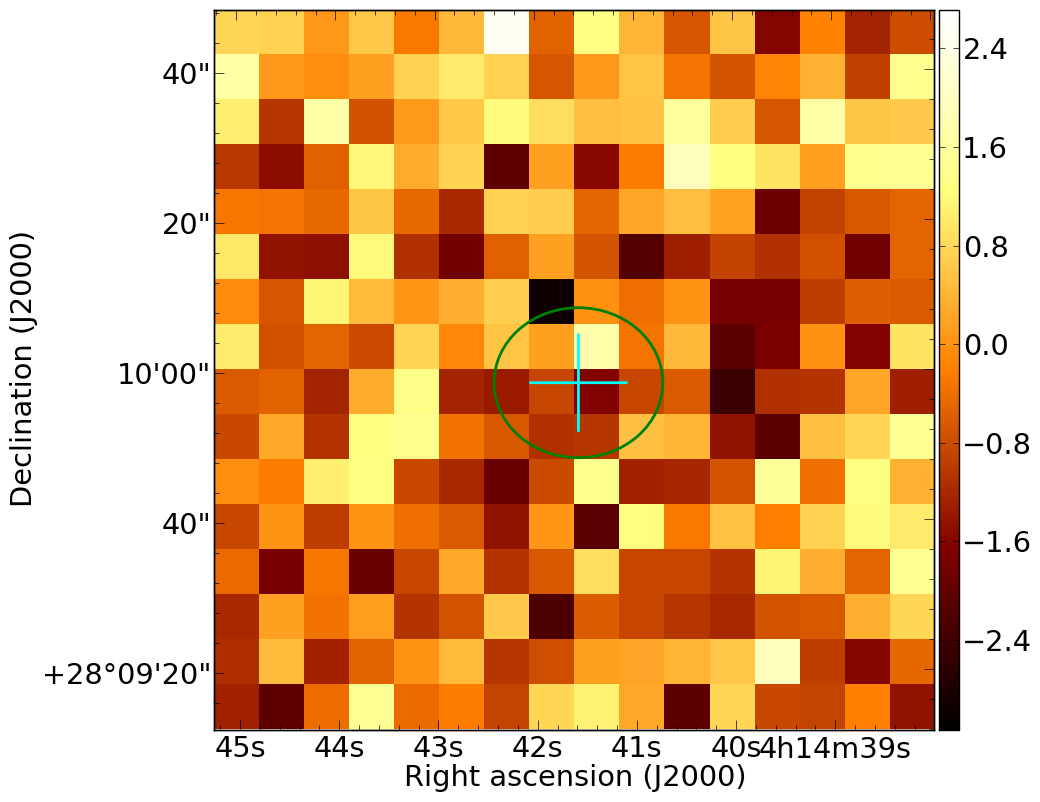}
\includegraphics[width=4cm]{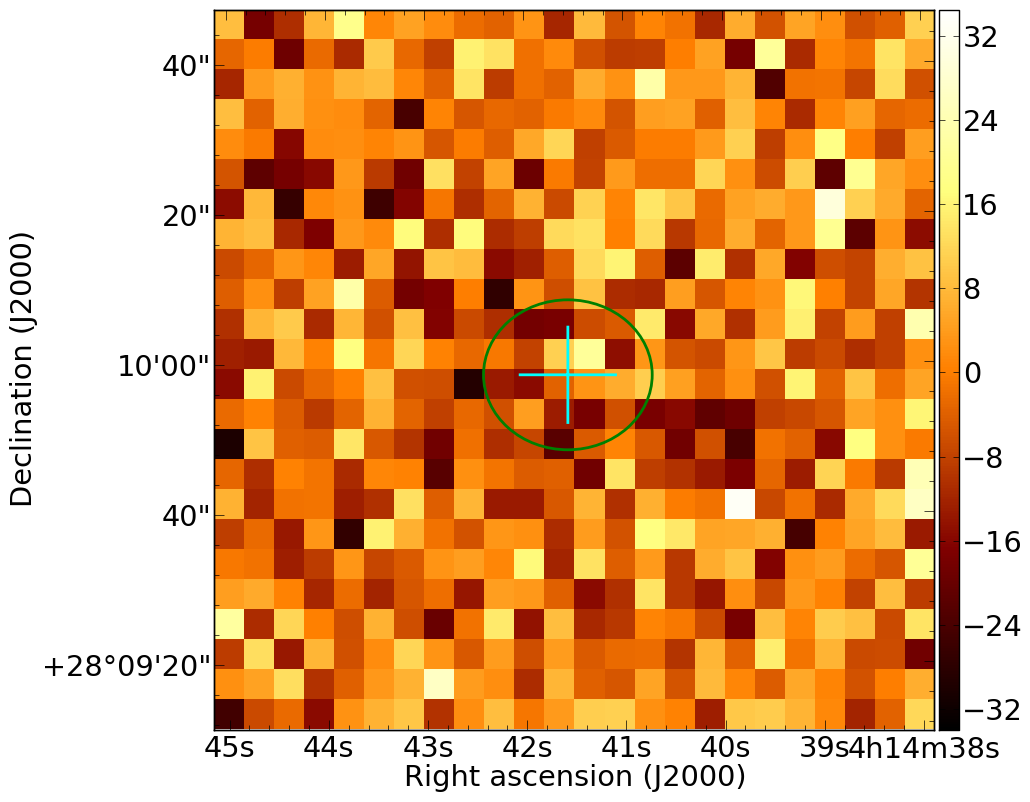}
\caption{as Fig.~\ref{diskfig0} J04144158+2809583 Class III \label{diskfig54}}
 \end{figure} 

\end{document}